%% file: main.tex
\RequirePackage{fix-cm}
\documentclass{svjour3}                     % onecolumn (standard format)
\smartqed  % flush right qed marks, e.g. at end of proof
\usepackage{graphicx}
\input{preembles/index.tex}

\begin{document}

\title{\tool: A Tool for End-to-End Python Library Migration%\thanks{Grants or other notes
}
\input{c-0/authors.tex}


\date{Received: date / Accepted: date}

\maketitle

\begin{abstract}
\input{c-0/abstract.tex}

%\subclass{MSC code1 \and MSC code2 \and more}
\end{abstract}

% ~\\
% \setcounter{tocdepth}{5}
% \tableofcontents

\input{c-0/intro.tex}
\input{c-0/background.tex}
\input{c4-llm/_index.tex}

\input{c5-tool/_index.tex}

\input{c-0/threats.tex}

\input{c-0/related-work.tex}

\input{c-0/conclusion.tex}

\input{c-0/acknowledgements.tex}
\input{c-0/declarations.tex}

\bibliographystyle{spbasic}      % basic style, author-year citations
\bibliography{refs.bib}   % name your BibTeX data base

\end{document}

%% file: preembles/index.tex
\input{preembles/packages.tex}

\input{preembles/settings.tex}

\input{preembles/notations.tex}
\input{preembles/commands--common-text.tex}

\input{preembles/commands--author-comments.tex}
\input{preembles/pymigtax-symbols.tex}

\input{preembles/commands-llmmig.tex}

\input{preembles/data-pymigbench.tex}
\input{preembles/layouts.tex}
\input{c4-llm/_llmmig-results-data.tex}

\input{c4-llm/_llmmig-unseen-results-data.tex}
\input{c5-tool/meta/results.tex} 
\input{c5-tool/meta/mig_complexity_stats.tex}
\input{c5-tool/meta/llm_token_stats.tex}

%% file: preembles/packages.tex
\usepackage{adjustbox}
\usepackage{algorithm}
\usepackage{algpseudocode}
\usepackage{amsmath}
\usepackage{amsfonts}
\usepackage{bookmark}
\usepackage{booktabs}
\usepackage[skip=0.3\baselineskip]{caption}
\usepackage{colortbl}
\usepackage{currfile}
\usepackage[inline]{enumitem}
\usepackage{fancyvrb}
\usepackage{graphicx}
\usepackage{hyperref}
\usepackage{listings}
\usepackage{makecell}
\usepackage{mathtools}
\usepackage{multicol}
\usepackage{multirow}
\usepackage{nameref}
\usepackage{namedef}
\usepackage{natbib}
\usepackage{pbox}
\usepackage{pgffor}
\usepackage{pgfplots}
\pgfplotsset{compat=1.18}
\usepackage{pgfplotstable} % <--- required for \pgfmathprintnumber
\usepackage{relsize}
\usepackage{soul}
\usepackage{subcaption}
\usepackage{tabularx}
\usepackage{tcolorbox}
\usepackage{textcomp}
\usepackage{tikz}
\usetikzlibrary{shapes.geometric, arrows, calc, fit, positioning, decorations.pathreplacing, intersections}
\usepackage{titlecaps}
\usepackage{titlesec}  % to customize section title formats
\usepackage{varioref}
\PassOptionsToPackage{dvipsnames}{xcolor}
\usepackage{xspace}
\usepackage{xstring}
\usepackage{tablefootnote}
\usepackage{fontawesome5}
\usepackage{sparklines}

\usepackage{bold-extra}
\usepackage{slantsc}% http://ctan.org/pkg/slantsc

%% file: preembles/settings.tex
\newcommand{\tablefontsize}{\footnotesize}

\DeclareFontShape{OT1}{cmr}{m}{scit}{<->ssub*cmr/m/sc}{}
\definecolor{backcolour}{rgb}{0.95,0.95,0.92}

\lstdefinestyle{nobottomframe}{
  frame=lrt,
}

\lstdefinestyle{notopframe}{
  frame=lrb,
}

\lstdefinestyle{nohorizframe}{
  frame=lr,
}

\renewcommand{\theparagraph}{\Alph{paragraph}.}
\titleformat{\paragraph}  % command
  [runin]                    % shape: run-in means title and text are on the same line
  {\bfseries}                 % format: bold
  {\theparagraph}         % label
  {.5em}                      % separation between label and title
  {}                         % before-code
  [:\hspace*{.5em}]                         % after-code

\renewcommand{\thesubparagraph}{(\roman{subparagraph})}
\titleformat{\subparagraph}  % command
  [runin]                    % shape: run-in means title and text are on the same line
  {\itshape}                 % format: italic
  {\thesubparagraph}         % label
  {.5em}                      % separation between label and title
  {}                         % before-code
  [:\ ]                         % after-code

  \renewcommand{\cite}{\citep} % make all citations parenthetical by default

%% file: preembles/notations.tex
\newcommand{\lib}[1]{\textit{#1}\xspace}
\newcommand{\aRepoLink}[1]{\href{https://github.com/#1}{#1}}

\newcommand{\aPair}[2]{\textit{#1{\textrightarrow}#2}}
\newcommand{\effortManual}{\ensuremath{\mathit{MigLOC}_\mathit{manual}^{\prime}}\xspace}
\newcommand{\effortAuto}{\ensuremath{\mathit{MigLOC}_{\mathit{auto}}^{\prime}}\xspace}
\newcommand{\migLOCManual}{\ensuremath{\mathit{MigLOC}_\mathit{manual}}\xspace}
\newcommand{\migLOCAuto}{\ensuremath{\mathit{MigLOC}_\mathit{auto}}\xspace}

\definecolor{rem}{HTML}{D55E00} % Magenta
\definecolor{add}{HTML}{009E73} % Teal
\definecolor{resultbg}{RGB}{240,248,255}
\definecolor{thesisstmtbg}{RGB}{228,247,239}

\newcommand{\segment}[1]{\texttt{#1}}

\let\oldttfamily\ttfamily
\renewcommand{\ttfamily}{\oldttfamily\small}

\newcommand{\HLC}[1]{\sethlcolor{red!10}{\textcolor{red}{\hl{#1}}}} % highlight code
\newcommand{\InvisibleCode}[0]{\textcolor{white}{{.}}} % invisible code for spacing

%% file: preembles/commands--common-text.tex
\makeatletter
\DeclareRobustCommand\onedot{\futurelet\@let@token\@onedot}
\def\@onedot{\ifx\@let@token.\else.\null\fi\xspace}

\def\eg{{e.g}\onedot,\xspace} 
\def\ie{{i.e}\onedot,\xspace}

\def\etal{{et al}\onedot}
\makeatother

\newcommand{\migbench}{\textsc{PyMigBench}\xspace}

\newcommand{\tool}[0]{\textsc{MigrateLib}\xspace}

\newcommand{\LLMVersion}{gpt-4o-mini-2024-07-18\xspace}

\newcommand{\PrepRound}{Pre-processing step\xspace}
\newcommand{\prepround}{pre-processing step\xspace}

\newcommand{\LLMMigRound}{LLM migration step\xspace}
\newcommand{\llmmiground}{LLM migration step\xspace}

\newcommand{\PostProcRound}{Post-processing step\xspace}

\newcommand{\MergeSkippedStep}{Merge skipped code\xspace}
\newcommand{\mergeSkippedStep}{merge skipped code\xspace}

\newcommand{\AsyncTransStep}[0]{{Infer async transformation}\xspace}
\newcommand{\asyncTransStep}[0]{{infer async transformation}\xspace}

\newcommand{\codechange}[0]{migration-related code change\xspace}

\newcommand{\codechanges}[0]{migration-related code changes\xspace}

\newcommand{\TestReportPre}{TestReport$_{premig}$\xspace}

\newcommand{\TestReportLLM}{TestReport$_{llmmig}$\xspace}
\newcommand{\TestReportMerge}{TestReport$_{merge}$\xspace}
\newcommand{\TestReportAsync}{TestReport$_{async}$\xspace}
\newcommand{\TestReportManual}{TestReport$_{manual}$\xspace}

\newcommand{\mysubcaption}[1]{\vspace{.5em}\centering{\footnotesize{#1}}}

\newcommand{\spaceOutPar}{\vspace*{1.5em}}

\newcommand{\MinCover}{80\%\xspace}

\newcommand{\ArtifactUrl}{\url{https://doi.org/10.6084/m9.figshare.29602517}}

%% file: preembles/pymigtax-symbols.tex
\newcommand{\fcall}[0]{{function call}\xspace}
\newcommand{\Fcall}[0]{{Function call}\xspace}

\newcommand{\Type}[0]{{Type}\xspace}
\newcommand{\type}[0]{{type}\xspace}
\newcommand{\im}[0]{{import}\xspace}
\newcommand{\imp}[0]{\im}

\newcommand{\Imp}[0]{{Import}\xspace}

\newcommand{\exc}[0]{{exception}\xspace}
\newcommand{\Exc}[0]{{Exception}\xspace}
\newcommand{\at}[0]{{attribute}\xspace}

\newcommand{\attr}[0]{{attribute}\xspace}
\newcommand{\Attr}[0]{{Attribute}\xspace}

\newcommand{\dec}[0]{{decorator}\xspace}
\newcommand{\Dec}[0]{{Decorator}\xspace}

\newcommand{\fref}[0]{{function reference}\xspace}

\newcommand{\oo}[0]{one-to-one\xspace}
\newcommand{\om}[0]{one-to-many\xspace}
\newcommand{\mo}[0]{many-to-one\xspace}
\newcommand{\mm}[0]{many-to-many\xspace}

\newcommand{\OO}[0]{One-to-One\xspace}
\newcommand{\OM}[0]{One-to-Many\xspace}
\newcommand{\MO}[0]{Many-to-One\xspace}
\newcommand{\MM}[0]{Many-to-Many\xspace}
\newcommand{\ZO}[0]{Zero-to-One\xspace}
\newcommand{\OZ}[0]{One-to-Zero\xspace}

\newcommand{\argAdd}[0]{{argument addition}\xspace}

\newcommand{\argDel}[0]{{argument deletion}\xspace}
\newcommand{\argTrans}[0]{{argument transformation}\xspace}
\newcommand{\enc}[0]{{element name change}\xspace}

\newcommand{\argNC}[0]{{argument name change}\xspace}
\newcommand{\ArgNC}[0]{{Argument name change}\xspace}

\newcommand{\ArgAddShort}[0]{{ArgAdd}\xspace}

\newcommand{\ArgDelShort}[0]{{ArgDel}\xspace}

\newcommand{\ArgTransShort}[0]{{ArgTrans}\xspace}

\newcommand{\EncShort}[0]{{ElemNC}\xspace}

\newcommand{\AsyncChangeShort}[0]{{AsyncTrans}\xspace}

\newcommand{\NoPropShort}[0]{{NoProps}\xspace}

\newcommand{\OutTransShort}[0]{{OutTrans}\xspace}

\newcommand{\ArgNCShort}[0]{{ArgNC}\xspace}

\newcommand{\ParamAddShort}[0]{{ParamAdd}\xspace}

\newcommand{\ArgTrans}[0]{{Argument transformation}\xspace}

\newcommand{\ElemNC}[0]{{Element name change}\xspace}

\newcommand{\noProps}[0]{{no properties}\xspace}

\newcommand{\paramAdd}[0]{{parameter addition to decorated function}\xspace}

%% file: preembles/commands-llmmig.tex
\newcommand{\hide}[1]{\ignorespaces}

\newcommand{\taxonomy}{\textsc{PyMigTax}\xspace}

\newcommand{\rqBench}[0]{\ref{rq:benchmark}\xspace}
\newcommand{\rqTest}[0]{\ref{rq:test}\xspace}

\newcommand{\ccs}[0]{code changes\xspace}
\newcommand{\cc}[0]{code change\xspace}

\newcommand{\pre}[1]{#1$_{pre}$\xspace}
\newcommand{\dev}[1]{#1$_{dev}$\xspace}
\newcommand{\llm}[1]{#1$_{llm}$\xspace}

\newcommand{\codePre}{\pre{Code}}
\newcommand{\codeDev}{\dev{Code}}
\newcommand{\codeLLM}{\llm{Code}}
\newcommand{\filePre}{\pre{File}}
\newcommand{\fileDev}{\dev{File}}
\newcommand{\fileLLM}{\llm{File}}

\newcommand{\changeDev}{\dev{Change}}
\newcommand{\changeLLM}{\llm{Change}}

\newcommand{\uChangeDev}{\dev{uChange}}
\newcommand{\uChangeLLM}{\llm{uChange}}

\newcommand{\MigStatusCorrect}{Correct\xspace}
\newcommand{\MigStatusExtraChange}{Correct with non-refactoring changes\xspace}
\newcommand{\MigStatusPartialCorrect}{Partially correct\xspace}
\newcommand{\MigStatusIncorrect}{Incorrect\xspace}
\newcommand{\MigStatusSyntaxError}{Syntax error\xspace}
\newcommand{\MigStatusResponseFailure}{Response failure\xspace}
\newcommand{\MigStatusResponseFailureLower}{response failure\xspace}

\newcommand{\CCStatusCorrect}{Correct\xspace}

\newcommand{\CCStatusIncorrect}{Incorrect\xspace}

\newcommand{\mini}[0]{GPT-4o mini\xspace}
\newcommand{\fouro}[0]{GPT-4o\xspace}
\newcommand{\llama}[0]{Llama 3.1\xspace}

\newcommand{\mn}[0]{Mini\xspace}
\newcommand{\fo}[0]{4o\xspace}
\newcommand{\lm}[0]{Llama\xspace}

\newcommand{\paramAddCell}{{\ParamAddShort}}
\newcommand{\argTransCell}{{\ArgTransShort}}
\newcommand{\argAddCell}{{\ArgAddShort}}
\newcommand{\argDelCell}{{\ArgDelShort}}
\newcommand{\argNCCell}{{\ArgNCShort}}
\newcommand{\asyncTransCell}{{\AsyncChangeShort}}
\newcommand{\outTransCell}{{\OutTransShort}}
\newcommand{\elemNCCell}{{\EncShort}}
\newcommand{\fRefCell}{Function ref}

\newcommand{\evalCountCell}{\#Eval}
\newcommand{\correctPercentCell}{\%Cor}

%% file: c4-llm/_llmmig-results-data.tex
\newcommand{\ArgAddLibPairCount}{49\xspace}
\newcommand{\ArgAddLibPairPercent}{37\%\xspace}
\newcommand{\ArgAddMigCount}{93\xspace}
\newcommand{\ArgAddMigPercent}{29\%\xspace}
\newcommand{\ArgAddTotalCount}{578\xspace}
\newcommand{\ArgAddTotalPercent}{19\%\xspace}
\newcommand{\ArgDelLibPairCount}{37\xspace}
\newcommand{\ArgDelLibPairPercent}{28\%\xspace}
\newcommand{\ArgDelMigCount}{66\xspace}
\newcommand{\ArgDelMigPercent}{21\%\xspace}
\newcommand{\ArgDelTotalCount}{349\xspace}
\newcommand{\ArgDelTotalPercent}{12\%\xspace}
\newcommand{\ArgNCLibPairCount}{11\xspace}
\newcommand{\ArgNCLibPairPercent}{8.2\%\xspace}
\newcommand{\ArgNCMigCount}{17\xspace}
\newcommand{\ArgNCMigPercent}{5.3\%\xspace}
\newcommand{\ArgNCTotalCount}{112\xspace}
\newcommand{\ArgNCTotalPercent}{3.7\%\xspace}
\newcommand{\ArgTransLibPairCount}{38\xspace}
\newcommand{\ArgTransLibPairPercent}{28\%\xspace}
\newcommand{\ArgTransMigCount}{81\xspace}
\newcommand{\ArgTransMigPercent}{25\%\xspace}
\newcommand{\ArgTransTotalCount}{317\xspace}
\newcommand{\ArgTransTotalPercent}{11\%\xspace}
\newcommand{\AsyncTransLibPairCount}{4\xspace}
\newcommand{\AsyncTransLibPairPercent}{3.0\%\xspace}
\newcommand{\AsyncTransMigCount}{13\xspace}
\newcommand{\AsyncTransMigPercent}{4.0\%\xspace}
\newcommand{\AsyncTransTotalCount}{50\xspace}
\newcommand{\AsyncTransTotalPercent}{1.7\%\xspace}
\newcommand{\AttrLibPairCount}{33\xspace}
\newcommand{\AttrLibPairPercent}{25\%\xspace}
\newcommand{\AttrMigCount}{54\xspace}
\newcommand{\AttrMigPercent}{17\%\xspace}
\newcommand{\AttrTotalCount}{183\xspace}
\newcommand{\AttrTotalPercent}{6.1\%\xspace}
\newcommand{\CCAllCount}{2,989\xspace}

\newcommand{\CCArgparseClickCorrectPercent}{94\%\xspace}

\newcommand{\CCExperimentCount}{2,910\xspace}
\newcommand{\DecLibPairCount}{13\xspace}
\newcommand{\DecLibPairPercent}{10\%\xspace}
\newcommand{\DecMigCount}{39\xspace}
\newcommand{\DecMigPercent}{12\%\xspace}
\newcommand{\DecTotalCount}{411\xspace}
\newcommand{\DecTotalPercent}{14\%\xspace}

\newcommand{\ElemNCLibPairCount}{88\xspace}
\newcommand{\ElemNCLibPairPercent}{66\%\xspace}
\newcommand{\ElemNCMigCount}{188\xspace}
\newcommand{\ElemNCMigPercent}{59\%\xspace}
\newcommand{\ElemNCTotalCount}{1,025\xspace}
\newcommand{\ElemNCTotalPercent}{34\%\xspace}
\newcommand{\ExcLibPairCount}{13\xspace}
\newcommand{\ExcLibPairPercent}{10\%\xspace}
\newcommand{\ExcMigCount}{16\xspace}
\newcommand{\ExcMigPercent}{5.0\%\xspace}
\newcommand{\ExcTotalCount}{29\xspace}
\newcommand{\ExcTotalPercent}{1.0\%\xspace}
\newcommand{\FCallLibPairCount}{105\xspace}
\newcommand{\FCallLibPairPercent}{78\%\xspace}
\newcommand{\FCallMigCount}{229\xspace}
\newcommand{\FCallMigPercent}{71\%\xspace}
\newcommand{\FCallTotalCount}{1,804\xspace}
\newcommand{\FCallTotalPercent}{60\%\xspace}
\newcommand{\FRefLibPairCount}{4\xspace}
\newcommand{\FRefLibPairPercent}{3.0\%\xspace}
\newcommand{\FRefMigCount}{4\xspace}
\newcommand{\FRefMigPercent}{1.2\%\xspace}
\newcommand{\FRefTotalCount}{12\xspace}
\newcommand{\FRefTotalPercent}{0.4\%\xspace}

\newcommand{\FouroAvgDiff}{4.8\%\xspace}

\newcommand{\FouroCCBMArgAddCorrectPercent}{88\%\xspace}
\newcommand{\FouroCCBMArgAddTotalCount}{244\xspace}

\newcommand{\FouroCCBMArgDelCorrectPercent}{89\%\xspace}
\newcommand{\FouroCCBMArgDelTotalCount}{234\xspace}

\newcommand{\FouroCCBMArgNCCorrectPercent}{84\%\xspace}
\newcommand{\FouroCCBMArgNCTotalCount}{58\xspace}

\newcommand{\FouroCCBMArgTransCorrectPercent}{77\%\xspace}
\newcommand{\FouroCCBMArgTransTotalCount}{220\xspace}

\newcommand{\FouroCCBMAsyncTransCorrectPercent}{92\%\xspace}
\newcommand{\FouroCCBMAsyncTransTotalCount}{50\xspace}

\newcommand{\FouroCCBMAttrCorrectPercent}{89\%\xspace}
\newcommand{\FouroCCBMAttrTotalCount}{103\xspace}

\newcommand{\FouroCCBMDecCorrectPercent}{87\%\xspace}
\newcommand{\FouroCCBMDecTotalCount}{353\xspace}

\newcommand{\FouroCCBMElemNCCorrectPercent}{90\%\xspace}
\newcommand{\FouroCCBMElemNCTotalCount}{771\xspace}

\newcommand{\FouroCCBMEvaluatedCount}{1,996\xspace}

\newcommand{\FouroCCBMExcCorrectPercent}{93\%\xspace}
\newcommand{\FouroCCBMExcTotalCount}{29\xspace}

\newcommand{\FouroCCBMExcludedCount}{914\xspace}

\newcommand{\FouroCCBMFCallCorrectPercent}{93\%\xspace}
\newcommand{\FouroCCBMFCallTotalCount}{1,092\xspace}

\newcommand{\FouroCCBMFRefCorrectPercent}{100\%\xspace}
\newcommand{\FouroCCBMFRefTotalCount}{12\xspace}

\newcommand{\FouroCCBMHigherCardinalityCorrectPercent}{84\%\xspace}
\newcommand{\FouroCCBMHigherCardinalityTotalCount}{164\xspace}

\newcommand{\FouroCCBMImpCorrectPercent}{98\%\xspace}
\newcommand{\FouroCCBMImpTotalCount}{593\xspace}

\newcommand{\FouroCCBMMMCorrectPercent}{89\%\xspace}
\newcommand{\FouroCCBMMMTotalCount}{27\xspace}

\newcommand{\FouroCCBMMOCorrectPercent}{91\%\xspace}
\newcommand{\FouroCCBMMOTotalCount}{34\xspace}

\newcommand{\FouroCCBMNoPropsCorrectPercent}{97\%\xspace}
\newcommand{\FouroCCBMNoPropsTotalCount}{150\xspace}

\newcommand{\FouroCCBMOMCorrectPercent}{80\%\xspace}
\newcommand{\FouroCCBMOMTotalCount}{103\xspace}

\newcommand{\FouroCCBMOOCorrectPercent}{91\%\xspace}
\newcommand{\FouroCCBMOOTotalCount}{894\xspace}

\newcommand{\FouroCCBMOZCorrectPercent}{100\%\xspace}
\newcommand{\FouroCCBMOZTotalCount}{290\xspace}

\newcommand{\FouroCCBMOutTransCorrectPercent}{92\%\xspace}
\newcommand{\FouroCCBMOutTransTotalCount}{49\xspace}

\newcommand{\FouroCCBMParamAddCorrectPercent}{90\%\xspace}
\newcommand{\FouroCCBMParamAddTotalCount}{127\xspace}

\newcommand{\FouroCCBMTotalCorrectCount}{1,867\xspace}
\newcommand{\FouroCCBMTotalCorrectPercent}{94\%\xspace}
\newcommand{\FouroCCBMTotalIncorrectCount}{129\xspace}
\newcommand{\FouroCCBMTotalIncorrectPercent}{6.5\%\xspace}

\newcommand{\FouroCCBMTypeCorrectPercent}{88\%\xspace}
\newcommand{\FouroCCBMTypeTotalCount}{34\xspace}

\newcommand{\FouroCCBMZOCorrectPercent}{89\%\xspace}
\newcommand{\FouroCCBMZOTotalCount}{55\xspace}

\newcommand{\FouroMigBMAtLeastOneCorrectCount}{296\xspace}
\newcommand{\FouroMigBMAtLeastOneCorrectPercent}{94\%\xspace}
\newcommand{\FouroMigBMCorrectCount}{179\xspace}
\newcommand{\FouroMigBMCorrectPercent}{57\%\xspace}
\newcommand{\FouroMigBMCorrectWithExtraChangesCount}{55\xspace}
\newcommand{\FouroMigBMCorrectWithExtraChangesPercent}{18\%\xspace}

\newcommand{\FouroMigBMIncorrectCount}{2\xspace}
\newcommand{\FouroMigBMIncorrectPercent}{0.6\%\xspace}
\newcommand{\FouroMigBMNoCorrectCount}{18\xspace}
\newcommand{\FouroMigBMNoCorrectPercent}{5.7\%\xspace}
\newcommand{\FouroMigBMPartiallyCorrectCount}{62\xspace}
\newcommand{\FouroMigBMPartiallyCorrectPercent}{20\%\xspace}
\newcommand{\FouroMigBMResponseFailureCount}{13\xspace}
\newcommand{\FouroMigBMResponseFailurePercent}{4.1\%\xspace}
\newcommand{\FouroMigBMSyntaxErrorCount}{3\xspace}
\newcommand{\FouroMigBMSyntaxErrorPercent}{1.0\%\xspace}
\newcommand{\FouroMigTestCorrectCount}{16\xspace}
\newcommand{\FouroMigTestCorrectPercent}{64\%\xspace}
\newcommand{\FouroMigTestIncorrectCount}{7\xspace}
\newcommand{\FouroMigTestIncorrectPercent}{28\%\xspace}
\newcommand{\FouroMigTestPartiallyCorrectCount}{2\xspace}
\newcommand{\FouroMigTestPartiallyCorrectPercent}{8.0\%\xspace}

\newcommand{\HigherCardinalityLibPairCount}{45\xspace}
\newcommand{\HigherCardinalityLibPairPercent}{34\%\xspace}
\newcommand{\HigherCardinalityMigCount}{91\xspace}
\newcommand{\HigherCardinalityMigPercent}{28\%\xspace}
\newcommand{\HigherCardinalityTotalCount}{210\xspace}
\newcommand{\HigherCardinalityTotalPercent}{7.0\%\xspace}
\newcommand{\ImpLibPairCount}{132\xspace}
\newcommand{\ImpLibPairPercent}{99\%\xspace}
\newcommand{\ImpMigCount}{318\xspace}
\newcommand{\ImpMigPercent}{99\%\xspace}
\newcommand{\ImpTotalCount}{732\xspace}
\newcommand{\ImpTotalPercent}{24\%\xspace}

\newcommand{\KappaOne}{0.73\xspace}
\newcommand{\KappaTwo}{0.83\xspace}

\newcommand{\LibPairAllCount}{134\xspace}
\newcommand{\LlamaAvgDiff}{6.6\%\xspace}

\newcommand{\LlamaCCBMArgAddCorrectPercent}{85\%\xspace}
\newcommand{\LlamaCCBMArgAddTotalCount}{150\xspace}

\newcommand{\LlamaCCBMArgDelCorrectPercent}{85\%\xspace}
\newcommand{\LlamaCCBMArgDelTotalCount}{175\xspace}

\newcommand{\LlamaCCBMArgNCCorrectPercent}{60\%\xspace}
\newcommand{\LlamaCCBMArgNCTotalCount}{20\xspace}

\newcommand{\LlamaCCBMArgTransCorrectPercent}{77\%\xspace}
\newcommand{\LlamaCCBMArgTransTotalCount}{115\xspace}

\newcommand{\LlamaCCBMAsyncTransCorrectPercent}{83\%\xspace}
\newcommand{\LlamaCCBMAsyncTransTotalCount}{35\xspace}

\newcommand{\LlamaCCBMAttrCorrectPercent}{83\%\xspace}
\newcommand{\LlamaCCBMAttrTotalCount}{63\xspace}

\newcommand{\LlamaCCBMDecCorrectPercent}{87\%\xspace}
\newcommand{\LlamaCCBMDecTotalCount}{197\xspace}

\newcommand{\LlamaCCBMElemNCCorrectPercent}{84\%\xspace}
\newcommand{\LlamaCCBMElemNCTotalCount}{398\xspace}

\newcommand{\LlamaCCBMEvaluatedCount}{911\xspace}

\newcommand{\LlamaCCBMExcCorrectPercent}{65\%\xspace}
\newcommand{\LlamaCCBMExcTotalCount}{17\xspace}

\newcommand{\LlamaCCBMExcludedCount}{1,999\xspace}

\newcommand{\LlamaCCBMFCallCorrectPercent}{88\%\xspace}
\newcommand{\LlamaCCBMFCallTotalCount}{518\xspace}

\newcommand{\LlamaCCBMFRefCorrectPercent}{33\%\xspace}
\newcommand{\LlamaCCBMFRefTotalCount}{3\xspace}

\newcommand{\LlamaCCBMHigherCardinalityCorrectPercent}{70\%\xspace}
\newcommand{\LlamaCCBMHigherCardinalityTotalCount}{94\xspace}

\newcommand{\LlamaCCBMImpCorrectPercent}{95\%\xspace}
\newcommand{\LlamaCCBMImpTotalCount}{276\xspace}

\newcommand{\LlamaCCBMMMCorrectPercent}{82\%\xspace}
\newcommand{\LlamaCCBMMMTotalCount}{17\xspace}

\newcommand{\LlamaCCBMMOCorrectPercent}{62\%\xspace}
\newcommand{\LlamaCCBMMOTotalCount}{16\xspace}

\newcommand{\LlamaCCBMNoPropsCorrectPercent}{84\%\xspace}
\newcommand{\LlamaCCBMNoPropsTotalCount}{43\xspace}

\newcommand{\LlamaCCBMOMCorrectPercent}{69\%\xspace}
\newcommand{\LlamaCCBMOMTotalCount}{61\xspace}

\newcommand{\LlamaCCBMOOCorrectPercent}{86\%\xspace}
\newcommand{\LlamaCCBMOOTotalCount}{408\xspace}

\newcommand{\LlamaCCBMOZCorrectPercent}{99\%\xspace}
\newcommand{\LlamaCCBMOZTotalCount}{99\xspace}

\newcommand{\LlamaCCBMOutTransCorrectPercent}{88\%\xspace}
\newcommand{\LlamaCCBMOutTransTotalCount}{25\xspace}

\newcommand{\LlamaCCBMParamAddCorrectPercent}{87\%\xspace}
\newcommand{\LlamaCCBMParamAddTotalCount}{110\xspace}

\newcommand{\LlamaCCBMTotalCorrectCount}{808\xspace}
\newcommand{\LlamaCCBMTotalCorrectPercent}{89\%\xspace}
\newcommand{\LlamaCCBMTotalIncorrectCount}{103\xspace}
\newcommand{\LlamaCCBMTotalIncorrectPercent}{11\%\xspace}

\newcommand{\LlamaCCBMTypeCorrectPercent}{75\%\xspace}
\newcommand{\LlamaCCBMTypeTotalCount}{12\xspace}

\newcommand{\LlamaCCBMZOCorrectPercent}{97\%\xspace}
\newcommand{\LlamaCCBMZOTotalCount}{34\xspace}

\newcommand{\LlamaMigBMAtLeastOneCorrectCount}{159\xspace}
\newcommand{\LlamaMigBMAtLeastOneCorrectPercent}{51\%\xspace}
\newcommand{\LlamaMigBMCorrectCount}{83\xspace}
\newcommand{\LlamaMigBMCorrectPercent}{26\%\xspace}
\newcommand{\LlamaMigBMCorrectWithExtraChangesCount}{28\xspace}
\newcommand{\LlamaMigBMCorrectWithExtraChangesPercent}{8.9\%\xspace}

\newcommand{\LlamaMigBMIncorrectCount}{7\xspace}
\newcommand{\LlamaMigBMIncorrectPercent}{2.2\%\xspace}
\newcommand{\LlamaMigBMNoCorrectCount}{155\xspace}
\newcommand{\LlamaMigBMNoCorrectPercent}{49\%\xspace}
\newcommand{\LlamaMigBMPartiallyCorrectCount}{48\xspace}
\newcommand{\LlamaMigBMPartiallyCorrectPercent}{15\%\xspace}
\newcommand{\LlamaMigBMResponseFailureCount}{132\xspace}
\newcommand{\LlamaMigBMResponseFailurePercent}{42\%\xspace}
\newcommand{\LlamaMigBMSyntaxErrorCount}{16\xspace}
\newcommand{\LlamaMigBMSyntaxErrorPercent}{5.1\%\xspace}
\newcommand{\LlamaMigTestCorrectCount}{9\xspace}
\newcommand{\LlamaMigTestCorrectPercent}{36\%\xspace}
\newcommand{\LlamaMigTestIncorrectCount}{15\xspace}
\newcommand{\LlamaMigTestIncorrectPercent}{60\%\xspace}
\newcommand{\LlamaMigTestPartiallyCorrectCount}{1\xspace}
\newcommand{\LlamaMigTestPartiallyCorrectPercent}{4.0\%\xspace}

\newcommand{\MMLibPairCount}{11\xspace}
\newcommand{\MMLibPairPercent}{8.2\%\xspace}
\newcommand{\MMMigCount}{21\xspace}
\newcommand{\MMMigPercent}{6.5\%\xspace}
\newcommand{\MMTotalCount}{27\xspace}
\newcommand{\MMTotalPercent}{0.9\%\xspace}
\newcommand{\MOLibPairCount}{24\xspace}
\newcommand{\MOLibPairPercent}{18\%\xspace}
\newcommand{\MOMigCount}{30\xspace}
\newcommand{\MOMigPercent}{9.3\%\xspace}
\newcommand{\MOTotalCount}{41\xspace}
\newcommand{\MOTotalPercent}{1.4\%\xspace}
\newcommand{\MigAllCount}{321\xspace}

\newcommand{\MigCoveredCount}{27\xspace}
\newcommand{\MigCoveredPassingCount}{25\xspace}
\newcommand{\MigErrorRunningTestCount}{172\xspace}
\newcommand{\MigExperimentCount}{314\xspace}
\newcommand{\MigHasTestsCount}{218\xspace}

\newcommand{\MigSyntaxErrorCount}{5\xspace}
\newcommand{\MigTestEvaluatedCount}{25\xspace}

\newcommand{\MigTestExecutedCount}{46\xspace}
\newcommand{\MiniAvgDiff}{4.0\%\xspace}

\newcommand{\MiniCCBMArgAddCorrectPercent}{80\%\xspace}
\newcommand{\MiniCCBMArgAddTotalCount}{244\xspace}

\newcommand{\MiniCCBMArgDelCorrectPercent}{87\%\xspace}
\newcommand{\MiniCCBMArgDelTotalCount}{234\xspace}

\newcommand{\MiniCCBMArgNCCorrectPercent}{59\%\xspace}
\newcommand{\MiniCCBMArgNCTotalCount}{58\xspace}

\newcommand{\MiniCCBMArgTransCorrectPercent}{76\%\xspace}
\newcommand{\MiniCCBMArgTransTotalCount}{201\xspace}

\newcommand{\MiniCCBMAsyncTransCorrectPercent}{90\%\xspace}
\newcommand{\MiniCCBMAsyncTransTotalCount}{50\xspace}

\newcommand{\MiniCCBMAttrCorrectPercent}{78\%\xspace}
\newcommand{\MiniCCBMAttrTotalCount}{103\xspace}

\newcommand{\MiniCCBMDecCorrectPercent}{90\%\xspace}
\newcommand{\MiniCCBMDecTotalCount}{335\xspace}

\newcommand{\MiniCCBMElemNCCorrectPercent}{85\%\xspace}
\newcommand{\MiniCCBMElemNCTotalCount}{773\xspace}

\newcommand{\MiniCCBMEvaluatedCount}{1,966\xspace}

\newcommand{\MiniCCBMExcCorrectPercent}{86\%\xspace}
\newcommand{\MiniCCBMExcTotalCount}{29\xspace}

\newcommand{\MiniCCBMExcludedCount}{944\xspace}

\newcommand{\MiniCCBMFCallCorrectPercent}{87\%\xspace}
\newcommand{\MiniCCBMFCallTotalCount}{1,093\xspace}

\newcommand{\MiniCCBMFRefCorrectPercent}{58\%\xspace}
\newcommand{\MiniCCBMFRefTotalCount}{12\xspace}

\newcommand{\MiniCCBMHigherCardinalityCorrectPercent}{79\%\xspace}
\newcommand{\MiniCCBMHigherCardinalityTotalCount}{164\xspace}

\newcommand{\MiniCCBMImpCorrectPercent}{94\%\xspace}
\newcommand{\MiniCCBMImpTotalCount}{580\xspace}

\newcommand{\MiniCCBMMMCorrectPercent}{78\%\xspace}
\newcommand{\MiniCCBMMMTotalCount}{27\xspace}

\newcommand{\MiniCCBMMOCorrectPercent}{82\%\xspace}
\newcommand{\MiniCCBMMOTotalCount}{34\xspace}

\newcommand{\MiniCCBMNoPropsCorrectPercent}{88\%\xspace}
\newcommand{\MiniCCBMNoPropsTotalCount}{150\xspace}

\newcommand{\MiniCCBMOMCorrectPercent}{78\%\xspace}
\newcommand{\MiniCCBMOMTotalCount}{103\xspace}

\newcommand{\MiniCCBMOOCorrectPercent}{84\%\xspace}
\newcommand{\MiniCCBMOOTotalCount}{877\xspace}

\newcommand{\MiniCCBMOZCorrectPercent}{99\%\xspace}
\newcommand{\MiniCCBMOZTotalCount}{290\xspace}

\newcommand{\MiniCCBMOutTransCorrectPercent}{78\%\xspace}
\newcommand{\MiniCCBMOutTransTotalCount}{49\xspace}

\newcommand{\MiniCCBMParamAddCorrectPercent}{95\%\xspace}
\newcommand{\MiniCCBMParamAddTotalCount}{127\xspace}

\newcommand{\MiniCCBMTotalCorrectCount}{1,744\xspace}
\newcommand{\MiniCCBMTotalCorrectPercent}{89\%\xspace}
\newcommand{\MiniCCBMTotalIncorrectCount}{222\xspace}
\newcommand{\MiniCCBMTotalIncorrectPercent}{11\%\xspace}

\newcommand{\MiniCCBMTypeCorrectPercent}{85\%\xspace}
\newcommand{\MiniCCBMTypeTotalCount}{34\xspace}

\newcommand{\MiniCCBMZOCorrectPercent}{87\%\xspace}
\newcommand{\MiniCCBMZOTotalCount}{55\xspace}

\newcommand{\MiniMigBMAtLeastOneCorrectCount}{291\xspace}
\newcommand{\MiniMigBMAtLeastOneCorrectPercent}{93\%\xspace}
\newcommand{\MiniMigBMCorrectCount}{154\xspace}
\newcommand{\MiniMigBMCorrectPercent}{49\%\xspace}
\newcommand{\MiniMigBMCorrectWithExtraChangesCount}{53\xspace}
\newcommand{\MiniMigBMCorrectWithExtraChangesPercent}{17\%\xspace}

\newcommand{\MiniMigBMIncorrectCount}{6\xspace}
\newcommand{\MiniMigBMIncorrectPercent}{1.9\%\xspace}
\newcommand{\MiniMigBMNoCorrectCount}{23\xspace}
\newcommand{\MiniMigBMNoCorrectPercent}{7.3\%\xspace}
\newcommand{\MiniMigBMPartiallyCorrectCount}{84\xspace}
\newcommand{\MiniMigBMPartiallyCorrectPercent}{27\%\xspace}
\newcommand{\MiniMigBMResponseFailureCount}{11\xspace}
\newcommand{\MiniMigBMResponseFailurePercent}{3.5\%\xspace}
\newcommand{\MiniMigBMSyntaxErrorCount}{6\xspace}
\newcommand{\MiniMigBMSyntaxErrorPercent}{1.9\%\xspace}
\newcommand{\MiniMigTestCorrectCount}{13\xspace}
\newcommand{\MiniMigTestCorrectPercent}{52\%\xspace}
\newcommand{\MiniMigTestIncorrectCount}{7\xspace}
\newcommand{\MiniMigTestIncorrectPercent}{28\%\xspace}
\newcommand{\MiniMigTestPartiallyCorrectCount}{5\xspace}
\newcommand{\MiniMigTestPartiallyCorrectPercent}{20\%\xspace}

\newcommand{\NoPropsLibPairCount}{36\xspace}
\newcommand{\NoPropsLibPairPercent}{27\%\xspace}
\newcommand{\NoPropsMigCount}{57\xspace}
\newcommand{\NoPropsMigPercent}{18\%\xspace}
\newcommand{\NoPropsTotalCount}{262\xspace}
\newcommand{\NoPropsTotalPercent}{8.8\%\xspace}

\newcommand{\OMLibPairCount}{23\xspace}
\newcommand{\OMLibPairPercent}{17\%\xspace}
\newcommand{\OMMigCount}{52\xspace}
\newcommand{\OMMigPercent}{16\%\xspace}
\newcommand{\OMTotalCount}{142\xspace}
\newcommand{\OMTotalPercent}{4.8\%\xspace}
\newcommand{\OOLibPairCount}{98\xspace}
\newcommand{\OOLibPairPercent}{73\%\xspace}
\newcommand{\OOMigCount}{200\xspace}
\newcommand{\OOMigPercent}{62\%\xspace}
\newcommand{\OOTotalCount}{1,552\xspace}
\newcommand{\OOTotalPercent}{52\%\xspace}
\newcommand{\OZLibPairCount}{25\xspace}
\newcommand{\OZLibPairPercent}{19\%\xspace}
\newcommand{\OZMigCount}{50\xspace}
\newcommand{\OZMigPercent}{16\%\xspace}
\newcommand{\OZTotalCount}{297\xspace}
\newcommand{\OZTotalPercent}{10\%\xspace}
\newcommand{\OutTransLibPairCount}{15\xspace}
\newcommand{\OutTransLibPairPercent}{11\%\xspace}
\newcommand{\OutTransMigCount}{16\xspace}
\newcommand{\OutTransMigPercent}{5.0\%\xspace}
\newcommand{\OutTransTotalCount}{49\xspace}
\newcommand{\OutTransTotalPercent}{1.6\%\xspace}
\newcommand{\ParamAddLibPairCount}{1\xspace}
\newcommand{\ParamAddLibPairPercent}{0.7\%\xspace}
\newcommand{\ParamAddMigCount}{13\xspace}
\newcommand{\ParamAddMigPercent}{4.0\%\xspace}
\newcommand{\ParamAddTotalCount}{127\xspace}
\newcommand{\ParamAddTotalPercent}{4.2\%\xspace}

\newcommand{\TypeLibPairCount}{13\xspace}
\newcommand{\TypeLibPairPercent}{10\%\xspace}
\newcommand{\TypeMigCount}{17\xspace}
\newcommand{\TypeMigPercent}{5.3\%\xspace}
\newcommand{\TypeTotalCount}{55\xspace}
\newcommand{\TypeTotalPercent}{1.8\%\xspace}
\newcommand{\ZOLibPairCount}{18\xspace}
\newcommand{\ZOLibPairPercent}{13\%\xspace}
\newcommand{\ZOMigCount}{35\xspace}
\newcommand{\ZOMigPercent}{11\%\xspace}
\newcommand{\ZOTotalCount}{198\xspace}
\newcommand{\ZOTotalPercent}{6.6\%\xspace}

%% file: c5-tool/meta/results.tex
\newcommand{\LibExpCount}{55\xspace}
\newcommand{\LibPairAnalogousCount}{60\xspace}
\newcommand{\LibPairCandidateCount}{4,856\xspace}
\newcommand{\LibPairExpCount}{48\xspace}
\newcommand{\LibPairLowFrequencyCount}{4,215\xspace}
\newcommand{\LibPairManuallyValidatedCount}{449\xspace}

\newcommand{\LibPairOldTargetCount}{192\xspace}
\newcommand{\MigAsyncTransformAppliedCount}{129\xspace}

\newcommand{\MigAsyncTransformCorrectCount}{233\xspace}
\newcommand{\MigAsyncTransformCorrectPercent}{32\%\xspace}
\newcommand{\MigAsyncTransformImprovedCount}{45\xspace}
\newcommand{\MigAsyncTransformImprovedPercent}{35\%\xspace}

\newcommand{\MigAsyncTransformMeanImprovement}{65\%\xspace}
\newcommand{\MigAsyncTransformMedianCorrectness}{86\%\xspace}

\newcommand{\MigExpCount}{717\xspace}
\newcommand{\MigLLMMigCorrectCount}{193\xspace}
\newcommand{\MigLLMMigCorrectPercent}{27\%\xspace}

\newcommand{\MigLLMMigMedianCorrectness}{65\%\xspace}

\newcommand{\MigManualFixApplied}{43\xspace}
\newcommand{\MigManualFixCorrected}{24\xspace}
\newcommand{\MigManualFixCorrectedPercent}{56\%\xspace}

\newcommand{\MigManualFixNotCorrected}{19\xspace}

\newcommand{\MigMergeSkippedAppliedCount}{260\xspace}

\newcommand{\MigMergeSkippedCorrectCount}{215\xspace}
\newcommand{\MigMergeSkippedCorrectPercent}{30\%\xspace}
\newcommand{\MigMergeSkippedImprovedCount}{88\xspace}
\newcommand{\MigMergeSkippedImprovedPercent}{34\%\xspace}

\newcommand{\MigMergeSkippedMeanImprovement}{46\%\xspace}
\newcommand{\MigMergeSkippedMedianCorrectness}{80\%\xspace}

\newcommand{\MigOverallCorrectCount}{233\xspace}
\newcommand{\MigOverallCorrectPercent}{32\%\xspace}

\newcommand{\MigOverallNeedManualFixCount}{484\xspace}

\newcommand{\MiglocAutoMedianPercent}{73\%\xspace}

\newcommand{\MiglocManualMedianPercent}{27\%\xspace}

\newcommand{\RepoAfterLargeFilterCount}{64,909\xspace}
\newcommand{\RepoAfterLowCoverageFilterCount}{800\xspace}
\newcommand{\RepoAfterLowPassingFilterCount}{1,357\xspace}
\newcommand{\RepoAfterManualFilterCount}{798\xspace}
\newcommand{\RepoAfterNoReqFileFilterCount}{24,993\xspace}
\newcommand{\RepoAfterNoTestFileByApiFilterCount}{12,435\xspace}
\newcommand{\RepoAfterTestNotRunFilterCount}{2,314\xspace}

\newcommand{\RepoExcludedLargeCount}{30,563\xspace}
\newcommand{\RepoExcludedLowCoverageCount}{557\xspace}
\newcommand{\RepoExcludedLowPassingCount}{957\xspace}
\newcommand{\RepoExcludedManualCount}{2\xspace}
\newcommand{\RepoExcludedNoReqFileCount}{39,916\xspace}
\newcommand{\RepoExcludedNoTestFileByApiCount}{12,558\xspace}
\newcommand{\RepoExcludedTestNotRunCount}{10,104\xspace}

\newcommand{\RepoExpCount}{175\xspace}
\newcommand{\RepoIncludedCount}{798\xspace}
\newcommand{\RepoManualFixApplied}{39\xspace}
\newcommand{\RepoNeedManualFixCount}{147\xspace}
\newcommand{\RepoSeartCount}{95,472\xspace}

\newcommand{\SeartFilterLastCommitAfter}{January 1, 2024\xspace}
\newcommand{\SeartFilterMinLines}{100\xspace}
\newcommand{\SourceLibAfterNoDescriptionFilterCount}{805\xspace}
\newcommand{\SourceLibAfterNotOnLibioFilterCount}{1,409\xspace}

\newcommand{\SourceLibCandidateCount}{1,421\xspace}
\newcommand{\SourceLibExcludedNoDescriptionCount}{2\xspace}
\newcommand{\SourceLibExcludedNotOnLibioCount}{12\xspace}
\newcommand{\SourceLibExcludedTooFewDependentsCount}{602\xspace}
\newcommand{\SourceLibExpCount}{24\xspace}
\newcommand{\SourceLibIncludedCount}{805\xspace}
\newcommand{\TargetLibExpCount}{37\xspace}

%% file: c5-tool/meta/mig_complexity_stats.tex
\newcommand{\MigHunksMin}{1\xspace}
\newcommand{\MigHunksMax}{1,738\xspace}
\newcommand{\MigHunksMean}{36\xspace}
\newcommand{\MigHunksMedian}{14\xspace}
\newcommand{\MigLinesMin}{1\xspace}
\newcommand{\MigLinesMax}{5,644\xspace}
\newcommand{\MigLinesMean}{217\xspace}
\newcommand{\MigLinesMedian}{64\xspace}
\newcommand{\MigImportOnlyCount}{5\xspace}
\newcommand{\MigFunctionCallPercent}{97\%\xspace}
\newcommand{\MigFunctionCallAvg}{67\xspace}
\newcommand{\MigExceptionPercent}{37\%\xspace}
\newcommand{\MigExceptionAvg}{2\xspace}
\newcommand{\MigDecoratorPercent}{33\%\xspace}
\newcommand{\MigDecoratorAvg}{9\xspace}
\newcommand{\MigAttributePercent}{81\%\xspace}
\newcommand{\MigAttributeAvg}{28\xspace}
\newcommand{\MigFullyCorrectCount}{229\xspace}
\newcommand{\MigFCFilesMin}{1\xspace}
\newcommand{\MigFCFilesMax}{17\xspace}
\newcommand{\MigFCFilesMean}{2\xspace}
\newcommand{\MigFCHunksMin}{1\xspace}
\newcommand{\MigFCHunksMax}{180\xspace}
\newcommand{\MigFCHunksMean}{13\xspace}
\newcommand{\MigFCLinesMin}{1\xspace}
\newcommand{\MigFCLinesMax}{3,889\xspace}
\newcommand{\MigFCLinesMean}{140\xspace}
\newcommand{\MigFCFunctionCallPercent}{97\%\xspace}
\newcommand{\MigFCExceptionPercent}{29\%\xspace}
\newcommand{\MigFCDecoratorPercent}{19\%\xspace}
\newcommand{\MigFCAttributePercent}{73\%\xspace}
\newcommand{\MigFilesMin}{1\xspace}
\newcommand{\MigFilesMax}{183\xspace}
\newcommand{\MigFilesMean}{4\xspace}
\newcommand{\MigFilesMedian}{2\xspace}

%% file: c5-tool/meta/llm_token_stats.tex
\newcommand{\tokensInTotal}{3,834,385\xspace}
\newcommand{\tokensOutTotal}{2,491,712\xspace}
\newcommand{\tokensTotal}{6,326,097\xspace}
\newcommand{\tokensMedianInPerMig}{2,704\xspace}
\newcommand{\tokensMedianOutPerMig}{2,110\xspace}

\newcommand{\tokenPriceDate}{2 June 2026\xspace}
\newcommand{\FouroCostPerMilIn}{US\$2.5\xspace}
\newcommand{\FouroCostPerMilOut}{US\$10\xspace}

\newcommand{\MedianCostPerProject}{US\$0.028\xspace}
\newcommand{\MaxCostPerProject}{US\$0.38\xspace}

%% file: c-0/authors.tex
\author{Mohayeminul Islam         \and
        Ajay Kumar Jha       \and
        May Mahmoud          \and
        Sarah Nadi        
}

\authorrunning{Islam, Jha, Mahmoud and Nadi} % if too long for running head

\institute{M. Islam \at
              University of Alberta \\
              \email{mohayemin@ualberta.ca} \and
           A. Jha \at
              North Dakota State University \\
              \email{ajay.jha.1@ndsu.edu} \and
            M. Mahmoud \at
            New York University Abu Dhabi \\
            \email{m.mahmoud@nyu.edu} \and
            S. Nadi \at
            New York University Abu Dhabi \\
            \email{sarah.nadi@nyu.edu}
}

%% file: c-0/abstract.tex
Library migration is the process of replacing a library with a similar one in a software project.
Manual library migration is time-consuming and error-prone,
as it requires developers to understand the Application Programming Interfaces (API) of both libraries,
map equivalent APIs, and perform the necessary code transformations. 
% Due to the difficulty of the library migration process, most of the existing automated techniques and tooling stop at the API mapping stage or support a limited set of libraries and code transformations.
This paper presents an LLM-based end-to-end solution that can automatically migrate code between any arbitrary pair of Python libraries that provide similar functionality.
Before building the technique, we first systematically study the capabilities of LLMs for library migration on PyMigBench, an existing benchmark of \MigExperimentCount real-world library migrations.
We find that the best performing LLM, \fouro, can correctly perform \FouroMigBMCorrectPercent of the migrations.
We also identify post-processing steps that can further improve the performance.
Based on this, we develop \tool, a command-line application that combines the power of LLMs, static analysis, and dynamic analysis to provide accurate library migration.
We evaluate \tool on \MigExpCount migration scenarios in \RepoExpCount real-world Python applications that are not from PyMigBench.
We find that \tool can migrate \MigOverallCorrectPercent of the migrations with complete correctness without any developer intervention.
We inspect a sample of the incomplete migrations to understand how many of the required changes have already been done by \tool and how many manual changes are still needed to complete the migration.
We find that for more than half of the inspected migrations, \tool completed  \MiglocAutoMedianPercent of the required migration-related changes, leaving only \MiglocManualMedianPercent of the required changes for manual intervention.

%% file: c-0/intro.tex
\section{Introduction}
\label{intro}

% Storyline:
% 1. We do an experiment using the migrations in the benchmark, do manual evaluation. In this phase, we do multiple runs per file, and do CC level granular evaluation.
% 2. We learn from this phase, and improve the process and build a tool.
% 3. We do a large scale "simulated" evaluation of the tool on client projects, and show that it works well. This evaluation checks runtime behavior of the tool using unit tests, and does not require manual evaluation.
% 4. For a portion of the failed migrations, we attempt to fix them manually, and show that the LLMs were close to the correct migration.

\textit{Library migration} is the process of replacing a library with an alternative one in a software project.
Once a developer decides on a new target library, library migration involves identifying the old library usage in the codebase, finding the corresponding Application Programming Interfaces (APIs) in the new library, and replacing the old API usage with the new API usage.
This can be tedious and error-prone, especially when the library is extensively used in a large codebase \cite{kula2018developers}.

Researchers have proposed various techniques to automate the library migration process. Most of these techniques, however, are limited to only finding the analogous APIs (\textit{API mapping}) \cite{teyton2013automatic, alrubaye2018automating, alrubaye2019use, miningAnalogicalAPIs, zhang2020deep, di2025deepmig}, but do not replace the old API usage in the codebase with the new one (\textit{code transformation}).
To find analogous APIs, some of these techniques mine existing migrations \cite{teyton2013automatic, alrubaye2018automating, alrubaye2019use} while others use machine learning models \cite{miningAnalogicalAPIs, di2025deepmig} and natural language processing techniques \cite{ni2021soar}.
To the best of our knowledge, there are two techniques that actually perform code transformation \cite{ni2021soar, nikolov2025google}, but both techniques evaluate their approaches on a small set of selected library pairs without providing a tool that can be used in practice.

In recent years, Large Language Models (LLMs) have shown that they can be effective in various software engineering tasks~\cite{ozkaya2023application, fan2023large, wang2024software}, 
including the necessary upstream tasks for library migration, such as code generation~\cite{nguyen2022empirical, peng2023impact, murali2023codecompose}, code comprehension~\cite{nam2024using}, and code transformation~\cite{wei2023copiloting, xia2023automated, zhou2023hybrid}.
This suggests that LLMs may be capable of performing library migration.
However, studies have also identified limitations of LLMs in other software engineering tasks, especially generating incorrect or incomplete code~\cite{fan2023large}.
These previous efforts suggest that LLMs may be capable of performing library migration and that LLM-based migration tooling could be effective.
However, we still need further empirical evidence to support this claim.

In this paper, our goal is to create an end-to-end solution for Python library migration that can potentially leverage the power of LLMs.
We do not consider the problem of deciding which library to migrate to, but rather focus on the actual migration process once the developer has already decided on the target library.
We choose Python as the programming language for our study, as it is one of the most popular programming languages and has a rich ecosystem of libraries.
We approach our goal in three phases as illustrated in Figure~\ref{fig:study-overview}.
In Phase 1, we first systematically investigate the capabilities and challenges of using LLMs for library migration. We conduct an experiment using \MigExperimentCount migrations  from \migbench~\cite{pymigbench}, a Python library migration benchmark we previously built.
We experiment with three LLMs (\llama, \mini, and \fouro) and measure the correctness of the migrations by comparing the LLM's migrations with the original developer's migrations.
We find that the three LLMs can correctly migrate \LlamaMigBMCorrectPercent, \MiniMigBMCorrectPercent, and \FouroMigBMCorrectPercent of the migrations, respectively.
Using an alternative test-based evaluation on a subset of the projects that have unit tests, we find that  \LlamaMigTestCorrectPercent, \MiniMigTestCorrectPercent, and \FouroMigTestCorrectPercent of the LLM migrations maintain the same test results as the original developer migrations for the three LLMs, respectively.
We also study the types of migration-related code changes LLMs excel at vs. struggle with.
Overall, we find that LLMs are capable of library migration, and as expected, the more powerful LLM (\fouro) performs best.
We also identify four major challenges that make solely relying on LLMs in a practical tool less effective: 
\begin{enumerate*}[label=(\arabic*)]
    \item The files that need migration are not always known beforehand.
	\item All LLMs often return only part of the code, skipping parts of the code that do not require migration. 
    \item For a given project, a file that we migrate may have other files that depend on it which may require changes during migration.
    Specifically, if any functions are made asynchronous during the migration, all parts of the code that use those functions must also be made asynchronous.
    \item The inconsistencies between the versions between the libraries used in the project and those assumed by the LLM can lead to incorrect migrations.
\end{enumerate*}

In Phase 2, we build an LLM-based CLI tool, \tool, that automates the full library migration process while addressing the previously identified limitations through pre- and post- processing using static and dynamic program analysis.
\tool takes in a complete Python project with an existing test suite and automatically performs the migration for the specified library pair.
\tool first identifies all files that require migration using static analysis and a dynamic call graph and sends those files to the LLM via a prompt that instructs the LLM to perform the migration.
\tool then verifies the results of the migration by comparing the test results before and after migration.
If any previously passing tests now fail, \tool performs a post-processing step where it checks if the LLM skipped any code and merges it back.
It also detects whether any functions were made async during the migration, and if so, finds other parts of the project that use those functions and makes them async as well.
Finally, it runs the unit tests again to determine the final migration status.

\input{c-0/img/overview.tex}

In Phase 3, we evaluate \tool using a new purely test-based evaluation setup, as opposed to the benchmark-based evaluation we did in Phase 1.
This allows us to conduct a larger-scale evaluation and also evaluate new migrations without the need for ground truth developer changes.
We run \tool on a set of \MigExpCount simulated migrations in \RepoExpCount unique projects.
We define a \textit{simulated migration} as one that is not retrieved from the project's history.
Specifically, we find source libraries used in a project and migrate the project to an analogous target library.
We ensure that the projects we choose have high code coverage (above \MinCover), so that the evaluation is reliable.
Given that \fouro had the best performance in Phase 1, we conduct this evaluation only using \fouro.
We find that \tool is able to successfully complete \MigOverallCorrectPercent migrations without any manual intervention (i.e., all tests that pass before migration still pass after migration).
To understand how much manual work is required to fix the remaining \MigOverallNeedManualFixCount migrations that the LLM left in an incomplete stage, we randomly select a subset of \MigManualFixApplied diverse migrations and attempt to manually fix them.
We measure the amount of manual work required to complete these migrations using the number of lines of code we need to manually edit.
We find that we are able to fix \MigManualFixCorrected (\MigManualFixCorrectedPercent) of our sampled migrations, where the manual code modification is only \MiglocManualMedianPercent of the total changes needed for the migration.
%The main reason we could not fix some of the remaining migrations is that the projects had complex logic that required a deep understanding of the project context, which was difficult to achieve.

Overall, our work provides a combination of empirical and engineering contributions that systematically address the challenges of library migration. This paper is an extended version of our prior conference paper~\cite{llmmig} that only focused on Phase 1. Phases 2 and 3 are novel contributions of this work.  
To summarize, the contributions of this paper are:
\begin{enumerate}
    \item An empirical study to explore the use of LLMs for library migration by comparing LLM changes to \MigExperimentCount benchmark developer changes. Our analysis systematically highlights the types of migration-related code changes LLMs excel at vs. struggle with.     
    \item We design and build an open-source end-to-end CLI tool, \tool, that automates the full library migration process using a combination of LLMs and pre- and post-processing using static and dynamic program analysis.
    \item We design a new evaluation setup that allows us to evaluate \tool on a large scale and also on newer projects. We run an extensive test-based evaluation of \tool on a set of \MigExpCount simulated migrations, and show that it works well with no or minimal manual editing for a diverse set of library pairs.
\end{enumerate}

\noindent
The source code of \tool, the experimental data, and the detailed results of our experiments are available at \ArtifactUrl.

%% file: c-0/img/overview.tex
\begin{figure}[t]
    \hypersetup{hidelinks}
    \centering
    \begin{tikzpicture}
        \tikzstyle{challengeBox} = [
            rectangle, dashed, draw=black, thick,
            inner sep=0.25cm,
        ]
        \tikzstyle{challengeText} = [                
            align=center,           
            text width=6.5cm,     
            inner xsep=-.40cm,
            dotted,
        ]
        \tikzstyle{contrib} = [
            rectangle, rounded corners, 
            draw=black,    
            align=center,     
            font=\footnotesize,  
        ]

        \tikzstyle{artifact} = [
            %rectangle, rounded corners, 
            %minimum width=3cm, 
            %minimum height=1cm,
            draw=none,    
            align=center,     
            font=\scriptsize,  
        ]

        \tikzstyle{conToConArrow} = [
            ->,>=triangle 45,
            font=\scriptsize
        ]

        \tikzstyle{model} = [
            rectangle, rounded corners, 
            draw=none, fill=gray!20,
            font=\scriptsize
        ]

        \def\xGapFirst{.4cm}
        \def\xgap{.75cm}
        \def\ygap{1cm}
        \def\contribOneTwoYGap{.50cm}

        \node (migbench) [artifact]{\migbench};

        \node (phase1) [contrib, anchor=north, yshift=-\ygap] at (migbench.south) {
            \hyperref[ch:llm]{Phase 1} \\
            \hyperref[ch:llm]{Explore LLMs for Library Migration}\\
            \tikz \node[model]{\llama};
            \tikz \node [model]{\mini};
            \tikz \node [model]{\fouro};
        };

        \node (phase2) [contrib, anchor=west, xshift=\xgap] at (phase1.east) {
            \hyperref[ch:tool]{Phase 2} \\
            \hyperref[ch:tool]{Build \tool}
        };

        \node (phase3) [contrib, anchor=west, xshift=\xgap] at (phase2.east) {
            \hyperref[ch:tool]{Phase 3} \\
            \hyperref[ch:tool]{Evaluate \tool}\\
            \tikz \node [model]{\fouro};
        };

        \node (simulated) [artifact, anchor=south, yshift=\ygap] at (phase3.north){Simulated migrations};

        \draw[conToConArrow] 
    (migbench.south) -- node[midway, right] {\MigExperimentCount migrations} (phase1.north);

        \draw[conToConArrow] (simulated.south)  -- node[midway, left] {\MigExpCount migrations}  (phase3.north);

        \draw[conToConArrow] (phase1.east) -> (phase2.west);
        \draw[conToConArrow] (phase2.east) -> (phase3.west);

    \end{tikzpicture}
    \caption{Overview of the contributions.}
    \label{fig:study-overview}
\end{figure}

%% file: c-0/background.tex
\section{Background and Terminology}
\label{sec:background}
In this section, we provide background information and terminology used in the rest of the paper.
\subsection{Library migration}

\textit{Library migration} is the process of updating a software project to replace a used library with another one that provides similar functionality \cite{teyton2012mining}.
The \textit{source} library is the one being replaced, and the \textit{target} library is the one that replaces it \cite{teyton2012mining}.
One \textit{migration} instance refers to a commit in a repository where a migration happened from a specific source library to a target library \cite{pymigbench}, denoted using the notation \aPair{source}{target}. 
For example, the commit \href{https://github.com/openstack/ironic/commit/b0607a26}{\texttt{b0607a26}} in repository \href{https://github.com/openstack/ironic}{\textit{openstack/ironic}} 
is a \aPair{retrying}{tenacity} migration.

A \textit{\codechange}, or \textit{\cc} for brevity, is a minimal replacement of source library APIs with target library APIs that cannot be meaningfully reduced further without losing the semantics of the change~\cite{pymigbench}.
A migration contains one or more \ccs.
The lines between the boxes in \autoref{fig:mig-example} show \ccs.
For example, segment P1 in \autoref{fig:mig-example-pre} is replaced by segment D1 in \autoref{fig:mig-example-dev}.

We use subscripts \textit{pre}, \textit{dev}, and \textit{llm} to denote data before migration, after developer's migration, and after an LLM's migration, respectively.
For example, \codePre, \codeDev, and \codeLLM denote three states of a piece of code.
\changeDev and \changeLLM denote \ccs by developers and an LLM, respectively.

\subsection{\migbench and \taxonomy}
We use two of our prior artifacts: \migbench \cite{pymigbench} and \taxonomy \cite{pymigtax}, in this work which we briefly describe next.
\migbench is a dataset of real-world Python library migrations, which we mined from open source GitHub repositories.
We use the latest version of \migbench, version 2.2.5, in our experiments in this paper.
The dataset has \MigAllCount migrations and \CCAllCount \codechanges.
%\migbench is that
It includes a detailed description of each \cc, including line numbers and API names, which facilitates our evaluation.

\taxonomy is a taxonomy of \codechanges \cite{pymigtax}, built using the first version of \migbench. 
In our prior work, we validated its generalizability on additional third-party data.
\taxonomy describes a code change based on three dimensions:
\textit{(1) Program elements}: the types of program elements involved in the change (\eg~\fcall and \attr);
\textit{(2) Cardinality}: how many source APIs are replaced with how many target APIs. For example, \om cardinality means one source API is replaced with multiple target APIs;
    \om, \mo, and \mm cardinalities are commonly referred to as \textit{higher-cardinality}; and
\textit{(3) Properties}: Additional properties to describe the \cc (\eg~\enc when the source and target APIs have different names.). %and \argAdd when the target API needs an additional argument to achieve the same behavior). 
The \cc data in \migbench is already annotated with the \taxonomy categories.
We use \taxonomy categories (shown in \autoref{tab:rq:cc-freq}) to understand the types of \ccs LLMs can handle.
The \taxonomy paper~\cite{pymigtax} has full category descriptions.
%Full descriptions available in the \taxonomy paper \cite{pymigtax}.

%% file: c4-llm/_index.tex
\section{Phase 1: Exploring LLMs for library migration}
\label{ch:llm}

\input{c4-llm/intro.tex}

\input{c4-llm/experiment-setup.tex}

\input{c4-llm/evaluation-benchmark--findings.tex}

\input{c4-llm/evaluation-unittest--findings.tex}

\input{c4-llm/discussion.tex}

%% file: c4-llm/intro.tex
We first systematically explore the capabilities of LLMs for library migration, while understanding how they handle different types of migration-related code changes.
We first explain our setup and methods, and then present the results.

%We evaluate the correctness of the migrations in two ways. First, we compare the LLM-migrated code with the developer-migrated code in \migbench and report correctness at two granularity levels: the full migration and the individual \ccs. 
%We use \taxonomy to identify which types of \ccs the LLMs can/cannot handle.
%Second, for the subset of migrations with available unit tests, we run the available tests to assess the correctness of the migrated code.

%% file: c4-llm/experiment-setup.tex
\subsection{Experiment setup}
\def\RQBenchmark{How similar are the LLM migrations to the benchmark migrations?\xspace}
\def\RQTest{How many migrations pass unit tests?\xspace}

We answer the following research questions in this phase:

\begin{enumerate}[label={RQ-1.\arabic*},leftmargin=*]
    \item \label{rq:benchmark} \textbf{\RQBenchmark} 
    We consider the \ccs stored in \migbench as the ground truth. We automatically check if the LLM was able to correctly perform all expected changes, while accounting for refactoring and alternative correct changes through manual review.
    
    \item \label{rq:test} \textbf{\RQTest} 
    To evaluate run-time correctness of the migrated code, we use a second evaluation strategy where we run any available unit tests.  
\end{enumerate}

We start with the \MigAllCount migrations in \migbench for our experiments. We discard 2 migrations whose repositories are no longer public, and \MigSyntaxErrorCount migrations where either the pre-migration or post-migration files recorded in the benchmark have syntax errors.
We conduct our experiments with the remaining \MigExperimentCount migrations containing \CCExperimentCount \codechanges.

\begin{figure}[t!]    
\begin{lstlisting}[
    label={fig:prompt-mig-expermiment},
    caption={LLM Migration Prompt},
    numbers=none,
    breakindent=0pt,
    xleftmargin=0pt,
]
The following Python code uses library <source-lib>. Migrate this code to use library <target-lib> instead. In the output, first explain the changes you made. Then, provide the modified code.

Do not make any changes to the code that are not related to migrating between these two libraries. Do not refactor. Do not reformat. Do not optimize. Do not
change coding style. 

Provided code:  
<premig-code>
\end{lstlisting}
\end{figure}

\subsubsection{Performing migration}
\label{sec:migration}
We use three off-the-shelf LLMs -- Llama 3.1-70B-Instruct \cite{llamapaper}, GPT-4o-mini-2024-07-18, and GPT-4o-2024-08-06 \cite{openaimodels} -- to perform migrations on the \migbench dataset.
We refer to these models as \lm, \mn, and \fo in the rest of the paper, respectively.

For each file having \ccs in a given migration, we use the LLM prompt template in \autoref{fig:prompt-mig-expermiment} to migrate the code.
% why this prompt is like this
The prompt is designed to ensure that the LLM performs focused, controlled migrations while maintaining transparency.
By asking the LLM to explain its changes, we ensure that the LLM justifies its modifications, which can be useful for a human evaluator reviewing the migration. 
The prompt also aims to restrict the LLM from making unrelated modifications such that we remain focused on the task of library migration.

Since LLMs are non-deterministic~\cite{atil2024llmstabilitydetailedanalysis}, we use a temperature of 0 and run each migration 10 times for each model.
To understand the extent of variability in the generated code, we calculate the difference between the runs.
We use git-diff to compute a diff between each pair of the 10 runs and normalize the number of different lines by dividing it by the total number of lines in \filePre.
We find that on average, two runs on the same file are only \LlamaAvgDiff, \MiniAvgDiff and \FouroAvgDiff different for \lm, \mn, and \fo, respectively.
Given this low variability and the high effort involved in manual validation, we randomly select one run for each model to evaluate the LLMs' results.
Our artifact contains all LLM migrations for all 10 runs.

\input{c4-llm/evaluation-benchmark--approach.tex}
\input{c4-llm/evaluation-unittest--approach.tex}

%% file: c4-llm/evaluation-benchmark--approach.tex
\subsubsection{Matching LLM migrations with developer migration}
In \rqBench, we compare the LLM migrations with the developer migrations in \migbench.
We do this for each \cc, and then aggregate the results at the migration level, while also considering the additional non-migration-related changes made by the developer and the LLM.
We now explain the process using the example in \autoref{fig:mig-example} that shows a migration from the library \textit{requests} to \textit{aiohttp}.

\input{c4-llm/img/mig-example.tex}

\paragraph{Auto change matching}
We first attempt to automatically match as many LLM changes (\changeLLM) as possible with developer change (\changeDev), while ensuring precision \ie not falsely marking an incorrect \changeLLM as correct.
Therefore, we consider only \textit{exact} syntactic matches between \changeDev and \changeLLM, ignoring formatting differences.
We find exact matches by identifying the AST nodes related to \changeDev in \fileDev and then matching them to corresponding nodes in \fileLLM.
If there are potentially multiple matches, we check the containing function and use proximity based heuristics demonstrated below.
We ensure that the matched node sets translate to the same code string using \texttt{ast.unparse}.

Consider \changeDev \segment{D1} in \autoref{fig:mig-example} which uses \texttt{ClientSession()} and \texttt{get()} from \textit{aiohttp}.
In \fileLLM, these APIs appear in multiple locations, but only Lines 15, 16, 21, and 22 are in the same containing function \texttt{fetch\_flights} as \changeDev.
Based on line number proximity, we identify two node sets \segment{L2} and \segment{L4} as potential matches for this \changeDev.
By comparing the code strings, we find that \segment{L2} exactly matches \changeDev.
By the end of the auto change matching step, we get a set of matched pairs of \changeDev and \changeLLM, a set of unmatched \changeDev (\uChangeDev), and a set of unmatched \changeLLM (\uChangeLLM).

\paragraph{Manually reviewing unmatched changes}
The automatic matching ensures that no change is incorrectly matched, but it may miss some correct matches due to syntactic differences, or alternative correct implementations.
The manual review step focuses on reviewing \uChangeDev and \uChangeLLM.
We use the online API documentation as well as the source code of the source and target libraries to understand the correct API usage in the context of the file.
For \autoref{fig:mig-example}, we manually find that \segment{L4} is a correct alternative to \segment{D2}, noting that the LLM omitted the default value (\texttt{True}) of the \texttt{allow\_redirects} argument, unlike the developer.
\segment{L8} is a correct replacement of \segment{D3}, where the LLM performed an extract variable refactoring (assigning \texttt{plan.id} to variable \texttt{pid} and then using \texttt{pid} in subsequent lines).
However, we find that the LLM used the function \texttt{get} instead of \texttt{post} at line 33,
making \segment{L0} an incorrect alternative of \segment{D4}.

After trying to match all changes in \changeDev, we may find that some \changeLLM remain unmatched.
Our aim is to categorize these as refactoring (not altering program semantics) or non-refactoring (potentially affecting code behavior).
If a \changeLLM is refactoring, we remove it from \uChangeLLM, as it does not impact migration correctness.
In the example, the LLM introduced handling of a \texttt{ValueError} (\segment{L3}), which changes the program behavior, so we keep it in \uChangeLLM.
Similarly, swapping variables in \segment{L5} is also non-refactoring.

Note that while \migbench records all migration-related \ccs, it does not record additional changes \textit{indirectly} related to the migration, \ie lines that do not have a target library API usage.
For example, adding \texttt{async} to function definitions (\segment{L1}, \segment{L6}, and \segment{L9}) due to async calls introduced by migration are not recorded.
We remove these changes from \uChangeLLM and mark them as correct.
Overall, \segment{L3} and \segment{L5} stay in \uChangeLLM, while \segment{L1}, \segment{L6}, \segment{L7}, and \segment{L9} get removed.

We perform a multi-reviewer manual review to ensure the quality of the matching process.
The first author first manually reviews 190 \ccs from one model.
Two other authors then review about half of those \ccs each independently.
We then measure the agreement using Cohen's Kappa \cite{cohen1960coefficient} to quantify the level of agreement between the reviewers.
We do not reach substantial agreement, so we discuss and resolve disagreements and update our coding guideline.
The two pairs of reviewers then independently review another 111 \ccs, achieving a Cohen's Kappa score of \KappaTwo (almost perfect) and \KappaOne (substantial)~ \cite{landis1977measurement}.
Accordingly, we proceed with only the first author reviewing the remaining changes.

\paragraph{Determining \cc status}
Based on the matching of \changeDev and \changeLLM, we determine each \textit{\cc status} as follows (example changes refer to \autoref{fig:mig-example}).

\begin{itemize}[leftmargin=*,topsep=0pt]
    \item \textit{Correct change}: LLM's change is correct, either exactly like the developer (\eg \segment{D1}--\segment{L2}), or with an alternative API (\eg \segment{D2}--\segment{L4}), or with same APIs but with some refactoring (\eg \segment{D3}--\segment{L8}).

    \item \textit{Incorrect change}: The LLM incorrectly implemented the migration change:
    Used an incorrect API (\eg \texttt{get()} instead of \texttt{post()} in \segment{L0}), did not attempt to migrate it at all, or incorrectly removed part of a code.
\end{itemize}

\paragraph{Determining migration status}
Based on the individual matched code changes, we now automatically determine an overall \textit{migration status}:
\begin{itemize}[leftmargin=*,topsep=0pt,parsep=0pt]
    \item \textit{\MigStatusResponseFailure}: The LLM did not generate a \fileLLM for at least one \filePre
%, because of limitations imposed by the model 
(e.g., due to token limit or API timeout).        

    \item \textit{\MigStatusSyntaxError}: The LLM generated a \fileLLM for all \filePre, but at least one \fileLLM has syntax errors. 

    \item \textit{\MigStatusIncorrect}: The LLM could not correctly migrate \textit{any} of the changes marked in \migbench.
    We assign this status when all \changeDev across the migration remain unmatched.

    \item \textit{\MigStatusPartialCorrect}: The LLM correctly migrated only \textit{some} of \changeDev.
    We assign this status when there are only some unmatched \changeDev.

    \item \textit{\MigStatusExtraChange}: The LLM correctly migrated \textit{all} \changeDev but also performed some non-refactoring changes. We assign this status when there are no unmatched \changeDev in any file, but there are some remaining unmatched \changeLLM.

    \item \textit{\MigStatusCorrect}: The LLM correctly migrated \textit{all} \changeDev for this migration, without any non-refactoring changes.    
    We assign this status when there are no unmatched \changeDev or \changeLLM left in any of the files after the matching process.
\end{itemize}

%% file: c4-llm/img/mig-example.tex
\begin{figure*}[t]    
    \includegraphics[width=.98\linewidth]{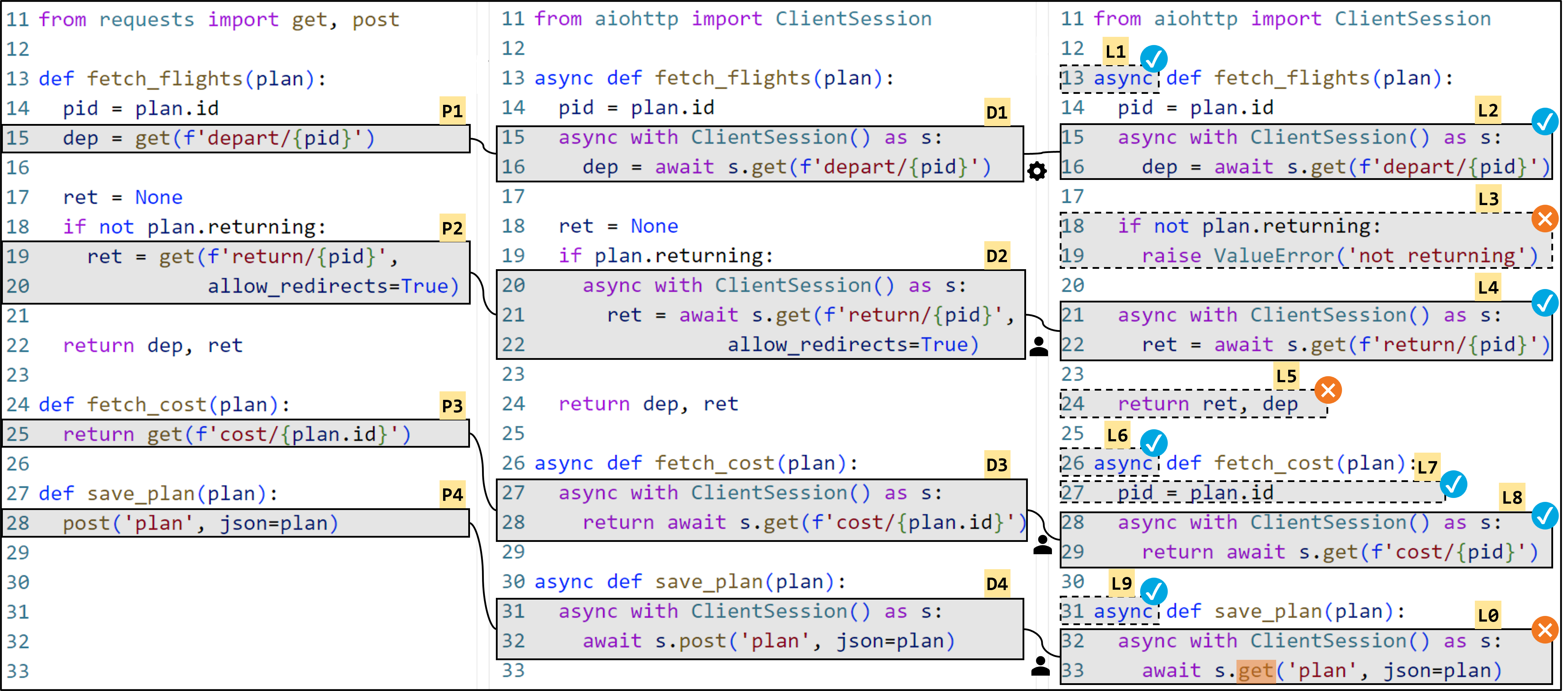}
    
    \begin{subfigure}[t]{.32\textwidth}        
        \vspace{-1.7em}
        \caption{\footnotesize{Before migration (\filePre)}}
        \label{fig:mig-example-pre}
    \end{subfigure}
    \begin{subfigure}[t]{.33\textwidth}            
        \vspace{-1.7em}
        \caption{\footnotesize{Developer's migration (\fileDev)}}
        \label{fig:mig-example-dev}
    \end{subfigure}
    \begin{subfigure}[t]{.33\textwidth}        
        \vspace{-1.7em}
        \caption{\footnotesize{LLM Migration (\fileLLM)}}
        \label{fig:mig-example-llm}
    \end{subfigure}
    % \begin{subfigure}[c]{.33\textwidth}
    %     \includegraphics[width=1\linewidth]{groups/mig-example-cc-description.png}
    %     \caption{\migbench \cc description}
    %     \label{fig:mig-example-cc-info}
    % \end{subfigure}    
    
    \caption{A sample \aPair{requests}{aiohttp} migration.} 
    \mysubcaption{
        The lines between the boxes show matching \ccs. 
        \faCog~denotes automatic matching, while \faUser~denotes manual matching.
        Dashed lines denote changes that are not recorded in \migbench.
        The blue check marks denote correct changes, while the orange crosses denote incorrect changes.
    }
    \label{fig:mig-example}
\end{figure*}

%% file: c4-llm/evaluation-unittest--approach.tex
\subsubsection{Evaluating LLM migrations using unit tests}
\label{sec:eval:test}
In \rqTest, we assess the same LLM \migbench migrations from \rqBench, but using the unit tests available in the corresponding repositories.
A test-based evaluation that runs the code ensures that the migration preserves the original behavior.
We now explain the process below.

We make a copy of the repository after the developers' migration (\codeDev).
We then make another copy of developer's migration, but replace all \fileDev with \fileLLM; we refer to this as \codeLLM.
The idea is that \codeLLM is basically the version of \codeDev where the LLM did the migration on behalf of the developer.

To set up a virtual environment, we need to identify the Python version used in the client repository.
If this information is available in the \texttt{setup.py} file, we use that version.
Otherwise, we resolve the version based on the migration commit date as follows.
We first identify the release dates of all minor versions of Python (3.6, 3.7 etc).
Then, we find the latest release date that is before the migration commit date.
This is the latest Python version that was available at the time of the migration; we use this version to create the virtual environment.
Next, we install the code dependencies using \texttt{pyproject.toml}, \texttt{setup.py} and requirements files.
In cases where specific dependency versions are not specified in those files, we look up the version history of the dependency on PyPI and install the latest version available at the migration commit date, similar to how we resolve the Python version. This ensures that the dependencies are compatible with the code at the time of migration.

Once the virtual environment is ready, we run all the unit tests while measuring coverage in the repository on \codeDev. %using the command \texttt{python -m pytest} on \codeDev.
%This command runs all the tests in the repository.
If the run has errors, we read the error log and try to fix the errors, and run the tests again. Note that we only fix configuration or environment-related errors, not code errors.
For example, if a dependency is missing, we install it; if the project requires a specific Python version, we set up the virtual environment with that version.
We maintain a configuration file where we record any project-specific configurations or commands we used. 
We include this information in our artifact.

We find that \MigHasTestsCount out of the \MigExperimentCount migrations have at least one test.
Among these, \MigErrorRunningTestCount migrations had unresolvable errors.
The failures are primarily due to missing dependencies, especially for older projects.
The remaining \MigTestExecutedCount migrations have tests that we are able to successfully run on \codeDev.

However, for the tests to be useful to validate migration, they must cover the \codechanges.
Using the coverage report, we find that only \MigCoveredCount migrations have at least one test that covers the code changes, and among them, \MigCoveredPassingCount migrations have at least one test that passes on \codeDev.
Thus, the evaluation in \ref{rq:test} focuses only on these \MigCoveredPassingCount migrations.
We run the entire test suite on \codeLLM for these \MigCoveredPassingCount migrations for each model using the same randomly selected LLM run used in \rqBench.

We compare the test results between \codeDev and \codeLLM and assign the following statuses to the migrations:
\begin{itemize}[leftmargin=*]
    \item \textit{\MigStatusCorrect}: All passing tests in \codeDev also pass in \codeLLM.
    \item \textit{\MigStatusPartialCorrect}: Some of the tests that pass in \codeDev pass in \codeLLM, but the others fail or raise run-time errors.
    \item \textit{\MigStatusIncorrect}: None of the passing tests in \codeDev pass in \codeLLM; they all fail or raise run-time errors.   
\end{itemize}

%% file: c4-llm/evaluation-benchmark--findings.tex
\subsection{\rqBench \RQBenchmark}

We now discuss the results of \rqBench. We first present the migration level results, then present the overall \cc level results, and finally analyze the correctness of \ccs by \taxonomy categories.

\subsubsection{Migration level correctness}

\autoref{tab:bm-eval-mig-level} shows the migration level results. 
We find that \fo performs the best with \FouroMigBMAtLeastOneCorrectPercent of the migrations having at least one correct \cc, closely followed by \mn with \MiniMigBMAtLeastOneCorrectPercent.
When considering fully correct migrations, \fo outperforms \mn considerably, with \FouroMigBMCorrectPercent compared to \MiniMigBMCorrectPercent.
\lm has the lowest performance, with only \LlamaMigBMCorrectPercent being fully correct and \LlamaMigBMAtLeastOneCorrectPercent having at least one correct \cc.
This is mainly due to a higher proportion of \textit{\MigStatusResponseFailureLower}s where \lm ran into token limits.
\lm also produces more syntax errors compared to the other two models (\LlamaMigBMSyntaxErrorPercent vs. \MiniMigBMSyntaxErrorPercent and \FouroMigBMSyntaxErrorPercent).

\begin{tcolorbox}[colback=resultbg, title={\rqBench: Migration level}]
%    LLMs can correctly migrate the majority of migrations, with 
  All three LLMs were able to perform correct migrations with \fo performing best: \FouroMigBMAtLeastOneCorrectPercent of its migrations were partially correct and \FouroMigBMCorrectPercent were fully correct.
\end{tcolorbox}

\subsubsection{Overall \cc level correctness}

\input{c4-llm/tab/bm-eval-mig-level.tex}

The \MigExperimentCount migrations we use for our evaluation contain a total of \CCExperimentCount \ccs.
However, when there is a response failure or syntax error in the LLM's migration, we cannot evaluate the correctness of the corresponding \ccs in that file.
Accordingly, in this analysis level, we exclude \LlamaCCBMExcludedCount \ccs from \lm, \MiniCCBMExcludedCount from \mn, and \FouroCCBMExcludedCount from \fo.
This leaves us with \LlamaCCBMEvaluatedCount, \MiniCCBMEvaluatedCount, and \FouroCCBMEvaluatedCount \ccs to evaluate for the three LLMs, respectively.

\input{c4-llm/tab/cc-level-summary.tex}

\autoref{tab:cc-level-summary} shows the overall correctness of the LLMs' migrations at the \cc level.
%We had a total of 4,883 \ccs for the three models combined. 
%For reference, we note that we automatically matched 2,828 (58\%) code changes, and the remaining 2,055 (42\%) changes required manual review.
\fo performs best, with \FouroCCBMTotalCorrectPercent of the \ccs being correct.
Interestingly, \mn and \lm both perform equally well at the \cc level, with both having \MiniCCBMTotalCorrectPercent correct \ccs.
This is in contrast to the migration level, where \mn performs better than \lm (\MiniMigBMCorrectPercent vs. \LlamaMigBMCorrectPercent).
This suggests that while \mn attempted more \ccs compared to \lm, the ones that \lm was able to attempt without failures or syntax errors were comparatively correct. 
%This is in contrast to the migration level, where \mn performs better than \lm (\MiniMigBMCorrectPercent vs. \LlamaMigBMCorrectPercent).
%This can be explained by the fact that \mn has attempted more \ccs compared to \lm (\MiniCCBMEvaluatedCount vs \LlamaCCBMEvaluatedCount).% as the latter has more syntax errors and response failures.

\begin{tcolorbox}[colback=resultbg, title={\rqBench: \cc level}]
  LLMs correctly migrate a high proportion of most individual \ccs, with \fo achieving \FouroCCBMTotalCorrectPercent correct \ccs. The other two models perform equally well with \MiniCCBMTotalCorrectPercent correct \ccs for both models.
\end{tcolorbox}

\subsubsection{Code change correctness by category}

\input{c4-llm/tab/cc-tax.tex}

To understand if some types of \ccs are more difficult to migrate, \autoref{tab:rq:cc-freq} shows the distribution of \cc correctness across the different \taxonomy categories.
The three columns in group ``{Frequency in \migbench}'' shows how often each \taxonomy category appears in all migrations (Migs), library pairs (LPs) and \ccs (CCs) in \migbench, which provides perspective on the category's impact in practice.
This information is the same for each LLM since it describes the existing data in \migbench.
For example, \FCallTotalCount(\FCallTotalPercent) of all \CCAllCount \ccs involve \fcall{s}.
The next three column groups show the performance of each model.
Since each LLM has a different number of evaluated code changes,
we show the number of evaluated code changes from each category (\evalCountCell) alongside the proportion of these code changes that are correct (\correctPercentCell).
For example, the \textit{\Fcall} row under the \textit{Program elements} group shows that \lm's syntactically correct migrations included \LlamaCCBMFCallTotalCount \ccs that involve \fcall{s}, and that \LlamaCCBMFCallCorrectPercent of these \ccs were correctly migrated.

\paragraph{Program elements}
\Fcall{s} are the most common program elements in terms of \ccs (\FCallTotalPercent), and second most common in terms of migrations (\FCallMigPercent) and library pairs (\FCallLibPairPercent).
All three LLMs correctly migrate a high proportion of these changes, with  \LlamaCCBMFCallCorrectPercent, \MiniCCBMFCallCorrectPercent, and \FouroCCBMFCallCorrectPercent of the \fcall{s} being correctly migrated  by the three LLMs. Almost all migrations and library pairs in \migbench require migrating \imp statements, a task that all three LLMs perform well, especially \fo correctly migrates nearly all of them (\FouroCCBMImpCorrectPercent).

\mn slightly outperforms the other models in migrating \dec{s} (\MiniCCBMDecCorrectPercent vs. \LlamaCCBMDecCorrectPercent and \FouroCCBMDecCorrectPercent); however, it does worst in \attr{s} (\MiniCCBMAttrCorrectPercent vs. \LlamaCCBMAttrCorrectPercent and \FouroCCBMAttrCorrectPercent).
\fo performs better than other models in migrating \type{s}, \exc{s} and \fref{s}, followed by \mn.
\lm and \mn both struggle with migrating \fref{s} (\LlamaCCBMFRefCorrectPercent and \MiniCCBMFRefCorrectPercent correct), while \fo correctly performs all such \ccs.
While this last category shows most discrepancy between the models, this program element is present in only \FRefTotalPercent of the \ccs, therefore does not affect the overall performance significantly.

\paragraph{Properties}
The property \textit{\noProps} indicates that the source and target APIs are identical, and hence no changes are required.
While \fo correctly leaves almost all of these APIs unchanged (\FouroCCBMNoPropsCorrectPercent),
\lm and \mn incorrectly changed several of them, resulting in \LlamaCCBMNoPropsCorrectPercent and \MiniCCBMNoPropsCorrectPercent correct, respectively.

\textit{\ElemNC} is the most common property in \migbench.
This indicates that the target APIs commonly bear different names from the source APIs. 
All LLMs perform well in making this change, with \lm, \mn, and \fo correctly migrating \LlamaCCBMElemNCCorrectPercent, \MiniCCBMElemNCCorrectPercent, and \FouroCCBMElemNCCorrectPercent changes, respectively.

All three LLMs performed better or equally well in \textit{\argDel} compared to those with \textit{\argAdd}, which is expected as adding an argument requires the LLM to find a suitable replacement, while deleting an argument is a simpler task.
The \textit{\ArgTrans} property indicates that an argument requires various changes, such as changing the type or value.
The three LLMs correctly migrate only 76-77\% of the \argTrans changes.

The property \textit{\paramAdd} is found in migrations in just one library, \lib{click} library, where some decorators require adding a parameter to the decorated function \cite{libclick}.
All three LLMs perform well in migrating these changes; interestingly, \mn performs better than \fo (\MiniCCBMParamAddCorrectPercent vs. \FouroCCBMParamAddCorrectPercent).
\mn also performed better than \fo in migrating \dec{s}, which are common in the click library.
These are the only two categories where \fo does not perform the best.

\textit{\ArgNC} applies when an argument representing the same semantics has different names in the source and target APIs.
While \fo performs reasonably well in migrating these changes (\FouroCCBMArgNCCorrectPercent), \lm and \mn fail frequently with them (only \LlamaCCBMArgNCCorrectPercent and \MiniCCBMArgNCCorrectPercent correct).

\paragraph{Cardinality} 
\OO \ccs are the most common in \migbench, and all three LLMs perform well in migrating them.
\OZ \ccs are the ones where a source API needs to be removed, but no target needs to be added.
Because it is a simple delete operation, all three LLMs migrate all or almost all of these changes correctly.
Its counterpart, \ZO \ccs, are the ones where there is no source library usage, but a target library usage needs to be added.
This is a more difficult change, as the LLM needs to find a suitable location to add the new library usage.
Interestingly, \lm performs the best in migrating these changes (\LlamaCCBMZOCorrectPercent correct), making this the only category where it performs better than the other models (\MiniCCBMZOCorrectPercent and \FouroCCBMZOCorrectPercent correct).
While previous research showed that higher-cardinality \ccs are generally difficult compared to \oo \ccs \cite{alrubaye2018automating,wang2016transforming, zhang2020deep,HuangMappingAPI2024},
the \migbench dataset shows that higher cardinality changes are frequent with \HigherCardinalityLibPairPercent of the library pairs requiring at least one higher-cardinality \cc.
Overall, we find that the LLMs are reasonably successful at migrating them, with \LlamaCCBMHigherCardinalityCorrectPercent, \MiniCCBMHigherCardinalityCorrectPercent, and \FouroCCBMHigherCardinalityCorrectPercent correctness rates.

\begin{tcolorbox}[colback=resultbg, title=\rqBench: Code change level by category]
    LLMs correctly migrate a high proportion of most \ccs, with \fo achieving \FouroCCBMTotalCorrectPercent correct \ccs.
    The models perform reasonably well in migrating difficult higher-cardinality \ccs (\fo gets \FouroCCBMHigherCardinalityCorrectPercent correct), but struggle relatively more with \argTrans and \argNC.
\end{tcolorbox}

%% file: c4-llm/tab/bm-eval-mig-level.tex
\begin{table*}[t]
    \renewcommand{\arraystretch}{1}
    \centering
    \caption{Correctness of \migbench migrations (\rqBench, migration level)}
    \footnotesize
    \label{tab:bm-eval-mig-level}
    {%
    \begin{tabular}{>{\quad}lrrr}
    \toprule
    ~                         & \multicolumn{3}{c}{Number (percentage) of migrations}\\ 
    \cmidrule{2-4}
    \hspace{-.8em}Status       & {\llama} & {\mini} & {\fouro} \\
    \midrule  
    \hspace{-.8em}\textit{At least partially correct} \\
    \MigStatusCorrect          & \LlamaMigBMCorrectCount~(\LlamaMigBMCorrectPercent) & \MiniMigBMCorrectCount~(\MiniMigBMCorrectPercent) & \FouroMigBMCorrectCount~(\FouroMigBMCorrectPercent) \\
    Correct w/ non-refactorings      & \LlamaMigBMCorrectWithExtraChangesCount~(\LlamaMigBMCorrectWithExtraChangesPercent) & \MiniMigBMCorrectWithExtraChangesCount~(\MiniMigBMCorrectWithExtraChangesPercent) & \FouroMigBMCorrectWithExtraChangesCount~(\FouroMigBMCorrectWithExtraChangesPercent) \\
    \MigStatusPartialCorrect   & \LlamaMigBMPartiallyCorrectCount~(\LlamaMigBMPartiallyCorrectPercent) & \MiniMigBMPartiallyCorrectCount~(\MiniMigBMPartiallyCorrectPercent) & \FouroMigBMPartiallyCorrectCount~(\FouroMigBMPartiallyCorrectPercent) \\    
    \textit{Subtotal}          & \textit{\LlamaMigBMAtLeastOneCorrectCount~(\LlamaMigBMAtLeastOneCorrectPercent)} & \textit{\MiniMigBMAtLeastOneCorrectCount~(\MiniMigBMAtLeastOneCorrectPercent)} & \textit{\FouroMigBMAtLeastOneCorrectCount~(\FouroMigBMAtLeastOneCorrectPercent)} \\
    \midrule
    \hspace{-.8em}\textit{Fully incorrect} \\
    \MigStatusIncorrect        & \LlamaMigBMIncorrectCount~(\LlamaMigBMIncorrectPercent) & \MiniMigBMIncorrectCount~(\MiniMigBMIncorrectPercent) & \FouroMigBMIncorrectCount~(\FouroMigBMIncorrectPercent) \\
    \MigStatusSyntaxError      & \LlamaMigBMSyntaxErrorCount~(\LlamaMigBMSyntaxErrorPercent) & \MiniMigBMSyntaxErrorCount~(\MiniMigBMSyntaxErrorPercent) & \FouroMigBMSyntaxErrorCount~(\FouroMigBMSyntaxErrorPercent) \\
    \MigStatusResponseFailure  & \LlamaMigBMResponseFailureCount~(\LlamaMigBMResponseFailurePercent) & \MiniMigBMResponseFailureCount~(\MiniMigBMResponseFailurePercent) & \FouroMigBMResponseFailureCount~(\FouroMigBMResponseFailurePercent) \\    
    \textit{Subtotal}          & \textit{\LlamaMigBMNoCorrectCount~(\LlamaMigBMNoCorrectPercent)} & \textit{\MiniMigBMNoCorrectCount~(\MiniMigBMNoCorrectPercent)} & \textit{\FouroMigBMNoCorrectCount~(\FouroMigBMNoCorrectPercent)} \\
    \midrule
    Total migrations    & \MigExperimentCount (100\%) & \MigExperimentCount (100\%) & \MigExperimentCount (100\%) \\ 
    \bottomrule
    \end{tabular}
    }
\end{table*}

%% file: c4-llm/tab/cc-level-summary.tex
\begin{table*}[t!]
    \renewcommand{\arraystretch}{1}
    \centering
    \caption{Correctness of \migbench \codechanges (\rqBench, overall code change level)}
    \footnotesize
    \label{tab:cc-level-summary}
    {%
    \begin{tabular}{lrrr}
    \toprule
    ~                         & \multicolumn{3}{c}{Number (percentage) of \ccs}\\ 
    \cmidrule{2-4}
    Status       & {\llama} & {\mini} & {\fouro} \\
    \midrule    
    \CCStatusCorrect            & \LlamaCCBMTotalCorrectCount~(\LlamaCCBMTotalCorrectPercent) & \MiniCCBMTotalCorrectCount~(\MiniCCBMTotalCorrectPercent) & \FouroCCBMTotalCorrectCount~(\FouroCCBMTotalCorrectPercent) \\
    \CCStatusIncorrect            & \LlamaCCBMTotalIncorrectCount~(\LlamaCCBMTotalIncorrectPercent) & \MiniCCBMTotalIncorrectCount~(\MiniCCBMTotalIncorrectPercent) & \FouroCCBMTotalIncorrectCount~(\FouroCCBMTotalIncorrectPercent) \\
    \midrule
    Evaluated \ccs & \LlamaCCBMEvaluatedCount~(100\%) & \MiniCCBMEvaluatedCount~(100\%) & \FouroCCBMEvaluatedCount~(100\%) \\
    Excluded \ccs  & \LlamaCCBMExcludedCount & \MiniCCBMExcludedCount & \FouroCCBMExcludedCount \\
    \midrule
    \hspace{-.8em}Total \ccs     & \CCExperimentCount & \CCExperimentCount & \CCExperimentCount \\
    \bottomrule
    \end{tabular}
    }
\end{table*}

%% file: c4-llm/tab/cc-tax.tex
\begin{table*}[t]
    \newcommand{\centercell}[1]{\multicolumn{1}{c}{#1}}
    \newcommand{\catHeader}{\multirow[b]{2}{*}{\makecell{\taxonomy\\ category}}}
    \newcommand{\freqTopHeader}{\multicolumn{3}{c}{Frequency in \migbench}}
    \newcommand{\modelHeader}[1]{\multicolumn{2}{c}{#1}}

    \centering
    \caption{Correctness of \ccs across \taxonomy categories (\rqBench, code change level by category)}
    \mysubcaption{
        \vspace{-1em}
        \textit{Migs} = Migrations, 
        \textit{LPs} = Library Pairs, 
        \textit{CCs} = Code Changes, 
        \textit{\evalCountCell} = Number of \ccs evaluated, 
        \textit{\correctPercentCell} = percentage of correct \ccs.\\        
    }
    \label{tab:rq:cc-freq}
    \tablefontsize
    \setlength{\tabcolsep}{1.8pt}
    \setlength{\cmidrulekern}{1pt}

    {
    \begin{tabular}{>{\hspace{.3em}}l@{~}rrrr@{~}rr@{~}rr@{~}r}
    \toprule
    % top heading row
    \catHeader             
    & \freqTopHeader 
    & \modelHeader{\llama}                                                                                    
    & \modelHeader{\mini}
    & \modelHeader{\fouro}\\
    \cmidrule(lr){2-4} \cmidrule(lr){5-6}\cmidrule(lr){7-8}\cmidrule(lr){9-10}
    % second heading row
    & \centercell{Migs}
    & \centercell{LPs}
    & \centercell{CCs} 
    & \evalCountCell          
    & \correctPercentCell
    & \evalCountCell 
    & \correctPercentCell                                         
    & \evalCountCell 
    & \correctPercentCell                                        
    \\
    \midrule
    \multicolumn{3}{l}{{\textit{Program elements}}}\\
    \Fcall
    %& \smallbar{FCall}    
    & \FCallMigCount (\FCallMigPercent)
    & \FCallLibPairCount (\FCallLibPairPercent)
    & \FCallTotalCount (\FCallTotalPercent)
    & \LlamaCCBMFCallTotalCount
    & \LlamaCCBMFCallCorrectPercent 
    & \MiniCCBMFCallTotalCount 
    & \MiniCCBMFCallCorrectPercent  
    & \FouroCCBMFCallTotalCount
    & \FouroCCBMFCallCorrectPercent 
    \\
    \Imp       
    %& \smallbar{Imp}       
    & \ImpMigCount (\ImpMigPercent)
    & \ImpLibPairCount (\ImpLibPairPercent)
    & \ImpTotalCount (\ImpTotalPercent)
    & \LlamaCCBMImpTotalCount   
    & \LlamaCCBMImpCorrectPercent  
    & \MiniCCBMImpTotalCount
    & \MiniCCBMImpCorrectPercent 
    & \FouroCCBMImpTotalCount
    & \FouroCCBMImpCorrectPercent 
    \\
    \Dec       
    % & \smallbar{Dec}       
    & \DecMigCount (\DecMigPercent)
    & \DecLibPairCount (\DecLibPairPercent)
    & \DecTotalCount (\DecTotalPercent)
    & \LlamaCCBMDecTotalCount
    & \LlamaCCBMDecCorrectPercent 
    & \MiniCCBMDecTotalCount
    & \MiniCCBMDecCorrectPercent 
    & \FouroCCBMDecTotalCount
    & \FouroCCBMDecCorrectPercent 
    \\
    \Attr      
    % & \smallbar{Attr}     
    & \AttrMigCount (\AttrMigPercent)
    & \AttrLibPairCount (\AttrLibPairPercent)
    & \AttrTotalCount (\AttrTotalPercent)
    & \LlamaCCBMAttrTotalCount
    & \LlamaCCBMAttrCorrectPercent 
    & \MiniCCBMAttrTotalCount
    & \MiniCCBMAttrCorrectPercent 
    & \FouroCCBMAttrTotalCount
    & \FouroCCBMAttrCorrectPercent 
    \\
    \Type      
    % & \smallbar{Type}
    & \TypeMigCount (\TypeMigPercent)
    & \TypeLibPairCount (\TypeLibPairPercent)
    & \TypeTotalCount (\TypeTotalPercent)
    & \LlamaCCBMTypeTotalCount
    & \LlamaCCBMTypeCorrectPercent 
    & \MiniCCBMTypeTotalCount
    & \MiniCCBMTypeCorrectPercent 
    & \FouroCCBMTypeTotalCount
    & \FouroCCBMTypeCorrectPercent 
    \\
    \Exc       
    % & \smallbar{Exc}
    & \ExcMigCount (\ExcMigPercent)
    & \ExcLibPairCount (\ExcLibPairPercent)
    & \ExcTotalCount (\ExcTotalPercent)
    & \LlamaCCBMExcTotalCount
    & \LlamaCCBMExcCorrectPercent 
    & \MiniCCBMExcTotalCount
    & \MiniCCBMExcCorrectPercent 
    & \FouroCCBMExcTotalCount
    & \FouroCCBMExcCorrectPercent 
    \\
    \fRefCell      
    %& \smallbar{Fref}
    & \FRefMigCount (\FRefMigPercent)
    & \FRefLibPairCount (\FRefLibPairPercent)
    & \FRefTotalCount (\FRefTotalPercent)
    & \LlamaCCBMFRefTotalCount
    & \LlamaCCBMFRefCorrectPercent 
    & \MiniCCBMFRefTotalCount
    & \MiniCCBMFRefCorrectPercent 
    & \FouroCCBMFRefTotalCount
    & \FouroCCBMFRefCorrectPercent 
    \\
    
    \midrule
    \multicolumn{3}{l}{{\textit{Properties}}}\\        
    \NoPropShort     
    %& \smallbar{NoProps}
    & \NoPropsMigCount (\NoPropsMigPercent)
    & \NoPropsLibPairCount (\NoPropsLibPairPercent)
    & \NoPropsTotalCount (\NoPropsTotalPercent)
    & \LlamaCCBMNoPropsTotalCount
    & \LlamaCCBMNoPropsCorrectPercent 
    & \MiniCCBMNoPropsTotalCount
    & \MiniCCBMNoPropsCorrectPercent 
    & \FouroCCBMNoPropsTotalCount
    & \FouroCCBMNoPropsCorrectPercent 
    \\
    \elemNCCell       
    %& \smallbar{ElemNC}
    & \ElemNCMigCount (\ElemNCMigPercent)
    & \ElemNCLibPairCount (\ElemNCLibPairPercent)
    & \ElemNCTotalCount (\ElemNCTotalPercent)
    & \LlamaCCBMElemNCTotalCount 
    & \LlamaCCBMElemNCCorrectPercent 
    & \MiniCCBMElemNCTotalCount 
    & \MiniCCBMElemNCCorrectPercent 
    & \FouroCCBMElemNCTotalCount 
    & \FouroCCBMElemNCCorrectPercent 
    \\
    \argAddCell       
    % & \smallbar{ArgAdd}
    & \ArgAddMigCount (\ArgAddMigPercent)
    & \ArgAddLibPairCount (\ArgAddLibPairPercent)
    & \ArgAddTotalCount (\ArgAddTotalPercent)
    & \LlamaCCBMArgAddTotalCount 
    & \LlamaCCBMArgAddCorrectPercent
    & \MiniCCBMArgAddTotalCount 
    & \MiniCCBMArgAddCorrectPercent 
    & \FouroCCBMArgAddTotalCount 
    & \FouroCCBMArgAddCorrectPercent 
    \\
    \argDelCell   
    %& \smallbar{ArgDel}
    & \ArgDelMigCount (\ArgDelMigPercent)
    & \ArgDelLibPairCount (\ArgDelLibPairPercent)
    & \ArgDelTotalCount (\ArgDelTotalPercent)
    & \LlamaCCBMArgDelTotalCount 
    & \LlamaCCBMArgDelCorrectPercent 
    & \MiniCCBMArgDelTotalCount 
    & \MiniCCBMArgDelCorrectPercent 
    & \FouroCCBMArgDelTotalCount 
    & \FouroCCBMArgDelCorrectPercent 
    \\
    \argTransCell     
    %& \smallbar{ArgTrans}
    & \ArgTransMigCount (\ArgTransMigPercent)
    & \ArgTransLibPairCount (\ArgTransLibPairPercent)
    & \ArgTransTotalCount (\ArgTransTotalPercent)
    & \LlamaCCBMArgTransTotalCount 
    & \LlamaCCBMArgTransCorrectPercent 
    & \MiniCCBMArgTransTotalCount 
    & \MiniCCBMArgTransCorrectPercent 
    & \FouroCCBMArgTransTotalCount 
    & \FouroCCBMArgTransCorrectPercent 
    \\
    \paramAddCell 
    %& \smallbar{ParamAdd}
    & \ParamAddMigCount (\ParamAddMigPercent)
    & \ParamAddLibPairCount (\ParamAddLibPairPercent)
    & \ParamAddTotalCount (\ParamAddTotalPercent)
    & \LlamaCCBMParamAddTotalCount 
    & \LlamaCCBMParamAddCorrectPercent 
    & \MiniCCBMParamAddTotalCount 
    & \MiniCCBMParamAddCorrectPercent 
    & \FouroCCBMParamAddTotalCount 
    & \FouroCCBMParamAddCorrectPercent 
    \\
    \argNCCell        
    % & \smallbar{ArgNC}
    & \ArgNCMigCount (\ArgNCMigPercent)
    & \ArgNCLibPairCount (\ArgNCLibPairPercent)
    & \ArgNCTotalCount (\ArgNCTotalPercent)
    & \LlamaCCBMArgNCTotalCount 
    & \LlamaCCBMArgNCCorrectPercent 
    & \MiniCCBMArgNCTotalCount 
    & \MiniCCBMArgNCCorrectPercent 
    & \FouroCCBMArgNCTotalCount 
    & \FouroCCBMArgNCCorrectPercent 
    \\
    \asyncTransCell   
    % & \smallbar{AsyncTrans}
    & \AsyncTransMigCount (\AsyncTransMigPercent)
    & \AsyncTransLibPairCount (\AsyncTransLibPairPercent)
    & \AsyncTransTotalCount (\AsyncTransTotalPercent)
    & \LlamaCCBMAsyncTransTotalCount 
    & \LlamaCCBMAsyncTransCorrectPercent 
    & \MiniCCBMAsyncTransTotalCount 
    & \MiniCCBMAsyncTransCorrectPercent 
    & \FouroCCBMAsyncTransTotalCount 
    & \FouroCCBMAsyncTransCorrectPercent 
    \\
    \outTransCell     
    %& \smallbar{OutTrans}
    & \OutTransMigCount (\OutTransMigPercent)
    & \OutTransLibPairCount (\OutTransLibPairPercent)
    & \OutTransTotalCount (\OutTransTotalPercent)
    & \LlamaCCBMOutTransTotalCount 
    & \LlamaCCBMOutTransCorrectPercent 
    & \MiniCCBMOutTransTotalCount 
    & \MiniCCBMOutTransCorrectPercent 
    & \FouroCCBMOutTransTotalCount 
    & \FouroCCBMOutTransCorrectPercent 
    \\

    \midrule
    \multicolumn{3}{l}{{\textit{Cardinality}}}\\
    \OO        
    % & \smallbar{OO}
    & \OOMigCount (\OOMigPercent)
    & \OOLibPairCount (\OOLibPairPercent)
    & \OOTotalCount (\OOTotalPercent)
    & \LlamaCCBMOOTotalCount 
    & \LlamaCCBMOOCorrectPercent 
    & \MiniCCBMOOTotalCount 
    & \MiniCCBMOOCorrectPercent 
    & \FouroCCBMOOTotalCount 
    & \FouroCCBMOOCorrectPercent 
    \\
    \OZ        
    % & \smallbar{OZ} 
    & \OZMigCount (\OZMigPercent)
    & \OZLibPairCount (\OZLibPairPercent)
    & \OZTotalCount (\OZTotalPercent)
    & \LlamaCCBMOZTotalCount 
    & \LlamaCCBMOZCorrectPercent 
    & \MiniCCBMOZTotalCount 
    & \MiniCCBMOZCorrectPercent 
    & \FouroCCBMOZTotalCount 
    & \FouroCCBMOZCorrectPercent 
    \\
    \ZO        
    % & \smallbar{ZO}
    & \ZOMigCount (\ZOMigPercent)
    & \ZOLibPairCount (\ZOLibPairPercent)
    & \ZOTotalCount (\ZOTotalPercent)
    & \LlamaCCBMZOTotalCount 
    & \LlamaCCBMZOCorrectPercent 
    & \MiniCCBMZOTotalCount 
    & \MiniCCBMZOCorrectPercent 
    & \FouroCCBMZOTotalCount 
    & \FouroCCBMZOCorrectPercent 
    \\
    \hspace{-.5em}\textit{Higher Card.} 
    % & \smallbar{HigherCardinality}
    & \textit{\HigherCardinalityMigCount (\HigherCardinalityMigPercent)}
    & \textit{\HigherCardinalityLibPairCount (\HigherCardinalityLibPairPercent)}
    & \textit{\HigherCardinalityTotalCount (\HigherCardinalityTotalPercent)}
    & \textit{\LlamaCCBMHigherCardinalityTotalCount} 
    & \textit{\LlamaCCBMHigherCardinalityCorrectPercent} 
    & \textit{\MiniCCBMHigherCardinalityTotalCount} 
    & \textit{\MiniCCBMHigherCardinalityCorrectPercent} 
    & \textit{\FouroCCBMHigherCardinalityTotalCount} 
    & \textit{\FouroCCBMHigherCardinalityCorrectPercent} 
    \\
    \OM
    % & \smallbar{OM} 
    & \OMMigCount (\OMMigPercent)
    & \OMLibPairCount (\OMLibPairPercent)
    & \OMTotalCount (\OMTotalPercent)
    & \LlamaCCBMOMTotalCount 
    & \LlamaCCBMOMCorrectPercent 
    & \MiniCCBMOMTotalCount 
    & \MiniCCBMOMCorrectPercent 
    & \FouroCCBMOMTotalCount     
    & \FouroCCBMOMCorrectPercent 
    \\
    \MO        
    % & \smallbar{MO}
    & \MOMigCount (\MOMigPercent)
    & \MOLibPairCount (\MOLibPairPercent)
    & \MOTotalCount (\MOTotalPercent)
    & \LlamaCCBMMOTotalCount 
    & \LlamaCCBMMOCorrectPercent 
    & \MiniCCBMMOTotalCount 
    & \MiniCCBMMOCorrectPercent 
    & \FouroCCBMMOTotalCount 
    & \FouroCCBMMOCorrectPercent 
    \\
    \MM        
    %& \smallbar{MM}
    & \MMMigCount (\MMMigPercent)
    & \MMLibPairCount (\MMLibPairPercent)
    & \MMTotalCount (\MMTotalPercent)
    & \LlamaCCBMMMTotalCount 
    & \LlamaCCBMMMCorrectPercent 
    & \MiniCCBMMMTotalCount 
    & \MiniCCBMMMCorrectPercent 
    & \FouroCCBMMMTotalCount 
    & \FouroCCBMMMCorrectPercent 
    \\
    \midrule
    Total 
    & \MigAllCount 
    & \LibPairAllCount 
    & \CCAllCount 
    & \LlamaCCBMEvaluatedCount 
    & \LlamaCCBMTotalCorrectPercent
    & \MiniCCBMEvaluatedCount  
    & \MiniCCBMTotalCorrectPercent 
    & \FouroCCBMEvaluatedCount  
    & \FouroCCBMTotalCorrectPercent 
    \\
    \bottomrule
    \end{tabular}
    }
\end{table*}

%% file: c4-llm/evaluation-unittest--findings.tex
\subsection{\ref{rq:test} \RQTest}
\label{sec:unittest-eval}

\input{c4-llm/tab/utest-eval-mig-level.tex}
\autoref{tab:utest-result} shows the results for the \MigTestEvaluatedCount migrations. %of the evaluation using unit tests for the \MigTestEvaluatedCount migrations that have tests covering the \codechanges.
\fo has the highest percentage of correct migrations, with \FouroMigTestCorrectPercent, followed by \mn with \MiniMigTestCorrectPercent, and \lm with \LlamaMigTestCorrectPercent.
\mn has more partially correct migrations than \fo and \lm, leading it to have an equal number of incorrect migrations as \fo (\MiniMigTestIncorrectPercent).
\lm, on the other hand, has a much higher proportion of incorrect migrations (\LlamaMigTestIncorrectPercent).%, which is significantly higher than the other two models.

\begin{tcolorbox}[colback=resultbg, title=\rqTest]
Out of \MigTestEvaluatedCount \migbench migrations with unit tests covering the \codechanges, the LLMs correctly migrated \LlamaMigTestCorrectPercent-\FouroMigTestCorrectPercent, with \fo being the highest.  
\end{tcolorbox}

%% file: c4-llm/tab/utest-eval-mig-level.tex
\begin{table}[t]
    \centering
    \caption{Correctness of \MigCoveredPassingCount \migbench migrations using their available unit tests (\ref{rq:test})}
    \footnotesize
    \label{tab:utest-result}
    %\begin{adjustbox}{width=.75\columnwidth}

    \begin{tabular}{lrrr}
    \toprule
    ~                         & \multicolumn{3}{c}{Number (percentage) of migrations}\\ 
    
    \cmidrule{2-4}
    Status                    & \llama                                                    & \mini & \fouro                                                                                \\
    \midrule  
    \MigStatusCorrect         & \LlamaMigTestCorrectCount                (\LlamaMigTestCorrectPercent)                & \MiniMigTestCorrectCount                 (\MiniMigTestCorrectPercent)                   & \FouroMigTestCorrectCount                 (\FouroMigTestCorrectPercent)                   \\
    \MigStatusPartialCorrect  & \LlamaMigTestPartiallyCorrectCount       (\LlamaMigTestPartiallyCorrectPercent)       & \MiniMigTestPartiallyCorrectCount        (\MiniMigTestPartiallyCorrectPercent)          & \FouroMigTestPartiallyCorrectCount        (\FouroMigTestPartiallyCorrectPercent)          \\
    \MigStatusIncorrect       & \LlamaMigTestIncorrectCount              (\LlamaMigTestIncorrectPercent)              & \MiniMigTestIncorrectCount               (\MiniMigTestIncorrectPercent)                 & \FouroMigTestIncorrectCount               (\FouroMigTestIncorrectPercent)                 \\    
    \midrule
    Evaluated       & \MigTestEvaluatedCount               & \MigTestEvaluatedCount & \MigTestEvaluatedCount \\    
    \bottomrule
    % \multicolumn{5}{l}{\textit{No tests: \MigNoTestsCount, Test errors: \MigErrorRunningTestCount, No coverage: \MigNotCoveredCount}}
    \end{tabular}
    %\end{adjustbox}
\end{table}

%% file: c4-llm/discussion.tex
\subsection{Strengths of LLM for library migration}
Overall, the results of this empirical study highlight the high overall correctness of LLM migrations.
We now discuss the key strengths we observe.

\paragraph*{Higher cardinality \ccs}
As shown in \rqBench, \lm, \mn, and \fo successfully migrated \LlamaCCBMHigherCardinalityCorrectPercent, \MiniCCBMHigherCardinalityCorrectPercent, and \FouroCCBMHigherCardinalityCorrectPercent of the higher cardinality \ccs, respectively, which the literature always viewed as complex migrations \cite{xu2019meditor,alrubaye2020learning}.
\autoref{fig:example-mm} shows an example of a \mm migration from \lib{twitter} to \lib{tweepy} done by \lm.
In addition to correctly replacing the functions \texttt{Twitter()}, and \texttt{OAuth()} with \texttt{OAuthHandler()}, \texttt{set\_access\_token()}, and \texttt{API()} with correct arguments,
\lm also splits the code into multiple statements, making the code more readable.
\input{c4-llm/lst/example-mm/_index.tex}

\vspace*{-1em}
\paragraph*{Different API styles}
We find that the LLMs are able to handle transformation between different API styles.
For example, the library \lib{argparse} provides API functions for parsing command line arguments, while \lib{click} provides decorators.
\autoref{fig:example-different-api-style} shows a migration example between these libraries.
Despite the completely different API styles, the three LLMs correctly migrate \CCArgparseClickCorrectPercent of \ccs between this library pair.
\input{c4-llm/lst/example-argparse-click/_index.tex}

\vspace*{-1em}
\paragraph*{Inferring changes outside of the source and target APIs}
We find that LLMs can infer some additional changes beyond the immediate APIs of the source or target library.
For example, the migration in \autoref{fig:example-correct-inferred-change} adds the \texttt{async\_mode} argument to the \texttt{SocketIO()} function.
This is interesting because the API is from the \lib{Flask-SocketIO} library, which is not the source (\lib{eventlet}) or target (\lib{gevent}) libraries.
However, the \texttt{SocketIO} function works with several libraries, \lib{eventlet} being the default \cite{flasksocketio}.
This means, if \fo did not explicitly set the \texttt{async\_mode} argument to \texttt{gevent}, the code would break after the migration.

\input{c4-llm/lst/example-4o-correct-inferred-change/_index.tex}

\subsection{Challenges of using LLMs for library migration}
From the results of the empirical study in Phase 1 and the above strengths, we see that LLMs have strong potential for library migration.
However, we also noted several challenges that would need to be addressed if LLMs are to be used for practical tooling for library migration.

\begin{enumerate}[label=\textbf{Challenge-\arabic*}, 
                  leftmargin=*,
                  itemindent=*]

\item \label{chal:migfile} \textbf{Files requiring migration may not be known upfront:}
In our experiment, we use \migbench to identify the files that require migration that we will send to the LLM.
However, in a real-world scenario, the files requiring migration may not be known upfront.
Therefore, a tool needs to first identify the files that require migration before sending them to the LLM for migration.

\item \label{chal:skipcode} \textbf{LLMs do not always return complete code:}
A common issue we observe across all three LLMs is that they often do not return the complete code.
Instead, they skips parts of the code, usually commenting that the next block of code remains unchanged.
This is especially when the skipped part does not require migration, despite our prompt explicitly asking them to return the entire code.
While this is not necessarily incorrect, as the skipped code may not require any changes, the resulting code is incomplete and cannot be provided directly to the developer.
    
\item \label{chal:dependentfiles} \textbf{Dependents of migrated files may also require changes:}
In our experiment, we provide the LLMs only the files that directly use the source library, as marked in \migbench.
However, other files that depend on the migrated files may also require changes.
This is especially true when functions are made asynchronous during the migration, as all parts of the code that use those functions must also be made asynchronous.

\item \label{chal:libversion}
\textbf{Library version compatibility:}
We do not mention any library version requirements in the prompt (\autoref{fig:prompt-mig-expermiment}), and we notice that the LLMs usually pick the APIs from the latest or recent versions of the target library.
In few cases, this caused runtime errors when running tests in LLM's code, which we report as incorrect in \rqTest.
This is because the application or the Python version used in the project may not be compatible with the latest version of the target library.
In an actual migration scenario, the developers may want to migrate to a specific version of the target library that is compatible with their project.
\end{enumerate}

%% file: c4-llm/lst/example-mm/_index.tex
\begin{figure}[!tp]
  \centering
  \begin{minipage}[t]{0.53\textwidth}
\begin{lstlisting}[
      language=Python, 
      title=Before migration (using \lib{twitter}), 
      xleftmargin=1em,
      firstnumber=4
      ]
tw_api = Twitter(
  auth = OAuth(oauth_token, oauth_secret,
         consumer_key, consumer_secret))

 ¤
\end{lstlisting}
  \end{minipage}
  \hfill
  \begin{minipage}[t]{0.45\textwidth}
    \vspace{0pt}%
\begin{lstlisting}[
      language=Python, 
      title=\lm{}'s migration (using \lib{tweepy}), 
      xleftmargin=1em,
      firstnumber=4
      ]
auth = OAuthHandler(
    consumer_key, consumer_secret)
auth.set_access_token(
    oauth_token, oauth_secret)
tw_api = API(auth)
\end{lstlisting}
  \end{minipage}   
  
  \caption{\lm correctly migrating a \mm code change}
  \label{fig:example-mm}
\end{figure}

%% file: c4-llm/lst/example-argparse-click/_index.tex
\begin{figure}[!tp]
  \centering
  \begin{minipage}[t]{0.55\textwidth}
\begin{lstlisting}[
      language=Python, 
      title=Before migration (using \lib{argparse}), 
      xleftmargin=1em,
      firstnumber=4
      ]
def parse_commands():
  p = ArgumentParser()
  p.add_argument("src", type=Path)
  p.add_argument("dst", type=Path)
  p.add_argument("usedir",
                 action = "store_true")
  return parser.parse_args()

def main():
  args = parse_commands()
  convert(args.src, args.dst, args.usedir)
\end{lstlisting}
  \end{minipage}
  \hfill
  \begin{minipage}[t]{0.43\textwidth}
    \vspace{0pt}%
\begin{lstlisting}[
      language=Python, 
      title=\fo{}'s migration (using \lib{click}), 
      xleftmargin=1em,
      firstnumber=4
      ]
@argument("src", 
          type=click.Path())
@argument("dst", 
          type=click.Path())
@argument("usedir", type=bool)
def main(src, dst, usedir):
  convert(src, dst, usedir)

¤
\end{lstlisting}
  \end{minipage}   
  
  \caption{\fo correctly migrating between libraries that use different API styles.}
  \label{fig:example-different-api-style}
\end{figure}

%% file: c4-llm/lst/example-4o-correct-inferred-change/_index.tex
\begin{figure}[!tp]
  \centering
  %\aMig{eventlet}{gevent}{stefal/rtkbase}{a4c347a2}\\
  \begin{minipage}[t]{0.40\textwidth}
\begin{lstlisting}[
      language=Python, 
      title=Before migration (using \lib{eventlet}), 
      xleftmargin=1em,
      firstnumber=3
      ]
socketio = SocketIO(app)
bootstrap = Bootstrap(app)
\end{lstlisting}
  \end{minipage}
  \hfill
  \begin{minipage}[t]{0.58\textwidth}
    \vspace{0pt}%
\begin{lstlisting}[
      language=Python, 
      title=\fo{}'s migration (using \lib{gevent}), 
      xleftmargin=1em,
      firstnumber=3
      ]
socketio = SocketIO(app, %\HLC{async\_mode="gevent"}%)
bootstrap = Bootstrap(app)
\end{lstlisting}
  \end{minipage}   
  
  \caption{\fo performing a correct inferred change}
  \label{fig:example-correct-inferred-change}
\end{figure}

%% file: c5-tool/_index.tex
\section{Phase 2: Building \tool}
\label{ch:tool}

\input{c5-tool/intro.tex}

\input{c5-tool/building-0-index.tex}

\input{c5-tool/eval-0-index.tex}

\input{c5-tool/discussion.tex}

%% file: c5-tool/intro.tex
% Storyline:
% 1. We do an experiment using the migrations in the benchmark, do manual evaluation. In this phase, we do multiple runs per file, and do CC level granular evaluation.
% 2. We learn from this phase, and improve the process and build a tool.
% 3. We do a large scale "simulated" evaluation of the tool on client projects, and show that it works well. This evaluation checks runtime behavior of the tool using unit tests, and does not require manual evaluation.
% 4. For a portion of the failed migrations, we attempt to fix them manually, and show that the LLMs were close to the correct migration.

In \autoref{ch:llm}, we show that LLMs can perform library migrations.
However, we also identify several challenges when using LLMs for library migration, especially when building an \textit{end-to-end} tool that developers can use.
In this section, we describe Phase 2 of our work, where we build \tool while addressing the identified challenges through pre- or post-processing steps as follows:
\begin{enumerate}
    \item Before migration, \tool identifies the files that use the source library, and thus require migration to address \ref{chal:migfile}.
    \item \tool allows the user to specify the source and target library version, and include this in the prompt, to address \ref{chal:libversion}.
    \item After migration, \tool automatically adds back any skipped code to address \ref{chal:skipcode}.
    \item After migration, \tool identifies the functions that become asynchronous due to the migration, and finds all other functions in the projects that use these functions, and makes them asynchronous as well to address \ref{chal:dependentfiles}.
  
\end{enumerate}

\input{c5-tool/img/tool-process.tex}

Figure~\ref{fig:migratelib-overview} shows an overview of how \tool works.
\tool works as a Command Line Interface (CLI) and takes the following as input:
\begin{enumerate*}[label=(\arabic*)]
    \item the path to a Python project that needs library migration,
    \item the source library name (the version is inferred),
    \item the target library name and version,
    \item the Python version in which the code should run,
    \item paths to one or more requirements files, 
    \item and an LLM to use for the migration.
\end{enumerate*}

\tool performs the migration in three main steps.
First, it sets up a virtual environment for the project and identifies all files that require migration (\textit{\PrepRound}).
In the next step, it prompts an LLM to migrate the identified files (\textit{\LLMMigRound}).
Finally, in the last step, it performs two post-processing steps to address \ref{chal:skipcode} and \ref{chal:dependentfiles} mentioned above to improve the migration results (\textit{\PostProcRound}).

We note that a key criteria for the input project is that it contains tests with high coverage,
as \tool uses the tests to verify the migration.
While \tool can perform the migration regardless of the test coverage, migration results without test validation may not be as reliable.
\tool also uses tests to complement the static analysis for identifying the files that require migration (Task 3, Section~\ref{subsec:prep-round} below), so the lack of tests may also impact the accuracy of the identified files.

%% file: c5-tool/img/tool-process.tex
\begin{figure}[t]
  \centering
  \begin{tikzpicture}
    \newcommand{\vSep}{1.5em}
    \newcommand{\hSep}{1.5em}
    \newcommand{\stepWidth}{.33\textwidth}
    \newcommand{\hideline}{none} % Change to 'none' when finished editing
    \newcommand{\headContentSep}{.25em}

    \tikzstyle{ioBox} = [
        trapezium, trapezium left angle=80, trapezium right angle=100,
        draw=black, thick,
    ];

    \tikzstyle{stepBox} = [
        rectangle, draw=black, thick,
        inner sep=.25em
    ];
    
    \tikzstyle{stepHead} = [
        draw=\hideline,
        rectangle, inner sep=0, text width=\stepWidth,
        align=center, font=\bfseries
    ];

    \tikzstyle{stepContent} = [
        draw=\hideline,
        rectangle, inner sep=0, text width=\stepWidth
    ];
    \tikzstyle{testResult} = [
        ioBox,
        text width=8em,
    ];

    \tikzstyle{compareBox} = [
        diamond, draw=black, thick, aspect=3, font=\scriptsize,
    ];

    \tikzstyle{dottedFrame} = [
        draw=black, dotted, thick,
    ];

    \tikzstyle{arrow} = [
        ->, thick
    ];

    %%%%% user input box start
    \node[stepHead](inputHead){User Input};
    \node[stepContent, text width=.55\textwidth, below=\headContentSep of inputHead, font=\scriptsize](inputContent){
            \textbullet~ Project path
            \textbullet~ Source lib
            \textbullet~ Target lib and version \\
            \textbullet~ Python version
            \textbullet~ requirements files
            \textbullet~ LLM
    };
    \node[stepBox, fit=(inputHead)(inputContent)] (input) {};
    
    %%%%% user input box end

    %%%%% Prep Round box start
    \node[stepHead, anchor=north west] (prepHead) at ([yshift=-\vSep]inputContent.south west) {\PrepRound};
    \node[stepContent, below=\headContentSep of prepHead] (prepContent){
        \textbullet~ Setup environment\\
        \textbullet~ Run tests\\
        \textbullet~ Identify files to migrate
    };
    \node[stepBox, fit=(prepHead)(prepContent)] (prep) {};
    \node[testResult, right=\hSep of prep] (testPre){\TestReportPre};

    %%%%% Prep Round box end

    %%%%% LLM Mig Round box start
    \node[stepHead, below=\vSep of prep] (llmMigHead){\LLMMigRound};
    \node[stepContent, below=\headContentSep of llmMigHead] (llmMigContent){
        \textbullet~ Prompt LLM\\
        \textbullet~ Run tests
    };

    \node[stepBox, fit=(llmMigHead)(llmMigContent)] (llmMig) {};
    \node[testResult, right=\hSep of llmMig] (testLLM){\TestReportLLM};
    \draw[arrow] (llmMig.east) -- (testLLM.west);

    \node[compareBox, right=\hSep of testLLM] (compareLLM){Has diff?};
    %\draw[arrow] (testPre.east) -- (compareLLM.north);
    \draw[arrow] (testLLM.east) -- (compareLLM.west);

    %%%%% LLM Mig Round box end

    %%%%% Merge Skipped Step box start
    \node[stepHead, below=\vSep of llmMig] (mergeSkippedHead){\MergeSkippedStep};
    \node[stepContent, below=\headContentSep of mergeSkippedHead] (mergeSkippedContent){
        \textbullet~ Find skipped code\\
        \textbullet~ Add skipped code back\\
        \textbullet~ Run tests
    };
    \node[stepBox, fit=(mergeSkippedHead)(mergeSkippedContent)] (mergeSkipped) {};
    \node[testResult, right=\hSep of mergeSkipped] (testMerge){\TestReportMerge};
    \draw[arrow] (mergeSkipped.east) -- (testMerge.west);

    \node[compareBox, right=\hSep of testMerge] (compareMerge){Has diff?};
    %\draw[arrow] (testPre.east) -- (compareMerge.north);
    \draw[arrow] (testMerge.east) -- (compareMerge.west);

    \draw[arrow] (compareLLM.south)  -- node[midway, below]{yes} (mergeSkipped.north east);

    %%%%% Merge Skipped Step box end

    %%%%% Async Transform Step box start
    \node[stepHead, below=\vSep of mergeSkipped] (asyncHead){\AsyncTransStep};
    \node[stepContent, below=\headContentSep of asyncHead] (asyncContent){
        \textbullet~ Find asynced functions\\
        \textbullet~ Find to-async functions\\
        \textbullet~ Find to-await calls\\
        \textbullet~ Apply transformations\\
        \textbullet~ Run tests
    };
    \node[stepBox, fit=(asyncHead)(asyncContent)] (async) {};
    \node[testResult, right=\hSep of async] (testAsync){\TestReportAsync};
    \draw[arrow] (async.east) -- (testAsync.west);

    \draw[arrow] (compareMerge.south)  -- node[midway, below]{yes} (async.north east);

    \node[dottedFrame, rounded corners, inner sep=.8em, fit=
        (mergeSkippedHead)(mergeSkippedContent)(testMerge)(compareMerge)
        (asyncHead)(asyncContent)(testAsync)
    ] (lastTwoRowsBox) {};
    \node[anchor=south east, font=\itshape] at (lastTwoRowsBox.south east) {Post-processing steps};
    
    %%%%% Async Transform Step box end

    % exit step beside compare box
    \node[ioBox, right=\hSep of testAsync] (output){
        Migration Report
    };

    %%% connections
    \draw[arrow] (input.south -| prep.north) -- (prep.north);
    \draw[arrow] (prep.east) -- (testPre.west);
    \draw[arrow] (prep.south) -- (llmMig.north);

    \draw[arrow] (compareLLM.south) to[out=-30,in=30,looseness=2] node[midway, right]{no} ($(output.north)+(0.6em,0)$);
    \draw[arrow] (compareMerge.south) -- node[midway, left]{no} (output.north);
    \draw[arrow] (testAsync.east) -- (output.west);

  \end{tikzpicture}
    \caption{Workflow of \tool}
    \mysubcaption{The \textit{Has diff?} nodes check whether a test report differs from \TestReportPre.}
  \label{fig:migratelib-overview}
\end{figure}

%% file: c5-tool/building-0-index.tex
\input{c5-tool/building-methods-prep.tex}
\input{c5-tool/building-methods-llmmig.tex}
\input{c5-tool/building-methods-post-processing.tex}

%% file: c5-tool/building-methods-prep.tex
\subsection{\PrepRound}
\label{subsec:prep-round}
The \prepround handles three main pre-processing tasks needed before the actual migration: 

\paragraph*{Task 1: Set up environment}
\tool creates a virtual environment for the provided Python version using the command \texttt{python -m venv .venv}, and then installs the required libraries using the \texttt{pip install -r <requirements-file-path>} command for each of the requirements files.
It then installs the target library at the provided version using the \texttt{pip install} command.
\tool requires some additional packages, such as \texttt{pytest}, \texttt{pytest-cov}, which it also installs during this step.
After the installation, \tool derives the source library version by querying the installed package using the \texttt{pip show <source-library-name>} command.

\paragraph*{Task 2: Run tests to get baseline}
Once the environment is set up, \tool runs the tests using the command \texttt{python -m pytest} on the project and saves the test report. 
We refer to this version of the test report as \textit{\TestReportPre}, which is used as the baseline for the migration.

\paragraph*{Task 3: Identify files that require migration}
\label{subsubsec:identifying-files}
This task addresses \ref{chal:migfile} from \autoref{ch:llm}, where we need to identify the files that require migration before sending them to the LLM.
One obvious way a file can use a library is by importing it.
For example, in \autoref{fig:lib-use-example}, the file \texttt{main.py} imports the \lib{geolib} library, thus it requires migration when migrating from \lib{geolib} to another library.
To identify such files, \tool reads and parses all Python files in the project and looks for import statements of the source library.
Note that a library can be imported using a different name than the library name; a library may also have multiple import names.
Therefore, \tool first finds all import names of the source library using the \textit{Johnnydep} library \cite{libjohnnydep}, which extracts the import names from the wheel file of the library.

A file may also use a library API without directly importing it.
An example is shown in \autoref{fig:lib-use-example}.
Notice that the \texttt{to\_dict()} function in the \texttt{serialize.py} takes as input a shape object from the \lib{geolib} library, even though it does not import the library.
It uses the \texttt{Shape.name} attribute and \texttt{Shape.area()} method, which are part of the \lib{geolib} library.
To identify such files, \tool builds a run-time call graph of the project by running the tests with profiling.
It then analyzes the call graph to find all files that call any function from the source library.

\input{c5-tool/lst/call-without-import-example}

Overall, in \autoref{fig:lib-use-example}, \tool will identify the file \texttt{main.py} requiring migration using static analysis, and \texttt{serialize.py} requiring migration using dynamic analysis.
We will mark both the files as requiring migration and send them to the LLM for migration in the next step.
There may be cases where the same file is identified by both static and dynamic analysis, which is not a problem as we will only send the file once to the LLM for migration.
Note that we search all Python files in the project, including test files, for both static and dynamic analysis.

%% file: c5-tool/lst/call-without-import-example.tex
\begin{figure*}[t]
   \centering
    \begin{minipage}[t]{0.4\textwidth}
        \vspace{0pt}%
        \begin{lstlisting}[language=Python, title=serialize.py]
def to_dict(shape):
  return {
    "name": %\HLC{shape.name}%,
    "area": %\HLC{shape.area()}%
  }
  ¤
\end{lstlisting}
\end{minipage}
\begin{minipage}[t]{0.4\textwidth}
\vspace{0pt}%
\begin{lstlisting}[language=Python, title=main.py]
from serialize import to_dict
%\HLC{from geolib import Square}%
import pprint

sq = %\HLC{Square(5)}%
pprint(to_dict(sq))
\end{lstlisting}
    \end{minipage}

\caption{Example of using a library without importing}
\label{fig:lib-use-example}
\end{figure*}

%% file: c5-tool/building-methods-llmmig.tex
\subsection{\LLMMigRound}
\label{subsec:llm-mig-round}

\tool can be configured to use any LLM that adheres to OpenAI's Chat Completions API format \cite{openaimodels}, which has become a standard API for most models.
For each file that requires migration, \tool generates a prompt using the template in \autoref{lst:prompt-tool} and sends the prompt to the LLM.
The prompt is based on our original prompt used in \autoref{ch:tool} (\autoref{fig:prompt-mig-expermiment}), but enhanced with more specific instructions.
For example, we address \ref{chal:libversion} by including the versions of the source and target libraries as an optional input that can be included in the prompt.
Rather than automatically infer potentially incorrect versions ourselves, we treat the library versions as input parameters that the user can specify (and which we then include in the prompt), which also gives the user more control over the migration process.

\input{c5-tool/lst/prompt.tex}

The LLM response contains the migrated code, which \tool extracts and saves.
When the LLM finishes migrating all files, \tool replaces the original files in the project with the migrated files.
It then runs the tests and generates a test report, which we refer to as \textit{\TestReportLLM}.

%% file: c5-tool/lst/prompt.tex
\begin{figure}[t]
\begin{lstlisting}[
    caption={Prompt template used by \tool}, 
    label={lst:prompt-tool},
    numbers=none,
    breakindent=0pt,
    xleftmargin=0pt,
]
The following Python code currently uses the library <source-lib> version
<source-ver>. Migrate this code to use the library <target-lib> version 
<target-ver> instead.

**Instructions:**
1. **Explain the Changes**: Begin the output with a brief explanation of the
   specific changes you made to migrate from "<source-lib>" to "<target-lib>".
2. **Provide the Modified Code**: After the explanation, present the modified
   code. Provide the entire code after migration even if only a part of it is
   changed.

**Important Guidelines**:
- Only make changes directly related to migrating between "<source-lib>" and 
  "<target-lib>".
- Do not refactor, reformat, optimize, or alter the original coding style.
- The code given to you is part of a larger application. Do not change the names 
  of classes, functions, or variables, because it can break the application.

Original code:
<premig-code>
\end{lstlisting}
\end{figure}

%% file: c5-tool/building-methods-post-processing.tex
%\subsection{\PostProcRound}
%\tool performs two post-processing steps we describe below.

\subsection{Post-processing Step 1: \MergeSkippedStep}
\label{subsec:merge-skipped-step}
We design this post-processing step to address \ref{chal:skipcode} from \autoref{ch:llm}.
The goal is to add back any unchanged code that LLM skipped returning in its response.
For each file that was migrated in the \llmmiground, 
\tool generates a diff between the original file and the migrated file using Python's \texttt{difflib} module.
Then it attempts to detect which of the hunks were entirely removed from the original file.
We make the following assumption to identify a hunk of skipped code:
\begin{enumerate}
    \item \label{no-src-api} The hunk should not use source library APIs such that it is not a migration-related change. Therefore, \tool checks if any of the removed lines in the hunk uses any source library API. If yes, the hunk is not considered skipped.
    
    \item \label{mostly-removal} The hunk must mostly include removal of lines. Our reasoning is that if the hunk has a similar number of additions and removals, it is likely that the hunk is replacing some code with other code, likely performing migration. If a hunk has more than 90\% of removed lines compared to total added+removed lines, we consider it mostly removal.

    \item \label{small-hunk} The hunk should not be too small. This assumption avoids incorrectly considering small migration-related removals as skipped code. We consider a hunk as too small if it has less than 9 lines of code.   
\end{enumerate}

We determine the above criteria thresholds through experimentation on a set of migration examples. 
If a hunk is identified as skipped code by the above criteria, \tool adds back the removed lines from the original file to the migrated file.
After finishing the merging of skipped code, \tool runs the tests again and generates a new test report, which we refer to as \textit{\TestReportMerge}.

\subsection{Post-processing Step 2: \AsyncTransStep}
\label{subsec:async-trans-step}
This post-processing step addresses \ref{chal:dependentfiles} from \autoref{ch:llm} by identifying functions that become asynchronous due to the migration, and making their transitive callers asynchronous as well.
Below we describe the steps in detail using the example in \autoref{fig:async-transform-example}.

\input{c5-tool/lst/async_transform_example/_index.tex}

\paragraph*{Identify asynced functions}
We call a function \textit{asynced} if it was synchronous before migration, but is asynchronous after the LLM migration.
For a given migrated file, \tool creates an abstract syntax tree (AST) of the code before migration and another AST after migration.
It then finds the function definitions that are not asynchronous in the AST before migration, but are asynchronous in the AST after migration.
Since the location (line number) of the function definition may change due to the migration, \tool matches the functions by their qualified name.
In \autoref{fig:async-transform-example}, the function \texttt{fetch\_data} is asynced.

\paragraph*{Identify to-async functions}
If a function \texttt{foo} calls a function \texttt{bar}, and \texttt{bar} calls a function \texttt{fizz}, then \texttt{foo} is said to transitively call \texttt{bar} and \texttt{fizz}.
For a given asynced function, \tool finds all its transitive callers using the call graph, and marks them as \textit{to-async}, meaning that the async keyword must be added to their function definition.
For example, in \autoref{fig:async-transform-example}, the function \texttt{fetch\_data} in \texttt{service.py} is asynced after the LLM migration. 
This function is called by the function \texttt{test\_fetch\_data} in \texttt{service\_test.py}, therefore, \tool marks \texttt{test\_fetch\_data} as to-async.

\paragraph*{Identify to-await calls}
\tool again uses the call graph to find all call sites of the asynced and to-async functions, and marks them as \textit{to-await}, meaning that an \texttt{await} keyword must be added to these calls.
For example, in the call to \texttt{fetch\_data} in the function \texttt{test\_fetch\_data} (line 5, after LLM migration) is marked as to-await, because it is a call to an asynced function.

\spaceOutPar
Upon finding all asynced functions, to-async functions, and to-await calls, \tool applies the necessary code transformations using the \lib{LibCST} library \cite{liblibcst}.
Additionally, \tool decorates any asynced and to-async test functions with the \texttt{pytest.mark.asyncio} decorator, and installs the \textit{pytest-asyncio} package to the virtual environment, which is required to run asynchronous tests.
After finishing the async transformations, \tool runs the tests again and generates a new test report, which we refer to as \textit{\TestReportAsync}.

%% file: c5-tool/lst/async_transform_example/_index.tex
\begin{figure*}[!htp]
  \centering
  \begin{minipage}[t]{0.53\textwidth} % top left
    \vspace{0pt}%
    \centering{{\texttt{\normalsize{service.py}}}}    
  \end{minipage}
  \hfill
  \begin{minipage}[t]{0.46\textwidth} % top right
    \vspace{0pt}%
    \centering{{\texttt{\normalsize{service\_test.py}}}}    
  \end{minipage}\\
  
  \vspace*{.5em}
  Before migration\\
  \vspace*{-.5em}
  
  \begin{minipage}[t]{0.53\textwidth} % top left
    \vspace{0pt}%
\begin{lstlisting}[language=Python]
import requests

def fetch_data(year):
  url = f"https://api.io/data/{year}"
  res = requests.get(url)
  res.raise_for_status()  
  data = req.json()
  return data
\end{lstlisting}
  \end{minipage}
  \hfill
  \begin{minipage}[t]{0.46\textwidth} % top right
    \vspace{0pt}%
\begin{lstlisting}[language=Python]
from service import fetch_data

def test_fetch_data():
  year = 2023
  data = fetch_data(year)
  assert isinstance(data, dict)
  assert "days" in data
  assert len(data["days"]) == 365
\end{lstlisting}
  \end{minipage}  \\
  
  \vspace*{.5em}
  After LLM migration\\
  \vspace*{-.5em}
  \begin{minipage}[t]{0.53\textwidth} % mid left
    \vspace{0pt}%
\begin{lstlisting}[language=Python]
%\HLC{from aiohttp import ClientSession}%

%\HLC{async}% def fetch_data(year):
  url = f"https://api.io/data/{year}"
    
  %\HLC{async with ClientSession() as session:}%
    %\HLC{async with session.get(url) as res:}%
      res.raise_for_status()
      data = %HLC{await}% res.json()
      return data

\end{lstlisting}
  \end{minipage}
  \hfill
  \begin{minipage}[t]{0.46\textwidth} % mid right
    \vspace{0pt}%
\begin{lstlisting}[language=Python]
from service import fetch_data

def test_fetch_data():
  year = 2023
  data = fetch_data(year)
  assert isinstance(data, dict)
  assert "days" in data
  assert len(data["days"]) == 365

%\InvisibleCode%
\end{lstlisting}
  \end{minipage}\\

  \vspace*{.5em}
  After \AsyncTransStep\\
  \vspace*{-.5em}

  \begin{minipage}[t]{0.53\textwidth} % bottom left
    \vspace{0pt}%
\begin{lstlisting}[language=Python]
from aiohttp import ClientSession

async def fetch_data(year):
  url = f"https://api.io/data/{year}"
    
  async with ClientSession() as session:
    async with session.get(url) as res:
      res.raise_for_status()
      data = await res.json()
      return data
\end{lstlisting}
  \end{minipage}
  \hfill
  \begin{minipage}[t]{0.46\textwidth} % bottom right
    \vspace{0pt}%
\begin{lstlisting}[language=Python]
%\HLC{import pytest}%
from service import fetch_data

%\HLC{@pytest.mark.asyncio}%
%\HLC{async}% def test_fetch_data():
  year = 2023
  data = %\HLC{await}% fetch_data(year)
  assert isinstance(data, dict)
  assert "days" in data
  assert len(data["days"]) == 365
\end{lstlisting}
  \end{minipage}
  \caption{Example of applying LLM migration and then \asyncTransStep}
  \mysubcaption{The first row presents the original code before migration. The second row displays the code after LLM migration. The third row shows the after applying \asyncTransStep. The left column contains the main code, while the right column contains its associated test. Changes introduced by each migration step are highlighted in red.}
  \label{fig:async-transform-example}
\end{figure*}

%% file: c5-tool/eval-0-index.tex
\section{Phase 3: Evaluating \tool}  
\label{sec:tool-eval}
In this phase, we want to run a large-scale evaluation of \tool.
While using \migbench in Phase 1 allowed us
to compare the LLM migrations to a developer-performed ground truth migration for systematic understanding of LLM capabilities, it has several limitations: (1) not all projects have tests, which makes an automated test-based evaluation infeasible/limited and (2) all the migrations in \migbench happened prior to
the known training cutoff date of the three models (Late
2023). Therefore, there is a possibility that the LLMs are
simply reciting the migrated code they have seen before (for
this particular codebase). 
Accordingly, in Phase 3, we come up with a different evaluation set up that allows us to evaluate \tool on a larger scale and on migrations that the LLMs is unlikely to have seen before, while still being able to automatically determine the correctness of the migration without manual evaluation.
We resort to the idea of \textit{simulated migrations} where we prioritize finding Python projects with high-coverage test suites. We then find source libraries used in the projects and migrate the project to use an analogous target library.

Our evaluation aims to answer the following research questions:
\newcommand{\rqcorrectness}{How many migrations can \tool perform correctly?}
\newcommand{\rqManualEffort}{How much further code changes are required to complete the incomplete migrations?}

\begin{enumerate}[label={RQ-2.\arabic*},leftmargin=*]
    \item \label{rq:correctness} \textbf{\rqcorrectness}  We measure how well \tool performs on the simulated migrations without any manual intervention.   
    \item \label{rq:manualEffort} \textbf{How much further code changes are required to complete the incomplete migrations?} While the best case scenario is that \tool can migrate the project without any issues, a more practical expectation is that \tool would do most of the task correctly, but there are some parts that require manual intervention. In this RQ, we measure the remaining manual work required based on the size of remaining code changes. 
\end{enumerate}

\input{c5-tool/eval-method.tex}

\input{c5-tool/eval-RQ-correctness.tex}
\input{c5-tool/eval-RQ-effort.tex}

%% file: c5-tool/eval-method.tex
\subsection{Data collection and selected model}
We want to find repositories that use some source library, and then migrate them to use an analogous target library.
To make the test-based evaluation reliable, we want to select repositories whose tests have good code coverage.
We start by downloading a list of \RepoSeartCount Python GitHub repository metadata using SEART~\cite{seart} service.
We use the filters to select Python repositories that have at least one commit after \SeartFilterLastCommitAfter to discard repositories that are not actively maintained.
We also filter out repositories with less than \SeartFilterMinLines lines of code to ensure that the repositories have some code to migrate.
We then filter out repositories using the following criteria before cloning them for further analysis:

\begin{enumerate}[label=\roman*.]    
    \item \textbf{Large repositories:} Extra-large repositories would take too long to clone and analyze, so we filter out \RepoExcludedLargeCount repositories larger than 10MB, leaving us with \RepoAfterLargeFilterCount repositories.
    \item \textbf{No requirements files:} We use requirements files to build the virtual environment to be able to run the tests. We filter out \RepoExcludedNoReqFileCount repositories that do not have a \texttt{requirements.txt} file in the root directory, leaving us with \RepoAfterNoReqFileFilterCount repositories.
    \item \textbf{No test files:} Python test file names usually contain the term \textit{test}. We use GitHub code search API\footnote{GitHub code search API: \url{https://docs.github.com/en/rest/search/search\#search-code}} to query files that contain the term \textit{test}, and filter out a repository if it does not have any such files. We filter out \RepoExcludedNoTestFileByApiCount repositories that do not have any test files, leaving us with \RepoAfterNoTestFileByApiFilterCount repositories. 
\end{enumerate}

When a repository passes the above heuristics, we clone it and attempt to run the tests.
In this process, we filter out the repositories with the following criteria:

\begin{enumerate}[label=\roman*., resume]
    \item \textbf{Could not run tests:} We attempt to set up a virtual environment for the repository using the requirements file and run the tests following \autoref{sec:eval:test} (\nameref{sec:eval:test}).        
    Due to various repository-specific requirements, \eg Python version, system requirements, and incomplete requirements file, we are often unable to run the tests.
    We filter out \RepoExcludedTestNotRunCount repositories that we could not run the tests for, leaving us with \RepoAfterTestNotRunFilterCount repositories.
    \item \textbf{Low passing rate:} We filter out the repositories that have less than 90\% of the tests passing, which filters out \RepoExcludedLowPassingCount repositories, leaving us with \RepoAfterLowPassingFilterCount repositories.
    \item \textbf{Low coverage:} We filter out repositories that have less than \MinCover line coverage by the tests, which filters out \RepoExcludedLowCoverageCount repositories, leaving us with \RepoAfterLowCoverageFilterCount repositories.
    \item \textbf{Tests require user interaction:} We identify \RepoExcludedManualCount repositories that require manual intervention and exclude them, leaving us with \RepoAfterManualFilterCount repositories, which we use for evaluation.
    We store the commit SHA for the version of the repository we use to ensure reproducibility.
\end{enumerate}

We then collect the dependencies of the \RepoIncludedCount selected repositories by parsing their requirements files.
This gives us an initial list of \SourceLibCandidateCount unique source libraries.
We exclude \SourceLibExcludedNotOnLibioCount libraries that are not found on libraries.io \cite{libraries.io}, leaving us with \SourceLibAfterNotOnLibioFilterCount libraries.
To ensure that the libraries are commonly known, we filter out \SourceLibExcludedTooFewDependentsCount libraries that have less than 100 dependents, and filter out \SourceLibExcludedNoDescriptionCount libraries that do not have a description, leaving us with \SourceLibAfterNoDescriptionFilterCount potential source libraries.

\begin{figure}[t!]
\begin{lstlisting}[
    caption={Prompt template for finding candidate analogous libraries},
    label={lst:prompt-candidate-target-lib},
    numbers=none,
    breakindent=0pt,
    xleftmargin=0pt,
]
Identify alternative Python libraries that offer similar functionality as
<lib_name>, allowing for potential replacement in applications. The alternative libraries should be available on PyPI. 
Produce the result as a JSON list, having only the library names. The produced library names should match the PyPI package names.
\end{lstlisting}
\end{figure}

For each of the \SourceLibIncludedCount source libraries, we find a list of candidate analogous libraries using \LLMVersion with the prompt template in \autoref{lst:prompt-candidate-target-lib}.
This gives us a total of \LibPairCandidateCount candidate library pairs.
All of these pairs may not actually be analogous, \ie they may not be suitable for migration, so we need to verify them.
This manual task can be very time-consuming; therefore, we first automatically filter out some of the pairs.
First, we filter out \LibPairLowFrequencyCount pairs where the source library is used in less than 10 of the selected repositories.
This helps us to keep the manual effort of evaluating analogous pairs manageable by only selecting pairs that appear in multiple projects in our data set.
Then, we filter out \LibPairOldTargetCount pairs with an unmaintained potential target library.
Specifically, we exclude the target libraries that have not been updated since January 1, 2024.
This leaves us with \LibPairManuallyValidatedCount pairs to manually verify.

During manual validation, we look into the library documentation and confirm \LibPairAnalogousCount library pairs that are indeed analogous and can be potentially migrated.
We exclude the remaining because they fall into one of the following categories \begin{enumerate*}[label=(\roman*)]
    \item \textbf{Not analogous:} Some libraries are not analogous, \eg they provide different functionality or have different APIs.
    \item \textbf{Test related libraries:} For example, \textit{pytest}, \textit{mock}, etc. We exclude these libraries because our evaluation process relies on the tests. Therefore, migrating the test libraries themselves may break our automated evaluation.
    \item \textbf{Non-API tools:} Some dependencies are command line interfaces. Since they are not used in the code, we ignore them. For example, \textit{black}, \textit{flake8}, etc.
    \item \textbf{Documentation tools:} Some dependencies are used for documentation generation. Since they are not used in the code, we ignore them. For example, \textit{sphinx}, \textit{mkdocs}, etc.    
\end{enumerate*}

Now that we have a list of \RepoIncludedCount repositories and \LibPairAnalogousCount analogous library pairs, we want to find which repositories actually use the source libraries in our list.
It could be the case that some repositories list a requirement that is not actually used in the code.  
Accordingly, we parse the source code of each repository and look for import statements that use the source libraries.
We find that \RepoExpCount repositories use at least one of the source libraries.
Note that each source library can have one or more analogous target libraries (e.g., \aPair{requests}{aiohttp} and \aPair{requests}{httpx}).
We use the term \textit{unique migration instance} to refer to the combination of a repository, a commit, a source library, and a target library.
We use the latest commit of each repository at the time of our experiment (January 2025) for this purpose.
We find a total of \MigExpCount unique migration instances, spanning \RepoExpCount repositories and \LibPairExpCount library pairs that we use for our experiments.
The \LibPairExpCount library pairs include \SourceLibExpCount unique source libraries and \TargetLibExpCount unique target libraries, and \LibExpCount unique libraries in total.
Across the migration instances, the minimum number of files with source library usages to migrate is \MigFilesMin and the maximum is \MigFilesMax, with a mean of \MigFilesMean and a median of \MigFilesMedian files to migrate per migration instance.

In our initial LLM empirical study (\autoref{ch:llm}), we see that \fouro performs better than the other two models.
Therefore, we use \fouro (gpt-4o-2024-11-20) to run \tool in this evaluation.

\subsection{Evaluation metrics}

Recall from \autoref{ch:tool} that \tool generates \TestReportPre before the migration, and three test reports after each migration step: \TestReportLLM after the \llmmiground, \TestReportMerge after the \mergeSkippedStep, and \TestReportAsync after the \asyncTransStep.
For a given post-migration test report, we define the \textit{correctness} of a migration after that step as the proportion of tests that pass in the test report compared to the tests that passed in the \TestReportPre.
For example, if 80 tests passed in \TestReportPre, and 60 of \textit{those} tests passed in \TestReportLLM, then the correctness after the \llmmiground is 75\% (60/80). 

It is possible that some migrations may not be fully correct even after all the \tool steps.
We want to see how much migration work is still needed after \tool migration to get a fully correct migration.
Therefore, we randomly select one migration instance from each of the library pairs and attempt to manually fix them.

After fixing a migration as much as possible, we run the tests again and generate a new test report, which we refer to as \textit{\TestReportManual}.
We then compute the correctness of the manually fixed version.
For the migrations which become fully correct, \ie 100\% correctness, we estimate the amount of manual work needed to fix them to answer \ref{rq:manualEffort}.
We use line number of modified lines to estimate the amount of manual change required to complete a migration.
For a migration, we compute a diff between the code before manual migration and the code after manual migration.
From the diff, we calculate the number of changed lines of code by adding number of removed lines and number of added lines.
We refer to this number as \migLOCManual.
Now, this number gives us the absolute number of lines changed,
therefore we cannot directly compare this number across different migrations. 
For example, a migration that required 50 changed lines where the tool performed 48 of them and we only have to edit 2 lines should not be considered the same as a migration that required 4 lines of code and we still have to edit 2 lines.
Therefore, we also measure the number of lines changed by \tool during the automated migration from the diff between the code before automatic migration and the code after migration by \tool.
We refer to this number as \migLOCAuto.
Accordingly, the total number of lines that needed to be changed for the migration is the sum of these two numbers.
We use these to normalize the \migLOCManual as follows:

\begin{equation}
\label{eq:manual-effort-metric}
    \effortManual = \frac{\migLOCManual}{\migLOCAuto + \migLOCManual} \times 100
\end{equation}

Similarly, normalized automatic migration changes is calculated as the ratio of automatic changes to the total changes.
\begin{equation}
\label{eq:auto-effort-metric}
    \effortAuto = \frac{\migLOCAuto}{\migLOCAuto + \migLOCManual} \times 100
\end{equation}

%% file: c5-tool/eval-RQ-correctness.tex
\subsection{\ref{rq:correctness}: \rqcorrectness}
\label{subsec:rq1}
\begin{figure}[t]
    \centering
    \includegraphics[width=.7\columnwidth]{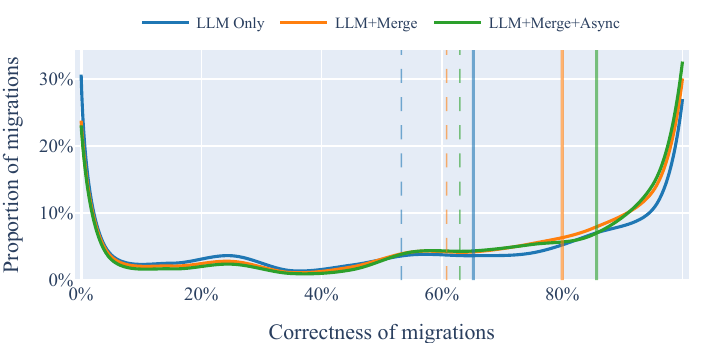}
    \caption{Distribution of correctness of migrations.}
    \mysubcaption{The dashed lines indicate the mean, the solid lines indicate the median.}
    \label{fig:mig-correctness-violin}
\end{figure}

\input{c5-tool/tab/ablation.tex}

\paragraph*{Correctness according to applied steps.}
\autoref{tab:ablation-comparison} summarizes the results of \ref{rq:correctness}, while comparing the different migration steps.
The table shows multiple metrics: 
the percentage of fully correct migrations is the proportion of migrations with 100\% correctness.
Median correctness is the median of the correctness value across all evaluated migrations.
Number of migrations applied to is the number of migrations where this step was applicable and thus got applied.
Number of migrations improved shows the number of migrations where correctness improved in comparison to the previous step.
Average improvement in correctness is the mean of improvement for the migrations where correctness improved after a post-processing step.
Note that the post-processing steps are applied only when the migrations are not fully correct and when that specific post-processing step is applicable.
For example, as shown in \autoref{tab:ablation-comparison}, \mergeSkippedStep is applied to \MigMergeSkippedAppliedCount migrations, among which \MigMergeSkippedImprovedCount (\MigMergeSkippedImprovedPercent) migrations have improved correctness with an average improvement of \MigMergeSkippedMeanImprovement per migration.
On the other hand, \asyncTransStep is applied to \MigAsyncTransformAppliedCount migrations, among which \MigAsyncTransformImprovedCount (\MigAsyncTransformImprovedPercent) migrations have improved correctness with an average improvement of \MigAsyncTransformMeanImprovement per migration.

The frequency curve chart in \autoref{fig:mig-correctness-violin} shows the distribution of correctness of migrations after each migration step.
All three distributions have concentrated values at 0\% and 100\%, indicating that most of the migrations either fail all tests or pass all tests. 
We see that the median correctness of migration increased from \MigLLMMigMedianCorrectness in the \llmmiground to \MigMergeSkippedMedianCorrectness after the \mergeSkippedStep, and further increased to \MigAsyncTransformMedianCorrectness after the \asyncTransStep.
Also note that out of the three migration steps, LLM-only has the highest proportion of migrations with 0\% correctness; this proportion decreases after each post-processing step.
Similarly, the proportion of migrations with 100\% correctness increases after each post-processing step.
The distributions show that after post-processing, there are fewer migrations with zero or low correctness, and more migrations with high correctness.
This indicates that the post-processing steps are effective in improving the correctness of migrations. 

To further understand when post-processing steps help or hinder the migration process, we manually analyze some of the migrations where the post-processing steps were applied.
As shown in Table~\ref{tab:ablation-comparison}, \mergeSkippedStep got applied to \MigMergeSkippedAppliedCount migrations, among which \MigMergeSkippedImprovedCount (\MigMergeSkippedImprovedPercent) migrations have improved correctness.
The most common scenario for the improved correctness cases is that the LLM skipped code at the end of long files, classes, or methods. In most cases, these parts did not even contain any source library API usage, and therefore, the LLM did not need to change them. Accordingly, adding back these parts as is ensured that the migrated code is complete and does not miss any code that may be required for the tests to pass.
We found one interesting case in a \aPair{requests}{httpx} migration in the \aRepoLink{bheinzerling/bpemb} repo.
Here, the LLM removed function \texttt{\_get\_project} from the middle of the file.
While this particular function did not require any migration, the LLM did edit parts of the file after the function was removed. Thus, the LLM did not simply stop migration like the common observed case; rather it intentionally removed this function.
The LLM likely removed this function because it was not used within the file; however, the function was used in the test suite and thus its removal causes tests to fail.
Our \mergeSkippedStep post-processing step added this function back to the migrated code, which improved the correctness of the migration.
Interestingly, we found that the consequent commit in the repo added source code that uses this function, which indicates that the LLM's decision to remove this function was not correct, and our post-processing step helped to fix this mistake.
As \autoref{tab:ablation-comparison} shows, applying \mergeSkippedStep did not change correctness in 131 (50\%) migrations.
This was usually because the migrated code itself had multiple problems that led most of the test suite to fail, with none of the failures particularly related to the skipped code. Consequently, restoring the skipped code did not improve correctness.
In 41 cases (16\%), applying \mergeSkippedStep reduced correctness. Our manual analysis revealed that alignment does not always work as expected when the skipped code occurs between other code fragments that were migrated. Specifically, reinserting the skipped code into the migrated code at positions shifted by the migration sometimes introduced syntax errors, causing all tests to fail. 

As Table~\ref{tab:ablation-comparison} shows, \asyncTransStep got applied to \MigAsyncTransformAppliedCount migrations, among which \MigAsyncTransformImprovedCount (\MigAsyncTransformImprovedPercent) migrations have improved correctness.
In these cases, our qualitative analysis shows that \tool correctly added the \texttt{async} or \texttt{await} keywords to/within the right functions, even when those functions appear in additional files that do not use the source library API and were accordingly not provided to the LLM.
% For the \asyncTransStep, we find that \tool always correctly applies the transformation for all the functions that need to be made asynchronous in files that have no source library API usage and thus were not seen by the LLM.
% In the case of \asyncTransStep, we find that when this step is a LLM always correctly applies the async transformation for the code that it migrated, but the step is still needed for the files where the source library API is not used.
% We have one unresolved limitation in the \asyncTransStep step.
We find only 17 cases where applying \asyncTransStep made the correctness worse.
All of these cases were due to very specific corner case usages of the usage of asynchronous functions in Python that \tool does not currently handle.
One example is when a function that got asynched during migration is called in a constructor of a class (i.e., \_\_init\_\_ method).
For example, assume \texttt{fetch\_year} from Figure~\ref{fig:async-transform-example} is called in a constructor of a class.
Since \tool identifies the constructor as a caller of the asynched function, it adds the \texttt{async} keyword to the constructor. However, asynchronous constructors are not allowed in Python, causing further tests to fail after the change.
Instead, the expected workaround is to use a asynchronous factory method to create the object.
There are only a few such corner cases, and handling them can be future feature enhancements to \tool.
In the remaining 67 (52\%) cases, the \asyncTransStep did not change correctness, which was usually because the migrated code itself had multiple problems that led most of the test suite to fail, with none of the failures particularly related to the async transformation.

\paragraph*{Difficulty of attempted migrations.}
To understand how difficult the migrations \tool attempted are, we also examine the number of diff hunks per migration, number of changed lines per migration (total of number added plus deleted), and the program elements that appear in the changes.
The number of hunks in a migration indicates how many separate code changes did \tool make in the migration, while the number of changed lines indicates the size of the migration.
We find that the minimum number of hunks in a migration is \MigHunksMin and the maximum is \MigHunksMax, with a mean of \MigHunksMean and a median of \MigHunksMedian hunks per migration.
As for size, the minimum number of changed lines in a migration is \MigLinesMin and the maximum is \MigLinesMax, with a mean of \MigLinesMean and a median of \MigLinesMedian lines changed per migration.

Automatically determining all code changes according to \taxonomy is difficult as it goes beyond simple AST diffing.
To still provide some proxy for the types of changes involved in the migration, we parse the ASTs to identify the program elements that appear in the changed lines of the migrations. We count the program elements that appear in the changed lines in both the before and after files.
Since there are many AST node types, we focus on a few key program elements that are commonly involved in library migrations, including function calls, function definitions, exceptions, decorators, and attributes. 
We find that there are only \MigImportOnlyCount migrations where only an import replacement happened.
In all remaining migrations, the changes involved actual code changes.
Function call replacements happen in \MigFunctionCallPercent of the migrations, with an average of \MigFunctionCallAvg function call replacements per migration. Exceptions are the second most common type of change, occurring in \MigExceptionPercent of the migrations, with an average of \MigExceptionAvg exception changed per migration.
Decorators and attributes also appear in the migration changes, occurring in \MigDecoratorPercent and \MigAttributePercent of the migrations, respectively, with an average of \MigDecoratorAvg and \MigAttributeAvg changes per migration, respectively.
Overall, this demonstrates that the migrations in our evaluation are not trivial, and involve a variety of code changes beyond just import replacement, which makes the migration process more complex and thus provides a good test for \tool's capabilities. 

\begin{figure}
    \centering
    \includegraphics[width=.7\columnwidth]{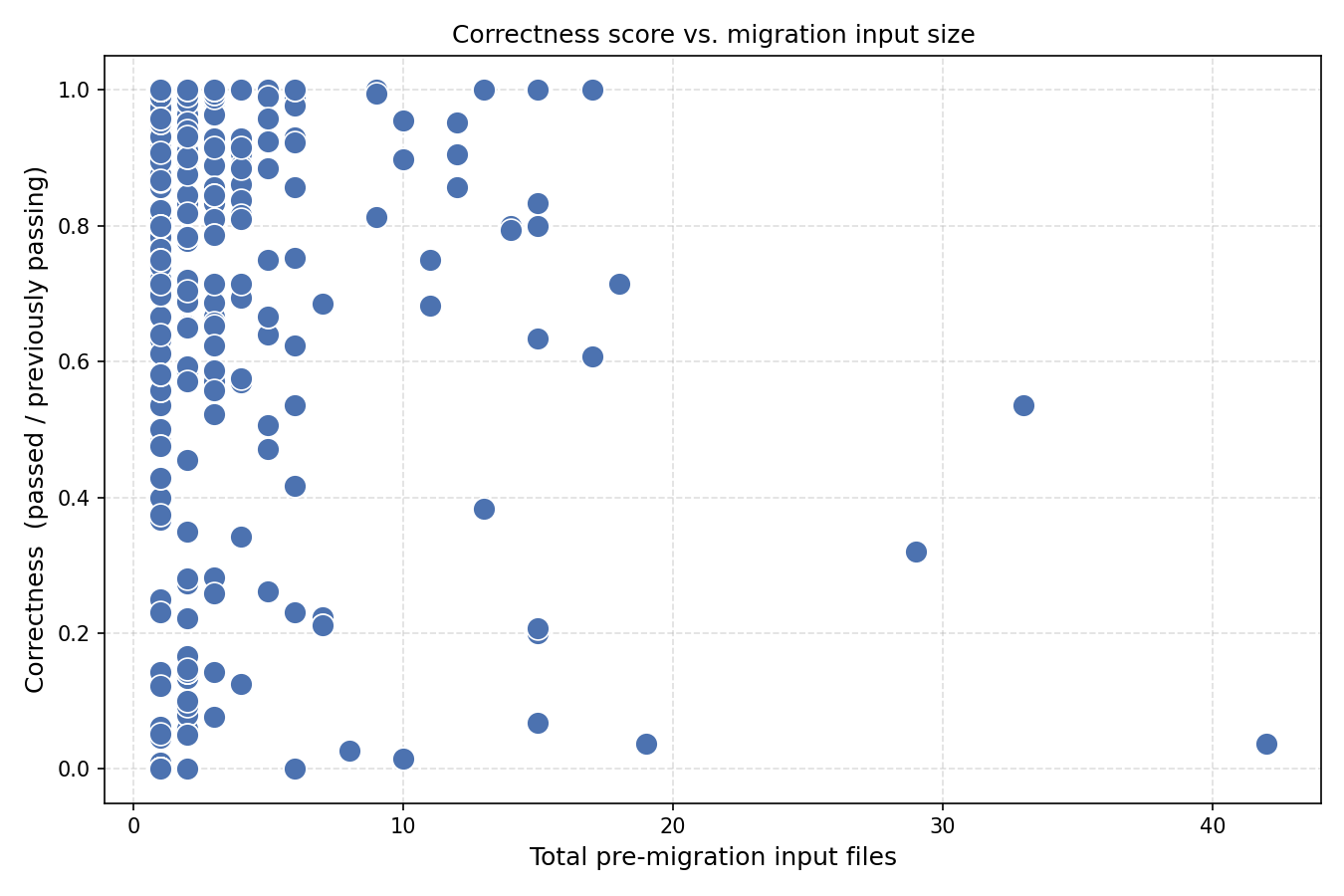}
    \caption{Correctness vs. number of files to migrate. One outlier with \MigFilesMax is excluded.}
    \label{fig:correctness-vs-size}
\end{figure}

Since we do not know the exact code changes required to complete each migration, we cannot directly compare the difficulty of the migrations that \tool successfully completed vs. those it did not complete.
The only pre-determined piece of information we have about the migrations is the number of files that require migration, which can be a proxy for the size of the migration.
Figure~\ref{fig:correctness-vs-size} shows the achieved correctness vs. number of input files to migrate for each migration. We can see that migrations with a high number of input files can still achieve high correctness, and there is no strong correlation between the number of files to migrate and the achieved correctness.

For the \MigFullyCorrectCount migrations that are fully correct, we have a ``ground truth'' of the required code changes to complete the migration.
Thus, if we specifically look at these \MigFullyCorrectCount fully correct migrations, we find that the number of files in those migrations ranges from \MigFCFilesMin to \MigFCFilesMax, with average \MigFCFilesMean. 
The number of hunks ranges from \MigFCHunksMin to \MigFCHunksMax, with average \MigFCHunksMean, and the number of changed lines ranges from \MigFCLinesMin to \MigFCLinesMax, with average of \MigFCLinesMean.
The code changes involved in these migrations are also not trivial, with \MigFCFunctionCallPercent of them involving function call replacements, \MigFCExceptionPercent involving exception changes, \MigFCDecoratorPercent involving decorator changes, and \MigFCAttributePercent involving attribute changes.
This gives us further confidence that the migrations \tool successfully completed are not trivial and vary in size and amount of required code changes.

\begin{tcolorbox}[colback=resultbg, title=\ref{rq:correctness}]
\tool is able to fully migrate \MigAsyncTransformCorrectPercent of the migrations without any manual intervention.
The median correctness of migrations improves from \MigLLMMigMedianCorrectness with only the \llmmiground to \MigAsyncTransformMedianCorrectness after the post-processing steps.
\end{tcolorbox}

%% file: c5-tool/tab/ablation.tex
% create a table

\begin{table*}[t]
    \caption{Comparison between migration steps}    
    \centering
    \label{tab:ablation-comparison}
    \tablefontsize
    \begin{tabular}{lcccc}
    \toprule
    Metric & LLM Only & LLM+Merge & LLM+Merge+Async \\
    \midrule
    \# migrations applied to    & \MigExpCount                  & \MigMergeSkippedAppliedCount      & \MigAsyncTransformAppliedCount \\
    \# (\%) fully correct migrations & \MigLLMMigCorrectCount (\MigLLMMigCorrectPercent)       & \MigMergeSkippedCorrectCount (\MigMergeSkippedCorrectPercent)    & \MigAsyncTransformCorrectCount (\MigAsyncTransformCorrectPercent)\\
    Median correctness          & \MigLLMMigMedianCorrectness   & \MigMergeSkippedMedianCorrectness & \MigAsyncTransformMedianCorrectness \\
   \# (\%) Migrations improved       & --                            & \MigMergeSkippedImprovedCount (\MigMergeSkippedImprovedPercent)     & \MigAsyncTransformImprovedCount (\MigAsyncTransformImprovedPercent) \\
   \# (\%) Migrations remained same       & --                            & 131 (50\%)     & 67 (52\%) \\
   \# (\%) Migrations got worse       & --                            & 41 (16\%)     & 17 (13\%) \\
    Average improvement in correctness & --                     & \MigMergeSkippedMeanImprovement   & \MigAsyncTransformMeanImprovement \\

    \bottomrule
    \end{tabular}
\end{table*}

%% file: c5-tool/eval-RQ-effort.tex
\subsection{\ref{rq:manualEffort}: How much further code changes are required to complete the incomplete migrations?}
\label{subsec:rqeffort}
\ref{rq:correctness} shows that \MigOverallCorrectCount migrations are completely correct after the automated migration process.
The remaining \MigOverallNeedManualFixCount migrations will therefore require the developers to fix/complete them.
To mimic what developers would do, we attempt to fix the migrations ourselves.
However, the \MigOverallNeedManualFixCount migrations are between \MigManualFixApplied unique library pairs in \RepoNeedManualFixCount repositories.
To reduce the amount of manual work needed, while keeping the dataset diverse, we attempt to manually fix one incomplete migration from each library pair.
Our selection results in a total of \MigManualFixApplied migrations between \MigManualFixApplied library pairs in \RepoManualFixApplied unique repositories, which we proceed to fix.

\begin{figure}[t]
\centering
\includegraphics[width=\linewidth]{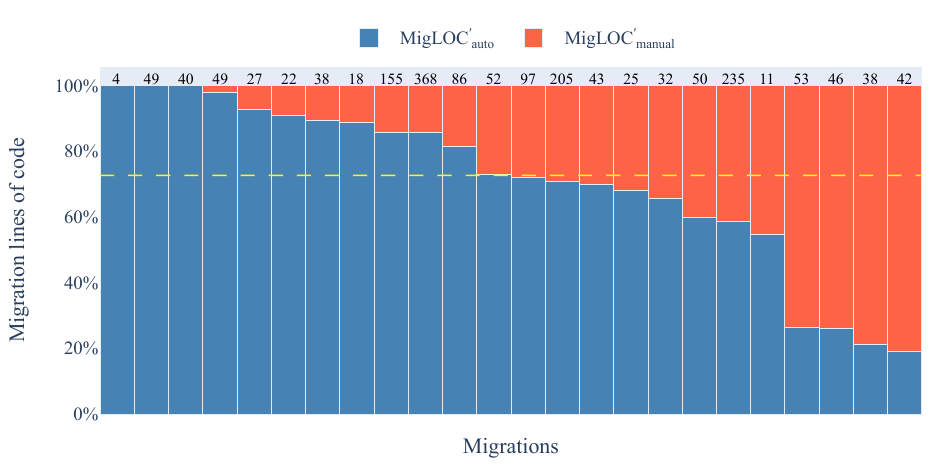}
\mysubcaption{The yellow dotted line shows median automatic change across all \MigManualFixCorrected migrations. The numbers on top of each bar show the total changed lines of code (automatic + manual) for that migration.}
\caption{Distribution of manual and automatic changes across manually fixed migrations.}
\label{fig:dev-effort-saving-stacked-bar}
\end{figure}

We are able to fix \MigManualFixCorrected out of \MigManualFixApplied migrations. We could not fix the remaining \MigManualFixNotCorrected due to one or more of the following reasons: 
(1) As outsiders to the project, some of the application logic was too complex for us to understand and fix the migration.
(2)
In one case, the LLM migrated to \lib{dataclasses} instead of \lib{cattrs}, which made the migration incorrect.
Fixing this migration would require undoing the entire migration and re-migrating it to the correct library, which is beyond the scope of fixing an incomplete migration.
% aio-libs@aiosmtpd__3615f5ef__attrs__cattrs
(3) In some cases, the tests themselves need to be updated to account for the behavior of the new target library. In several cases, we could not figure the correct expectation in the test oracle (again due to the complexity of the application) or because the original tests had complex mocking of the source library APIs.
(4) We did observe cases where additional dependent libraries require migration, beyond the source-target pair.
We consider such migrations beyond scope of this work, thus we did not attempt to fix such migrations.

For the \MigManualFixCorrected migrations we were able to completely fix, we measure \effortManual and \effortAuto using \autoref{eq:manual-effort-metric} and \autoref{eq:auto-effort-metric}, respectively.
\autoref{fig:dev-effort-saving-stacked-bar} shows the proportion of automatic and manual changes made for each migration.
The blue part of the stacked bar is the \effortAuto, while the red part is \effortManual.
The yellow dotted line shows the median of \effortAuto~ across all \MigManualFixCorrected migrations, which is \MiglocAutoMedianPercent.
This means that for a median migration, \tool performed \MiglocAutoMedianPercent of the changes, while the developers would only need to perform \MiglocManualMedianPercent of the changes.
Note that 3 bars in the figure have no red part, this is because the manual fixes required to successfully complete the migration were not source code changes but some environment change, such as installing a missing dependency.

\begin{tcolorbox}[colback=resultbg, title=\ref{rq:manualEffort}]
For the migrations that required manual fixes, \tool performed a median of \MiglocAutoMedianPercent of the migration changes, while a developers would only need to perform \MiglocManualMedianPercent of the changes to complete the migration.
\end{tcolorbox}

%On average, the automated migration by \tool modified \MiglocAutoAvg lines of code (\MiglocAutoMedian median, \MiglocManualMin--\MiglocAutoMax range).
%In contrast, the manual fixes requires an average of \MiglocManualAvg lines of changes (\MiglocManualMedian median, \MiglocManualMin--\MiglocManualMax range).
%This suggests that although the tool performs the majority of the migration, some manual intervention is still needed for correctness.

%These results suggest that although the tool may not always produce fully correct migrations, it significantly reduces the amount of work required to reach correctness. This validates the practicality of using \tool in real-world migration scenarios.

%% file: c5-tool/discussion.tex
%\subsection{Implications and discussion}
%\label{sec:too:discussion}

%\subsubsection{Limitations of \tool migrations}

To understand where \tool fell short, we qualitatively analyze the manual fixes we had to perform. 
We categorize them into the following observations. 
We provide (adapted) examples from the actual migrations.
%To understand the current limitations of \tool, we look at the manual fixes we had to perform.
%We discuss the migrations produced by \tool and what we had to fix manually, and discuss the frequently reoccurring problems of \tool migrations as well as propose possible improvements.
%The examples we show here are from the original migrations from our experiments, but they are often adopted for clarity and conciseness.
%In the left-to-right or top-to-bottom comparisons, the changes are highlighted in red.

%\input{c5-tool/lst/rename_public_api/_index.tex}

\begin{itemize}%[label={\textbf{Limitation-\arabic*}}, leftmargin=*, itemindent=6em]

\item \label{tool:issue:incorrectmig} \textbf{The LLM often incorrectly uses target APIs:}
Sometimes the target API call used by the LLM does not correctly replace the source API call.
This includes incorrect import statements, incorrect arguments to API calls, missing API calls, incorrect exception types, incorrect use of \texttt{async}/\texttt{await}, missing non-default argument, and other API related mistakes.
For example, in a \aPair{click}{typer} migration, the LLM did not use a name for the \texttt{Typer()} call, which was required.
The equivalent \texttt{Click} API does not require a name, so \tool missed adding it.
% alanhamlett@pip-update-requirements__e407b929__click__typer
% alexferl@vyper__a0a6ea7f__toml__tomli
% apple@app-store-server-library-python__1202058d__requests__requests_futures
% avinassh@haxor__8c0cb6be__aiohttp__httpx
% bheinzerling@bpemb__a75008c1__requests__httpx
% blueprints-org@blueprints__b3bb6987__matplotlib__plotly
% carbonblack@carbon-black-cloud-sdk-python__0a3b0cfb__jsonschema__cerberus
% cmccomb@trussme__0c7316cf__matplotlib__altair
% colour-science@colour-datasets__aa4ae7be__tqdm__progressbar2
% dwave-examples@3d-bin-packing__abb0ebd3__tabulate__prettytable
% github@automatic-contrib-prs__3c357a2e__python-dotenv__dynaconf
% lipoja@urlextract__3dec042b__filelock__portalocker

\item \label{tool:issue:skippedcode} \textbf{The LLM skips part of the code:}
Despite explicitly mentioning in the prompt not to skip parts of the code (\autoref{lst:prompt-tool}), the LLM still skips part of the code in migrations, especially when the code pattern is repetitive or does not have source API usage.
We observe the same behavior in \autoref{ch:llm}, 
which inspired us to add the \textit{\MergeSkippedStep} post-processing step in \tool to merge the skipped code with the migrated code (see \autoref{subsec:merge-skipped-step}).
However, the heuristics used in \MergeSkippedStep still fail to identify some skipped code so we had to manually add it back in the fix.
% 12932@cf-speedtest__78d9ee1d__requests__pycurl
% ankane@mapkick.py__7d625d80__flask__bottle
% apple@app-store-server-library-python__1202058d__requests__requests_futures
% blueprints-org@blueprints__b3bb6987__matplotlib__plotly
% colour-science@colour-datasets__aa4ae7be__tqdm__progressbar2
% dave-howard@vsdx__ca31a2c9__jinja2__mako
% france-travail@words_n_fun__3bf4b159__tqdm__alive-progress

\item \label{tool:issue:complexasync} \textbf{\tool struggles with complex async migrations:}
Some sync to async migrations require complex changes in the code structure, such as changing class constructors to async methods, changing sync context managers to async context managers.
The LLM often struggles to perform such complex changes correctly.
Our \asyncTransStep in tool only performs simple transformations, such as adding \texttt{async}/\texttt{await} keywords and changing context managers.
We manually fixed some of these complex async migrations.
% codewithemad@apyrat__de1852d8__requests__aiohttp - did not fix

\item \label{tool:issue:refactor} \textbf{The LLM refactors public program elements:}
In the \llmmiground, \tool migrates each file in isolation, which can lead to incorrect migrations. 
For example, In the \aPair{sqlalchemy}{tortoise-orm} migration, \tool renames a public class \texttt{EncryptedAlchemyBinaryField} defined in the client project to 
\texttt{EncryptedTortoiseBinaryField} to match the target library's name.
This class is used in other parts of the code.
Renaming the file prevents its callers from finding it, causing the errors in tests.
Since \tool does not pass the code that contains usage of the class in other files to the LLM, the LLM does not understand that it should not rename the class.
Note that the prompt (\autoref{lst:prompt-tool}) does ask not to change any API names, however, the LLMs do not always strictly follow the instructions.

\input{c5-tool/lst/manually-add-api-call/_index.tex}

\item \label{tool:issue:context} \textbf{The LLM does not have the dependent code:}
\autoref{fig:manually-add-api-call} shows an example where the underlying LLM was not able to add an additional API call (\texttt{settings.reload()}) required for correct migration.
This is required because the target library, \lib{dynaconf}, loads the environment variables when the library is imported for the first time, and a reload is required if the environment variables are changed.
The original code uses \lib{python-dotenv} to load the environment variables, which reads the environment variables at runtime.
Notice how the test, which is in another file that does not get passed to the LLM, updates the environment variables before running the tests. 
This updating happens after the library is imported, 
causing the tests to fail without the explicit \texttt{settings.reload()} call in the migrated code.
If the LLM knew about the dependent code, it may be able to better understand the context of the migration.

\item \label{tool:issue:config} \textbf{The LLM is unaware of configuration requirements:}
In some cases, the migration requires changes in the project configuration rather than code changes.
For example, in a migration from \aPair{chardet}{cchardet}, adjusting the Python version from \texttt{3.11} to \texttt{3.9} resolved the test failures.
Similarly, we made a migration from \aPair{flask}{sanic} successful by installing an additional dependency, \lib{sanic-ext}, with no changes to the migrated code.
The LLM is unaware of such configuration requirements, and therefore, it cannot perform such changes.
% cmccomb@trussme__0c7316cf__matplotlib__altair
% abinthomasonline@repopack-py__dc2b9243__chardet__cchardet
% cmccomb@trussme__0c7316cf__matplotlib__altair
% cskorpion@vmprof-firefox-converter__e1f6bb79__flask__sanic
% fidelity@selective__7e07abc8__seaborn__altair

\item \label{tool:issue:testupdate} \textbf{Often test expectations need to be updated:}
There are cases where the tests require changes due to the target library's different behavior.
For example, often the source library raises a generic exception, which the tests expect, while the target library raises a more specific exception.
Another example is \lib{coloroma} uses integers to represent colors (green is \texttt{32}), while \lib{rich} uses strings (green is \texttt{"[green]"}).
% dmtf@redfish-protocol-validator__e2f28669__colorama__rich
% funilrys@pyfunceble__a2756c4f__python-dotenv__environs
% dwave-examples@3d-bin-packing__abb0ebd3__tabulate__prettytable
% seamile@jsonfmt__c6f6d17f__toml__tomlkit
% giuliorossetti@ash__1106b609__seaborn__plotly

\item \label{too:issue:mocking} \textbf{\tool does not handle mock migrations:}
Some tests mock the source library's API calls in the tests, and these mocks need to be migrated to the target library's API calls.
This is something we did not consider when designing \tool, and therefore, it does not handle such cases.
% bigcommerce@bigcommerce-api-python__599b27e6__requests__treq
% dmtf@redfish-protocol-validator__e2f28669__colorama__rich
% electronic-mango@simple-justwatch-python-api__5524c491__httpx__aiohttp
% growthbook@growthbook-python__699fba51__urllib3__requests
% ozgurkara@fastapi-pydiator__7d70a90a__httpx__urllib3

\item \label{tool:issue:multilibmigration} \textbf{\tool does not support multi-library migrations:}
We design \tool to migrate from just one source library to just one target library, which may not always be sufficient.
The project \aRepoLink{miguelgrinberg/microblog-api} is such an example, which uses \lib{flask} and some other libraries that depend on \lib{flask}.
To fully migrate to \lib{fastapi} (or any other target library), we also need to migrate all these dependent libraries.
% miguelgrinberg@microblog-api__e250d711__flask__cherrypy
% miguelgrinberg@microblog-api__e250d711__flask__fastapi
% miguelgrinberg@microblog-api__e250d711__flask__tornado

\end{itemize}

\section{Discussion}
\label{sec:tool-discussion}
Overall, we find that \tool automatically completes migration for a large portion of cases (\MigOverallCorrectPercent migrations) 
and does most of the required code changes (\MiglocAutoMedianPercent changes) for the remaining migrations,
often leaving few edits for the developers to perform (\MiglocManualMedianPercent changes).
While further improvements in the migration process may potentially help in reducing the manual effort, the remaining effort seems fairly minimal in most cases, specially for the developers who are familiar with the libraries and the project.
We now discuss some potential improvement and future research and development directions for \tool based on our findings, as well as its potential usage in practice.

\subsection{Potential Improvements}

\paragraph{Handling Refactorings} The two post-processing steps currently included in \tool show that simple automated steps can significantly improve the correctness of migrations.
However, Section \autoref{subsec:rqeffort} also discussed that that our current implementation has limitations and can therefore be further improved. For example, to handle problems introduced by refactorings, we can follow ideas from refactoring-aware merging~\cite{refmerge} where \tool can detect \cite{RefactoringMiner2} and revert refactorings before applying the migration changes, and then re-apply the refactoring changes on the whole codebase.

\paragraph{Balancing Context and Re-prompting} We provide the LLM with a fairly straight-forward context, the file that uses the source library APIs.
This is often insufficient, and on the contrary, it is often too much if the file is large but only a small part of it needs migration.
Identifying and sending only relevant parts of the code may help the LLM to perform better migration.
This however, comes with the cost of more pre-processing to identify the relevant code, as well as more post-processing to replace old code with the new code in the project.

Instead, another direction is to consider iteratively re-prompting the LLM with additional context, such as test error messages after running the tests on the migrated code.
The challenge here is to identify meaningful error messages and generate a prompt that can help the LLM to understand the context better.
Trying out different re-prompting strategies for this may be an interesting future research direction.

\paragraph{Handling Dependent Libraries} 
\tool currently supports migrating from one source library to one corresponding target library.
However, migrating from one library often requires also replacing related auxilary libraries. 
In \autoref{subsec:rqeffort}, we did indeed observe some migrations that required additional dependent libraries to be migrated.

Since it is unrealistic to expect the developer to know all auxiliary libraries that need to be used with the target library, 
one potential way to handle this is to add a separate step before the migration where the LLM is instructed to check the existing dependency list and decide if other auxilary libraries need to be migrated as well. 
While this approach offloads the responsibility of identifying combined analagous libraries to the LLM, it also introduces additional complexity in managing and maintaining the migration process.
The actual migration step will require changing the prompts and instructions to the LLM to handle multiple source and target libraries, which may be more challenging for the LLM.
This is an interesting avenue for future research, likely requiring some experimentation to find the right balance between providing enough context and not overwhelming the LLM with too much information. 
We do note though that our post-processing steps do not depend on the (number of) libraries involved and can therefore continue to be used as is in a multi-library scenario.

\paragraph{Adapting \tool for Other Languages} While \tool is built specifically for Python, its general approach can be applied to other programming languages as well. The language-specific parts of the tool relate to the pre- and post- processing steps, such as building the call graph and identifying skipped code, which can be easily adapted for other languages through some engineering effort. The LLM usage part itself is language-agnostic, with minimum changes required to the prompt to adjust to the input code's language.

\subsection{Using \tool in Practice}

\paragraph{Test Coverage Requirement}
\tool depends on the existence of high-coverage tests in the project.
While a developer is free to use \tool on a project without tests, they need to be aware that the migration results may not be reliable. 
Specifically, without tests, the preprocessing step (\autoref{subsec:prep-round}) will miss identifying files that require migration but do not directly import the source library.
After the LLM migration, \tool will also not be able to validate the migration results due to the absences of tests.
Finally, the post-processing steps will not run because there will be no difference between \TestReportPre and \TestReportLLM.
With these limitations, a developer who uses \tool on a project without tests will only perform the LLM migration, with no way to verify the migration.
The developer will then need to verify the functionality themselves, outside \tool's pipeline.

\paragraph{Using \tool vs. a General-purpose Agent}
AI coding agents, such as GitHub copilot, Cursor, Devstral, have recently gained attention for performing various software engineering tasks, including code transformation tasks\cite{react2023}.
This naturally leads to the question of whether AI agents can be used to perform automated library migration, without the need for a specialized tool.
The main advantage of AI agents is that they can autonomously perform the given task by iteratively deciding what code to change, what tests to run, and how to fix the migration based on the test results.
In other words, using agents can potentially replace the pre- and post-processing steps we specifically developed for the library migration task.
To understand the potential of using AI agents for library migration, we run a small experiment with 15 migration tasks from our dataset.
To ensure diversity, we used stratified sampling to select migrations between 15 different library pairs with varying correctness levels of \tool's migrations (1 migration with 0\% correctness, 4 with $(0,30\%]$, 4 with $(30,60\%]$, 3 with $(60,80\%]$, and 2 with $(80,100\%]$).
We use the Github Copilot Local agent with GPT-4o as the underlying LLM.
We choose to use the same LLM as our experiments in \autoref{ch:tool} to have a fair comparison between the agentic and non-agentic approaches.
We find that the agent achieved a higher correctness score for 6 out of the 15 migrations (median improvement of 27\%), 7 were relatively worse (median 25\% worse), and 2 were the same at 0\% and 14\% correctness.
These results suggest that while AI agents can be helpful for library migration, they are not necessarily better than a targeted LLM-based pipeline like \tool.
This aligns with the findings of Xia \etal \cite{xia2025demystifying} in the context of automatic program repair, where they find that an ``agentless'' approach through a targeted LLM-based pipeline can outperform agentic approaches.

\paragraph{Cost of using \tool}
The cost of using LLMs can be a concern for the developers.
Therefore, we try to estimate the cost of using \tool for library migration.
The total number of tokens sent to the LLM across all migrations in our dataset is \tokensInTotal, and the total number of tokens received from the LLM is \tokensOutTotal, with a total of \tokensTotal tokens.
The median number of tokens sent to the LLM per migration is \tokensMedianInPerMig and the median number of tokens received from the LLM per migration is \tokensMedianOutPerMig.
As of \tokenPriceDate, the cost of using GPT-4o is \FouroCostPerMilIn per million input tokens and \FouroCostPerMilOut per million output tokens.
Based on these numbers, the estimated cost of using \tool for migrating a median migration in our dataset is \MedianCostPerProject.
The migration in our dataset that consumed the most tokens has an estimated cost of \MaxCostPerProject.
This is a fairly low cost for the value that \tool can provide in automating a large portion of the migration work.

%% file: c5-tool/lst/manually-add-api-call/_index.tex
% github@automatic-contrib-prs__3c357a2e__python-dotenv__dynaconf: fixed with alternative API

\begin{figure*}[!tp]
  \centering
  \begin{minipage}[t]{0.54\textwidth}
\begin{lstlisting}[language=Python, title=Original code]
gh_actor = os.getenv("GH_ACTOR", "nb")
gh_app_id = get_int_env_var("GH_APP_ID")
\end{lstlisting}
    \vspace*{-1.3em}%
\begin{lstlisting}[language=Python, title=\tool migration]
gh_actor = %\HLC{settings.get}%("GH_ACTOR", "nb")
gh_app_id = get_int_env_var("GH_APP_ID")
\end{lstlisting}
    \vspace*{-1.3em}%
\begin{lstlisting}[language=Python, title=Manual fix]
%\HLC{settings.reload()}%
gh_actor = settings.get("GH_ACTOR", "nb")
gh_app_id = get_int_env_var("GH_APP_ID")
\end{lstlisting}
  \end{minipage}
  \hfill
  \begin{minipage}[t]{0.44\textwidth}
    \vspace{0pt}%
\begin{lstlisting}[language=Python, title=Test code]
@patch.dict(
  os.environ, 
  {"GH_ACTOR": "testactor"}
)
def test_get_env_vars(self):
  expected_var = EnvVars(
    "testactor",
  )
  result = get_env_vars(True)
  actual = str(result)
  expected = str(expected_var)
  assert actual == expected
\end{lstlisting}
  \end{minipage}

  \caption{Example of requiring to call an additional API in a \aPair{python-dotenv}{dynaconf} migration.}    
  \label{fig:manually-add-api-call}
\end{figure*}

%% file: c-0/threats.tex
\section{Threats to validity}
\label{sec:tool:threats}

\subsection{Internal validity}
There were several manual steps involved in our evaluations.
Manual activities are inherently subjective and may contain errors.
In Phase 1, we manually validated the correctness of \ccs and labeled any non-migration related change as refactoring or non-refactoring.
To minimize subjectivity, two authors independently perform these steps and resolved disagreements through discussion.
In Phase 3, only one author performs all manual migrations.
However, we rely on objective test outcomes as the primary measure of correctness.
%The second issue is inherent to the fact that none of the authors are direct contributors to the project
%Some migrations initially fail due to environment or dependency issues rather than code errors.
%We address these by adjusting configurations when possible, but environment-specific factors, such as Python and library version, operating system, may still affect the results.

All the migrations in \migbench happened prior to
the known training cutoff date of the three models (Late
2023). Therefore, there is a possibility that the LLMs are
simply reciting the migrated code they have seen before (for
this particular code base). This means that the reported results in Phase 1 may be an upper bound or optimistic estimate of the LLMs' migration capabilities.
While this setup was essential in Phase 1 to allow us to compare the LLM migrations to a developer-performed ground truth migration for systematic evaluation and to identify the types of migration-related code changes LLMs excel at vs. struggle with, we designed a different evaluation approach in Phase 3 to overcome this threat, and thus provide a more realistic estimate of the LLMs' migration capabilities.
To further quantify the extent of data leakage in phase 3, we search the history of the repositories we use to see if the target library ever appeared there (e.g., the reverse migration happened before). We find that only 54/717 (7.5\%) migrations had the target library in their dependencies before. While not absolutely guaranteed since we do not have access to the training data, this strongly suggests that the majority of our evaluation is on code that the LLM has not seen the target library being used for before. 

The \mergeSkippedStep uses some thresholds to identify skipped code hunks, which we determine through experimentation.
Different thresholds may yield different results.
However, we believe our chosen thresholds strike a reasonable balance between adding back genuinely skipped code and avoiding incorrect additions.
If the assumptions do not hold, the resulting migration may potentially fail, therefore, the results we have may be a lower bound of the actual migration correctness.

\subsection{External validity}
We use three off the shelf LLMs in Phase 1, and the best of them in Phase 3.
Using an LLM specifically tuned for the library migration task could potentially yield better results.
Therefore, our findings in this paper may be viewed as a lower bound of the performance of potentially more powerful and specialized models.

In Phase 3, we focus on Python projects that are well-tested and actively maintained, as selected through specific filters.
This focus may limit the generalizability of our findings to projects with less test coverage, different maintenance status, or other programming languages.
The set of library pairs and migration scenarios is also limited by the availability of analogous libraries and the manual validation process.
As a result, the observed effectiveness of \tool may not fully generalize to all migration contexts.
That said, the size of the dataset (\MigExpCount migrations) and the diversity of selected projects (\RepoExpCount repositories) provide a reasonable basis for evaluating the tool's effectiveness in a variety of migration scenarios.

The libraries we consider in Phase 1 and Phase 3 are relatively popular and well-established.
This familiarity with APIs and API usage is precisely our hope when using LLMs for migration.
However, being familiar with the APIs alone does not always guarantee being able to map them to each other or apply the right transformations on all code bases, which is why our evaluation is necessary.
For newer libraries, additional information, \eg API documentation, may be needed to guide the LLM in the migration process.
This can be an interesting avenue for future study, and thus improvement of \tool.

\subsection{Construct validity}
In \rqBench, we analyze the \cc categories based on \changeDev, not on \changeLLM.
The categories can differ if the LLM produces a different \cc than the developer.
\migbench already provides us with the \changeDev categories; categorizing the LLM changes according to \taxonomy \cite{pymigtax} requires additional manual effort that is beyond the scope of this work.
That said, our observation is that the code/API differences in most cases are minor, and would not yield in a different \taxonomy category.
% That said, these problem causes migration to fail, therefore, the results we have may be lower bound of the actual migration correctness.

We use automated test suites to assess migration correctness in Phase 3, but these tests may not cover all possible behaviors or edge cases in the migrated projects.
To minimize this, we select projects with at least \MinCover lines of code covered by tests, which helps ensure that the tests are comprehensive enough to provide a reliable measure of correctness.

We use migration lines of code (\autoref{eq:manual-effort-metric}) to estimate the amount of migration-related code changes left for the developer to make. This metric only captures the quantity of changes and does not capture all aspects of migration work, such as time spent understanding and comparing the source and target APIs, or debugging the migration changes.
We choose to use this metric since it is objective and measurable.
We additionally accompany it with a qualitative analysis of the types of changes we had to make, which can provide more insights into the nature of the remaining work.

%% file: c-0/related-work.tex
\section{Related Work}
\label{ch:relatedwork}

Prior work studied how developers compare analogous libraries~\cite{selectingLibraries,Mor18,Tan24} as well as the factors that lead them to decide to replace one library with another~\cite{he2021large}. Such reasons include project integration needs, lack of maintenance of the current library, better usability or features of the new library, and dependency simplification. In contrast, most library migration techniques (discussed below) start their work with the assumption that the target library has already been identified and focus on automating the migration process itself. This is also our focus in this paper. 

The majority of the existing library migration techniques focus on identifying API mappings between two analogous libraries.
Some of these techniques mine existing library migrations to identify the API mappings \cite{teyton2012mining, teyton2013automatic, alrubaye2019use, alrubaye2019migrationminer, alrubaye2020learning}.
Some techniques see the requirement of existing migrations as a limitation, and therefore use other techniques, such as machine learning \cite{alrubaye2018automating, miningAnalogicalAPIs} and natural language processing \cite{ni2021soar}.
These techniques only identify the API mappings, but do not perform client code transformation, which is a separate challenge.

SOAR \cite{ni2021soar} uses API documentation to identify API mappings, and then uses the mappings and unit tests to guide program synthesis for client code transformation.
SOAR has several limitations that prevent it from being a general solution for library migration.
SOAR is evaluated on only two pairs of deep learning libraries, focusing on their migration of neural network models.
These APIs are usually called in a sequence, where the output of one call is used as input to the next call.
The nature of these APIs allows SOAR to check the program state after each API call, which is not the case for most libraries.
Additionally, given the limited number of libraries, their technique relies on the library-specific error message format to further guide the synthesis process, requiring extensive changes and technique adaptations for running SOAR on other libraries.
SOAR supports \om \ccs for a small set of pre-determined APIs, whose information is hard-coded for the transformation process.
SOAR also times out for 20\% of the evaluated migrations, even for small code snippets, the largest being 183 lines of code.

Zhou \etal \cite{hapim} proposed HaPiM, which first trains an unsupervised machine learning model capable of API mapping, and then uses the API mappings to guide an LLM in code transformation.
Similar to Chen \etal \cite{miningAnalogicalAPIs}, HaPiM learns from API usage patterns of source and target libraries in their respective client projects, therefore does not require any existing migration examples.
For the code transformation, they use the API mappings to prompt an LLM to replace source API calls with target API calls.
Their approach outperforms MigrationMapper \cite{alrubaye2019use} and GPT-3.5 Turbo on the BLEU \cite{bleu} and CodeBLEU \cite{codebleu} metrics evaluated on 5 Java library pairs.
A significant limitation of HaPiM is that it works with small independent code snippets that use the source library APIs, and results in an equivalent snippet that uses the target library APIs.
This is as opposed to real-life code where the source API usages are interleaved with other types of code, such as variable declarations, control statements, and other library usages.
This means that to use HaPiM, developers need to first extract code snippets that contains only the source library API usages from the client code, then use HaPiM to replace them with target library APIs, and then re-integrate the transformed code back into the client code, which is not practical for large-scale library migrations.

Nikolov \etal \cite{nikolov2025google} describe Google's experiences in using LLMs for several internal code transformation tasks, including \aPair{Joda Time}{Java Time} migrations.
They do not attempt to provide a general solution for library migration; rather, they use LLMs to perform the migration they need.
Therefore, their prompts contains instructions and contextual information tailored specifically for \aPair{Joda Time}{Java Time} migrations.
Nonetheless, they show that using LLMs reduced migration time by at least 50\% compared to manual migration.

There are also a few studies that focus on migrating between different versions of the \textit{same library}.
The study of Nikolov \etal \cite{nikolov2025google} we describe above also includes \aPair{JUnit 3}{JUnit 4} migrations.
Almeida \etal \cite{almeida2024automatic} evaluate the performance of ChatGPT for \aPair{SQAlchemy 1} {SQLAlchemy 2} migrations.
While similar, migrating between different versions of the same library is different from migrating between different libraries, because the former can leverage the evolution or release history of the libraries to identify API changes.

The above shows that there have been different approaches to automate the library migration process.
However, these techniques come with one or more limitations that prevent them from being a general solution for library migration.
Our tool, \tool, resolves these limitations and can automate the entire library migration process for any arbitrary analogous library pair.
We evaluate \tool on a large and diverse dataset of \MigExpCount migrations between \LibPairExpCount analogous library pairs, and show that \tool can successfully perform most of the migration with no or minimal manual intervention from the developer.

%% file: c-0/conclusion.tex
\section{Conclusion}
\label{sec:tool:conclusion}

Our goal in this paper is to fully automate the Python library migration process.
We systematically approached this goal by first conducting an empirical study to understand the capabilities of LLMs for automating Python library migration.
Overall, we find that LLMs can handle complex code changes but we also identify some challenges that need to be addressed to make LLM-based migration practical.
%We find that the best performing LLM, \fo, correctly migrates \FouroCCBMTotalCorrectPercent of the individual code changes.
%At the migration level, it perfectly migrates \FouroMigBMCorrectPercent of the migrations, and \FouroMigBMAtLeastOneCorrectPercent of them have at least one correctly migrated \cc.
%We also identify some challenges that need to be addressed to make LLM-based migration practical.
Based on these findings and insights, we built \tool, a practical CLI tool that combines LLMs with static and dynamic program analysis to automate the full migration process.
Our evaluation on \MigExpCount simulated migrations across \RepoExpCount diverse projects and \LibPairExpCount library pairs shows that \tool can fully automate a large portion of migrations (\MigOverallCorrectPercent), 
with most remaining cases requiring only minor manual intervention (median \MiglocManualMedianPercent changes).
The tool's layered approach, including post-processing steps, improves LLM migration correctness (from \MigLLMMigMedianCorrectness to \MigAsyncTransformMedianCorrectness) and reduces developer effort.
Overall, \tool lowers the barrier for large-scale library migration in real-world projects, demonstrating accessible and effective automated library migration.

%In the first phase of the work, we conduct an empirical study using three LLMs on the \migbench dataset to understand how well LLMs can perform library migration.
%We also run unit tests for a subset of the migrations, and find that \FouroMigTestCorrectPercent of the migrations by \fo have the same sets of tests passing in the developer's migration and the LLM's migration.
%A much lower cost LLM \mn and the free \lm also perform relatively well, with both correctly migrating \MiniCCBMTotalCorrectPercent of \ccs.
%Overall, we find that LLMs can handle complex code changes, therefore, we conclude that LLMs are a promising approach for automating library migration.

%
%We use these insights to build \tool, a CLI tool that automates the full library migration process in Phase 2.

%% file: c-0/acknowledgements.tex
\begin{acknowledgements}
We thank Dr. Ildar Akhmetov, Northeastern University, for his contributions to the early ideas of this work.
\end{acknowledgements}

%% file: c-0/declarations.tex
\section*{Declarations}
\paragraph*{Funding}
This work was supported by the Canada Research Chairs Program and NSERC CREATE.

\paragraph*{Author contributions}
Below are the contributions of each author --
\begin{itemize}
    \item \textit{Mohayeminul Islam} contributed to idea conceptualization, design and implementation of the \tool, conducting the experiments, writing script for the experiments, manual evaluations in Phase 1, manual migration fix in Phase 3, and writing the manuscript.
    \item \textit{Ajay Kumar Jha} contributed to idea conceptualization, design of the \tool, manual evaluations in Phase 1, and writing the manuscript.
    \item \textit{May Mahmoud} manual evaluations in Phase 1, and revising the manuscript.
    \item \textit{Sarah Nadi} contributed to idea conceptualization, design of the \tool, manual evaluations in Phase 1, and writing the manuscript. She is the Principal Investigator of the project.
\end{itemize}

\paragraph*{Data availability}
The source code of \tool, the experimental data, and the scripts and the detailed results of our experiments are available at {\ArtifactUrl}.

\paragraph*{Conflict of interest} No conflict.

\paragraph*{Ethical approval/informed concent/clinical trial number}
Not applicable.

%% file: refs.bib
@inproceedings{Mor18,
  author    = {Fernando L{\'o}pez de la Mora and Sarah Nadi},
  title     = {An Empirical Study of Metric-based Comparisons of Software Libraries},
  booktitle = {Proceedings of the 14th International Conference on Predictive Models and Data Analytics in Software Engineering (PROMISE)},
  year      = {2018},
  publisher = {Association for Computing Machinery},
  doi       = {10.1145/3273934.3273937}
}

@inproceedings{Tan24,
  author    = {Minaoar Hossain Tanzil and Gias Uddin and Ann Barcomb},
  title     = {{How do people decide?}: A Model for Software Library Selection},
  booktitle = {2024 IEEE/ACM 17th International Conference on Cooperative and Human Aspects of Software Engineering (CHASE)},
  year      = {2024},
  publisher = {ACM},
  doi       = {10.1145/3641822.3641865}
}

@inproceedings{pymigbench,
  author       = {Islam, Mohayeminul and Jha, Ajay Kumar and Nadi, Sarah and Akhmetov, Ildar},
  title        = {PyMigBench: A Benchmark for Python Library Migration},
  booktitle    = {2023 IEEE/ACM 20th International Conference on Mining Software Repositories (MSR)},
  pages        = {511--515},
  year         = {2023},
  organization = {IEEE},
  doi          = {10.1109/MSR59073.2023.00075}
}

@inproceedings{pymigtax,
  title     = {Characterizing Python Library Migrations},
  author    = {Mohayeminul Islam and Ajay Kumar Jha and Ildar Akhmetov and Sarah Nadi},
  booktitle = {Proceedings of the ACM International Conference on the Foundations of Software Engineering (FSE)},
  year      = {2024},
  articleno = {5},
  numpages  = {23},
  doi       = {10.1145/3643731}
}

@inproceedings{llmmig,
  title     = {An Empirical Study of Python Library Migration Using Large Language Models},
  author    = {Islam, Md Mohayeminul and Jha, Ajay Kumar and Mahmoud, May and Akhmetov, Ildar and Nadi, Sarah},
  booktitle = {Proceedings of the 40th IEEE/ACM International Conference on Automated Software Engineering (ASE 2025)},
  year      = {2025}
}

@misc{atil2024llmstabilitydetailedanalysis,
  title         = {LLM Stability: A detailed analysis with some surprises},
  author        = {Berk Atil and Alexa Chittams and Liseng Fu and Ferhan Ture and Lixinyu Xu and Breck Baldwin},
  year          = {2024},
  eprint        = {2408.04667},
  archiveprefix = {arXiv},
  primaryclass  = {cs.CL},
  url           = {https://arxiv.org/abs/2408.04667}
}

@article{HuangMappingAPI2024,
  author    = {Huang, Zhenfei and Chen, Junjie and Jiang, Jiajun and Liang, Yihua and You, Hanmo and Li, Fengjie},
  title     = {Mapping APIs in Dynamic-typed Programs by Leveraging Transfer Learning},
  year      = {2024},
  publisher = {Association for Computing Machinery},
  address   = {New York, NY, USA},
  issn      = {1049-331X},
  url       = {https://doi.org/10.1145/3641848},
  doi       = {10.1145/3641848},
  note      = {Just Accepted},
  journal   = {ACM Trans. Softw. Eng. Methodol.}
}

@inproceedings{teyton2012mining,
  title        = {Mining library migration graphs},
  author       = {Teyton, Cedric and Falleri, Jean-Remy and Blanc, Xavier},
  booktitle    = {2012 19th Working Conference on Reverse Engineering},
  pages        = {289--298},
  year         = {2012},
  organization = {IEEE}
}

@inproceedings{teyton2013automatic,
  title        = {Automatic discovery of function mappings between similar libraries},
  author       = {Teyton, C{\'e}dric and Falleri, Jean-R{\'e}my and Blanc, Xavier},
  booktitle    = {2013 20th Working Conference on Reverse Engineering (WCRE)},
  pages        = {192--201},
  year         = {2013},
  organization = {IEEE}
}

@inproceedings{wang2016transforming,
  author    = {Chenglong Wang and Jiajun Jiang and Jun Li and Yingfei Xiong and Xiangyu Luo and Lu Zhang and Zhenjiang Hu},
  title     = {{Transforming Programs between APIs with Many-to-Many Mappings}},
  booktitle = {30th European Conference on Object-Oriented Programming (ECOOP 2016)},
  pages     = {25:1--25:26},
  series    = {Leibniz International Proceedings in Informatics (LIPIcs)},
  isbn      = {978-3-95977-014-9},
  issn      = {1868-8969},
  year      = {2016},
  volume    = {56},
  editor    = {Shriram Krishnamurthi and Benjamin S. Lerner},
  publisher = {Schloss Dagstuhl--Leibniz-Zentrum fuer Informatik},
  address   = {Dagstuhl, Germany},
  url       = {http://drops.dagstuhl.de/opus/volltexte/2016/6119},
  urn       = {urn:nbn:de:0030-drops-61195},
  doi       = {10.4230/LIPIcs.ECOOP.2016.25}
}

@inproceedings{he2021large,
  title     = {A large-scale empirical study on Java library migrations: prevalence, trends, and rationales},
  author    = {He, Hao and He, Runzhi and Gu, Haiqiao and Zhou, Minghui},
  booktitle = {Proceedings of the 29th ACM Joint Meeting on European Software Engineering Conference and Symposium on the Foundations of Software Engineering},
  pages     = {478--490},
  year      = {2021}
}

@article{kula2018developers,
  title     = {Do developers update their library dependencies?},
  author    = {Kula, Raula Gaikovina and German, Daniel M and Ouni, Ali and Ishio, Takashi and Inoue, Katsuro},
  journal   = {Empirical Software Engineering},
  volume    = {23},
  number    = {1},
  pages     = {384--417},
  year      = {2018},
  publisher = {Springer}
}

@inproceedings{alrubaye2018automating,
  title     = {Automating the detection of third-party Java library migration at the function level.},
  author    = {Alrubaye, Hussein and Mkaouer, Mohamed Wiem},
  booktitle = {CASCON},
  pages     = {60--71},
  year      = {2018}
}

@inproceedings{alrubaye2019use,
  title        = {On the use of information retrieval to automate the detection of third-party java library migration at the method level},
  author       = {Alrubaye, Hussein and Mkaouer, Mohamed Wiem and Ouni, Ali},
  booktitle    = {2019 IEEE/ACM 27th International Conference on Program Comprehension (ICPC)},
  pages        = {347--357},
  year         = {2019},
  organization = {IEEE}
}

@inproceedings{alrubaye2019migrationminer,
  title        = {Migrationminer: An automated detection tool of third-party java library migration at the method level},
  author       = {Alrubaye, Hussein and Mkaouer, Mohamed Wiem and Ouni, Ali},
  booktitle    = {2019 IEEE international conference on software maintenance and evolution (ICSME)},
  pages        = {414--417},
  year         = {2019},
  organization = {IEEE}
}

@article{alrubaye2020learning,
  title     = {Learning to recommend third-party library migration opportunities at the API level},
  author    = {Alrubaye, Hussein and Mkaouer, Mohamed Wiem and Khokhlov, Igor and Reznik, Leon and Ouni, Ali and Mcgoff, Jason},
  journal   = {Applied Soft Computing},
  volume    = {90},
  pages     = {106140},
  year      = {2020},
  publisher = {Elsevier}
}

@inproceedings{xu2019meditor,
  title        = {Meditor: inference and application of API migration edits},
  author       = {Xu, Shengzhe and Dong, Ziqi and Meng, Na},
  booktitle    = {2019 IEEE/ACM 27th International Conference on Program Comprehension (ICPC)},
  pages        = {335--346},
  year         = {2019},
  organization = {IEEE}
}

@inproceedings{ni2021soar,
  title        = {Soar: a synthesis approach for data science api refactoring},
  author       = {Ni, Ansong and Ramos, Daniel and Yang, Aidan ZH and Lynce, In{\^e}s and Manquinho, Vasco and Martins, Ruben and Le Goues, Claire},
  booktitle    = {2021 IEEE/ACM 43rd International Conference on Software Engineering (ICSE)},
  pages        = {112--124},
  year         = {2021},
  organization = {IEEE}
}

@inproceedings{selectingLibraries,
  title     = {Selecting third-party libraries: The practitioners’ perspective},
  author    = {Larios Vargas, Enrique and Aniche, Maur{\'\i}cio and Treude, Christoph and Bruntink, Magiel and Gousios, Georgios},
  booktitle = {Proceedings of the 28th ACM joint meeting on european software engineering conference and symposium on the foundations of software engineering},
  pages     = {245--256},
  year      = {2020}
}

@article{miningAnalogicalAPIs,
  title     = {Mining likely analogical apis across third-party libraries via large-scale unsupervised api semantics embedding},
  author    = {Chen, Chunyang and Xing, Zhenchang and Liu, Yang and Xiong, Kent Ong Long},
  journal   = {IEEE Transactions on Software Engineering},
  volume    = {47},
  number    = {3},
  pages     = {432--447},
  year      = {2019},
  publisher = {IEEE}
}

@inproceedings{zhang2020deep,
  title     = {Deep-diving into documentation to develop improved java-to-swift api mapping},
  author    = {Zhang, Zejun and Pan, Minxue and Zhang, Tian and Zhou, Xinyu and Li, Xuandong},
  booktitle = {Proceedings of the 28th International Conference on Program Comprehension},
  pages     = {106--116},
  year      = {2020}
}

@article{cohen1960coefficient,
  title     = {A coefficient of agreement for nominal scales},
  author    = {Cohen, Jacob},
  journal   = {Educational and psychological measurement},
  volume    = {20},
  number    = {1},
  pages     = {37--46},
  year      = {1960},
  publisher = {Sage Publications Sage CA: Thousand Oaks, CA}
}

@article{landis1977measurement,
  title     = {The measurement of observer agreement for categorical data},
  author    = {Landis, J Richard and Koch, Gary G},
  journal   = {biometrics},
  pages     = {159--174},
  year      = {1977},
  publisher = {JSTOR}
}

@inproceedings{nguyen2022empirical,
  title     = {An empirical evaluation of GitHub copilot's code suggestions},
  author    = {Nguyen, Nhan and Nadi, Sarah},
  booktitle = {Proceedings of the 19th International Conference on Mining Software Repositories},
  pages     = {1--5},
  year      = {2022}
}

@misc{libraries.io,
  doi          = {10.5281/ZENODO.808272},
  howpublished = {https://zenodo.org/record/808272},
  author       = {Katz,  Jeremy},
  language     = {en},
  title        = {Libraries.io Open Source Repository and Dependency Metadata},
  publisher    = {Zenodo},
  year         = {2020},
  copyright    = {Creative Commons Attribution 4.0 International}
}

@inproceedings{hapim,
  title     = {Hybrid API migration: A marriage of small API mapping models and large language models},
  author    = {Zhou, Bingzhe and Wang, Xinying and Xu, Shengbin and Yao, Yuan and Pan, Minxue and Xu, Feng and Ma, Xiaoxing},
  booktitle = {Proceedings of the 14th Asia-Pacific Symposium on Internetware},
  pages     = {12--21},
  year      = {2023}
}

@inproceedings{almeida2024automatic,
  title     = {Automatic Library Migration Using Large Language Models: First Results},
  author    = {Almeida, Aylton and Xavier, Laerte and Valente, Marco Tulio},
  booktitle = {Proceedings of the 18th ACM/IEEE International Symposium on Empirical Software Engineering and Measurement},
  pages     = {427--433},
  year      = {2024}
}

@article{nikolov2025google,
  title   = {How is Google using AI for internal code migrations?},
  author  = {Nikolov, Stoyan and Codecasa, Daniele and Sjovall, Anna and Tabachnyk, Maxim and Chandra, Satish and Taneja, Siddharth and Ziftci, Celal},
  journal = {arXiv preprint arXiv:2501.06972},
  year    = {2025}
}

@article{ozkaya2023application,
  title     = {Application of large language models to software engineering tasks: Opportunities, risks, and implications},
  author    = {Ozkaya, Ipek},
  journal   = {IEEE Software},
  volume    = {40},
  number    = {3},
  pages     = {4--8},
  year      = {2023},
  publisher = {IEEE}
}

@article{fan2023large,
  title   = {Large language models for software engineering: Survey and open problems},
  author  = {Fan, Angela and Gokkaya, Beliz and Harman, Mark and Lyubarskiy, Mitya and Sengupta, Shubho and Yoo, Shin and Zhang, Jie M},
  journal = {arXiv preprint arXiv:2310.03533},
  year    = {2023}
}

@article{wang2024software,
  title     = {Software testing with large language models: Survey, landscape, and vision},
  author    = {Wang, Junjie and Huang, Yuchao and Chen, Chunyang and Liu, Zhe and Wang, Song and Wang, Qing},
  journal   = {IEEE Transactions on Software Engineering},
  year      = {2024},
  publisher = {IEEE}
}

@article{peng2023impact,
  title   = {The impact of ai on developer productivity: Evidence from github copilot},
  author  = {Peng, Sida and Kalliamvakou, Eirini and Cihon, Peter and Demirer, Mert},
  journal = {arXiv preprint arXiv:2302.06590},
  year    = {2023}
}

@article{murali2023codecompose,
  title   = {CodeCompose: A large-scale industrial deployment of AI-assisted code authoring},
  author  = {Murali, Vijayaraghavan and Maddila, Chandra and Ahmad, Imad and Bolin, Michael and Cheng, Daniel and Ghorbani, Negar and Fernandez, Renuka and Nagappan, Nachiappan},
  journal = {arXiv preprint arXiv:2305.12050},
  year    = {2023}
}

@inproceedings{nam2024using,
  title        = {Using an LLM to Help With Code Understanding},
  author       = {Nam, Daye and Macvean, Andrew and Hellendoorn, Vincent and Vasilescu, Bogdan and Myers, Brad},
  booktitle    = {2024 IEEE/ACM 46th International Conference on Software Engineering (ICSE)},
  pages        = {881--881},
  year         = {2024},
  organization = {IEEE Computer Society}
}

@inproceedings{wei2023copiloting,
  title     = {Copiloting the copilots: Fusing large language models with completion engines for automated program repair},
  author    = {Wei, Yuxiang and Xia, Chunqiu Steven and Zhang, Lingming},
  booktitle = {Proceedings of the 31st ACM Joint European Software Engineering Conference and Symposium on the Foundations of Software Engineering},
  pages     = {172--184},
  year      = {2023}
}

@inproceedings{xia2023automated,
  title        = {Automated program repair in the era of large pre-trained language models},
  author       = {Xia, Chunqiu Steven and Wei, Yuxiang and Zhang, Lingming},
  booktitle    = {2023 IEEE/ACM 45th International Conference on Software Engineering (ICSE)},
  pages        = {1482--1494},
  year         = {2023},
  organization = {IEEE}
}

@inproceedings{zhou2023hybrid,
  title     = {Hybrid API migration: A marriage of small API mapping models and large language models},
  author    = {Zhou, Bingzhe and Wang, Xinying and Xu, Shengbin and Yao, Yuan and Pan, Minxue and Xu, Feng and Ma, Xiaoxing},
  booktitle = {Proceedings of the 14th Asia-Pacific Symposium on Internetware},
  pages     = {12--21},
  year      = {2023}
}

@article{codebleu,
  title   = {Codebleu: a method for automatic evaluation of code synthesis},
  author  = {Ren, Shuo and Guo, Daya and Lu, Shuai and Zhou, Long and Liu, Shujie and Tang, Duyu and Sundaresan, Neel and Zhou, Ming and Blanco, Ambrosio and Ma, Shuai},
  journal = {arXiv preprint arXiv:2009.10297},
  year    = {2020}
}

@inproceedings{bleu,
  title     = {Bleu: a method for automatic evaluation of machine translation},
  author    = {Papineni, Kishore and Roukos, Salim and Ward, Todd and Zhu, Wei-Jing},
  booktitle = {Proceedings of the 40th annual meeting of the Association for Computational Linguistics},
  pages     = {311--318},
  year      = {2002}
}

@article{llamapaper,
  title   = {The llama 3 herd of models},
  author  = {Dubey, Abhimanyu and Jauhri, Abhinav and Pandey, Abhinav and Kadian, Abhishek and Al-Dahle, Ahmad and Letman, Aiesha and Mathur, Akhil and Schelten, Alan and Yang, Amy and Fan, Angela and others},
  journal = {arXiv preprint arXiv:2407.21783},
  year    = {2024}
}

@misc{openaimodels,
  author       = {{OpenAI}},
  title        = {{OpenAI} models},
  howpublished = {https://platform.openai.com/docs/models},
  lastaccessed = {March 22, 2024},
  year         = {2024}
}

@misc{libclick,
  author       = {Pallets},
  title        = {click},
  howpublished = {https://pypi.org/project/click},
  lastaccessed = {March 22, 2024},
  year         = {2024}
}

@misc{flasksocketio,
  author       = {{Flask}},
  tilte        = {Flask-socketio documentation},
  howpublished = {https://flask-socketio.readthedocs.io/en/latest/api.html},
  year         = {2024}
}

@inproceedings{seart,
  author    = {O. Dabic and E. Aghajani and G. Bavota},
  booktitle = {2021 2021 IEEE/ACM 18th International Conference on Mining Software Repositories (MSR) (MSR)},
  title     = {Sampling Projects in GitHub for MSR Studies},
  year      = {2021},
  pages     = {560-564},
  doi       = {10.1109/MSR52588.2021.00074},
  url       = {https://doi.ieeecomputersociety.org/10.1109/MSR52588.2021.00074},
  publisher = {IEEE Computer Society},
  address   = {Los Alamitos, CA, USA},
  month     = {5}
}

@online{liblibcst,
  author  = {The LibCST contributors},
  title   = {LibCST},
  year    = {2025},
  url     = {https://libcst.readthedocs.io},
  urldate = {March 24, 2025}
}

@online{libjohnnydep,
  author  = {Wim Jeantine-Glenn},
  title   = {johnnydep},
  year    = {2025},
  url     = {https://pypi.org/project/johnnydep},
  urldate = {March 24, 2025}
}

@article{RefactoringMiner2,
  author  = {Tsantalis, Nikolaos and Ketkar, Ameya and Dig, Danny},
  title   = {RefactoringMiner 2.0},
  journal = {IEEE Transactions on Software Engineering},
  year    = {2022},
  volume  = {48},
  number  = {3},
  pages   = {930-950},
  doi     = {10.1109/TSE.2020.3007722}
}

@article{refmerge,
  title     = {Operation-based refactoring-aware merging: An empirical evaluation},
  author    = {Ellis, Max and Nadi, Sarah and Dig, Danny},
  journal   = {IEEE Transactions on Software Engineering},
  volume    = {49},
  number    = {4},
  pages     = {2698--2721},
  year      = {2022},
  publisher = {IEEE}
}

@article{di2025deepmig,
  title     = {DeepMig: A transformer-based approach to support coupled library and code migrations},
  author    = {Di Rocco, Juri and Nguyen, Phuong T and Di Sipio, Claudio and Rubei, Riccardo and Di Ruscio, Davide and Di Penta, Massimiliano},
  journal   = {Information and Software Technology},
  volume    = {177},
  pages     = {107588},
  year      = {2025},
  publisher = {Elsevier}
}

@article{xia2025demystifying,
  title={Demystifying llm-based software engineering agents},
  author={Xia, Chunqiu Steven and Deng, Yinlin and Dunn, Soren and Zhang, Lingming},
  journal={Proceedings of the ACM on Software Engineering},
  volume={2},
  number={FSE},
  pages={801--824},
  year={2025},
  publisher={ACM New York, NY, USA}
}

@misc{react2023,
  title         = {ReAct: Synergizing Reasoning and Acting in Language Models},
  author        = {Yao, Shunyu and Zhao, Jeffrey and Yu, Dian and Du, Nan and Shafran, Izhak and Narasimhan, Karthik and Cao, Yuan},
  year          = {2023},
  eprint        = {2210.03629},
  archiveprefix = {arXiv},
  primaryclass  = {cs.CL},
  doi           = {10.48550/arXiv.2210.03629},
  url           = {https://arxiv.org/abs/2210.03629}
}
